\documentclass[a4paper,oneside,11pt]{article}
\usepackage{cprform}
\usepackage{amsmath,amstext,amsfonts,amsbsy,amssymb,amscd,epsfig,lscape,graphics}
\usepackage{appendix}
\usepackage{hyperref}

\usepackage{subfigure}
\usepackage{times}
\usepackage{float}

\newcommand{\Dslash}{\relax{\kern+.25em / \kern-.70em D}}

\newcommand{\Real}{\relax{\mathsf{\Gamma\kern-.35em R}}}
\newcommand{\Int}{\relax{\mathsf{Z\kern-.40em Z}}}

\newcommand{\obar}[1]{\kern3pt\overline{\kern-2pt #1\kern-0pt}\kern1pt}

\newcommand{\corrbar}[1]{\kern3pt\overline{\kern-2pt #1\kern-0pt}\kern1pt}

\newcommand{\oVApAVren}[1]{\kern3pt\overline{\kern-2pt #1\kern-0pt}\kern1pt_{\rm\scriptscriptstyle VA+AV;s}}

\newcommand{\zbar}{\kern3pt\overline{\kern-2pt Z\kern-0pt}\kern1pt}

\newcommand{\zbarVApAV}[1]{\kern3pt\overline{\kern-2pt Z\kern-0pt}\kern1pt_{\rm\scriptscriptstyle VA+AV #1}}

\newcommand{\LSB}{\raisebox{-0.3ex}{\mbox{\LARGE$\left[\right.$}}}
\newcommand{\RSB}{\raisebox{-0.3ex}{\mbox{\LARGE$\left.\right]$}}}

\def\gsim{\mathrel{\raise2pt\hbox to 8pt{\raise -5pt\hbox{$\sim$}\hss{$>$}}}}
\def\rsim{\mathrel{\raise2pt\hbox to 8pt{\raise -5pt\hbox{$\sim$}\hss{$>$}}}}
\def\lsim{\mathrel{\raise2pt\hbox to 8pt{\raise -5pt\hbox{$\sim$}\hss{$<$}}}}
\newcommand{\be}{\begin{equation}}
\newcommand{\ee}{\end{equation}}

\begin{document}

\bibliographystyle{mybibstyle}


\begin{titlepage}

\vspace*{-30truemm}
\begin{flushright}
\hspace*{-1cm} CERN-PH-TH/2013-173, IFIC/13-47, LPT-ORSAY 13-59, \\ LTH 982, MITP/13-045, RM3-TH/13-6, ROM2F/2013/12
 
\end{flushright}

\vspace*{5truemm}
\centerline{\Large \bf  B-physics from $\mathbf{N_f=2}$ tmQCD: the Standard Model and beyond} 

\vskip 5 true mm
\centerline{\bigrm  N. Carrasco$^{(a)}$, M. Ciuchini$^{(b)}$, P. Dimopoulos$^{(c,d)}$, R. Frezzotti$^{(d,e)}$, V. Gim\'enez$^{(a)}$, 
 }  
\centerline{\bigrm  G. Herdoiza$^{(f)}$, V. Lubicz$^{(g,b)}$,  C. Michael$^{(h)}$, E. Picca$^{(g)}$, 
G.C. Rossi$^{(d,e)}$,}
\centerline{\bigrm  F. Sanfilippo$^{(i)}$, A.~Shindler$^{(j)}$, L. Silvestrini$^{(k)}$, 
S. Simula$^{(b)}$, C. Tarantino$^{(g,b)}$}

\vspace*{1truemm}
\begin{figure}[!h]
  \begin{center}
    \includegraphics[scale=0.70]{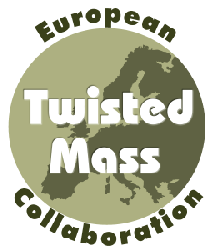}
 \end{center}
\end{figure}

\vskip 1 true mm
\vskip 0 true mm
\centerline{\it $^{(a)}$  Departament de F\'{\i}sica Te\`orica and IFIC, Univ. de Val\`encia-CSIC}
\centerline{\it Dr.~Moliner 50, E-46100 Val\`encia, Spain,}
\vskip 2 true mm
\centerline{\it $^{(b)}$ INFN, Sezione di Roma Tre}
\centerline{\it c/o Dipartimento di Fisica, Universit\`a  Roma Tre}
\centerline{\it Via della Vasca Navale 84, I-00146 Rome, Italy,}
\vskip 2 true mm
\centerline{\it $^{(c)}$ Centro Fermi - Museo Storico della Fisica e Centro Studi e Ricerche Enrico Fermi}
\centerline{\it Compendio del Viminale,  Piazza del Viminale 1 I–00184 Rome, Italy, }
\vskip 2 true mm
\centerline{\it $^{(d)}$ Dipartimento di Fisica, Universit\`a di Roma ``Tor Vergata''}
\centerline{\it Via della Ricerca Scientifica 1, I-00133 Rome, Italy,}
\vskip 2 true mm
\centerline{\it $^{(e)}$ INFN, Sezione di ``Tor Vergata"}
\centerline{\it c/o Dipartimento di Fisica, Universit\`a di Roma ``Tor Vergata''}
\centerline{\it Via della Ricerca Scientifica 1, I-00133 Rome, Italy,}
\vskip 2 true mm
\centerline{\it $^{(f)}$ PRISMA Cluster of Excellence, Institut f{\"u}r Kernphysik}
\centerline{\it Johannes Gutenberg-Universit{\"a}t, D-55099 Mainz, Germany,}
\vskip 2 true mm
\centerline{\it $^{(g)}$ Dipartimento di Fisica, Universit\`a  Roma Tre}
\centerline{\it Via della Vasca Navale 84, I-00146 Rome, Italy,}
\vskip 2 true mm
\centerline{\it $^{(h)}$ Theoretical Physics Division, Department of Mathematical Sciences}
\centerline{\it The University of Liverpool, Liverpool L69 3BX, UK,}
\vskip 2 true mm
\centerline{\it $^{(i)}$ Laboratoire de Physique Th\'eorique (B\^{a}t. 210),}
\centerline{\it Universit\'e Paris Sud, F-91405 Orasay-Cedex, France,}
\vskip 2 true mm
\centerline{\it $^{(j)}$ CERN, Physics Department, }
\centerline{\it 1211 Geneva 23, Switzerland,}
\vskip 2 true mm
\centerline{\it $^{(k)}$ INFN, Sezione di Roma,}
\centerline{\it c/o Dipartimento di Fisica, Sapienza, Universit\`a di Roma}
\centerline{\it Piazzale A. Moro, I-00185 Rome, Italy}



\newpage

\thicktablerule
\vskip 1 true mm
\noindent{\tenbf Abstract}
\vskip 1 true mm
\noindent
{\tenrm  
We present a lattice QCD computation of the $b$-quark mass, the $B$ and
$B_s$ decay constants, the $B$-mixing bag parameters for the full four-fermion operator basis as well as determinations for 
$\xi$ and $f_{Bq}\sqrt{B_i^{(q)}}$ extrapolated to the continuum limit and to the physical pion mass.  
We used $N_f = 2$ twisted mass Wilson fermions at four values of the lattice spacing with pion masses ranging from 
280 to 500 MeV. Extrapolation in the heavy quark mass from the charm to the bottom quark region has been carried out on ratios 
of physical quantities computed at nearby quark masses, exploiting the fact that they have an exactly known infinite mass limit.  
Our results are $m_b(m_b, \overline{\rm{MS}})=4.29(12)$ GeV, $f_{Bs}=228(8)$ MeV,  $f_{B}=189(8)$ MeV and $f_{Bs}/f_B=1.206(24)$. 
Moreover with our results for the bag-parameters we find $\xi=1.225(31)$, $B_1^{(s)}/B_1^{(d)}=1.01(2)$, 
$f_{Bd}\sqrt{\hat{B}_{1}^{(d)}} = 216(10)$ MeV  and 
$f_{Bs}\sqrt{\hat{B}_{1}^{(s)}} = 262(10)$ MeV. 
We also computed the bag parameters for the complete basis of the
four-fermion operators which are required in beyond the SM theories. By using these results for the bag parameters
we are able to provide a refined Unitarity Triangle analysis in the presence of New Physics, improving the bounds coming
from $B_{(s)}-\bar B_{(s)}$ mixing.
}
\vskip 3 true mm
\thicktablerule
\eject
\end{titlepage}

\section{Introduction}
\label{sec:intro}
Physical processes in the $B$-sector are crucial to perform accurate
tests of the Standard Model (SM) and search for possible signals of
New Physics (NP). The experimental accuracy in flavour processes has
recently been significantly increased by the $B$ factories, and new
measurements are being performed, mainly thanks to the remarkable
performance of the dedicated experiment LHCb. On the theoretical side,
lattice computations have entered during the past few years a
precision era, in which the target per cent precision for some of the
relevant hadronic parameters in Flavour Physics is becoming
accessible. In particular, in the study of $B$-physics processes,
there has been substantial progress thanks to alternative lattice
methods and techniques aimed at treating the heavy quarks on the
lattice with controlled systematic uncertainties. For a recent review
of lattice results see Ref.~\cite{Tarantino:2012iw}.  Lattice methods
are irreplaceable for the calculation of the so called golden plated
processes since hadronisation effects are fully under control. They
lead to accurate determinations of decay constants, form factors and
bag-parameters.  For example, the leptonic decays $B \rightarrow \tau
\nu_{\tau}$ and $B_{(s)}^0 \rightarrow \mu^{+} \mu^{-}$ receive
precise input information from lattice determinations of the B-mesons
decay constants, that are necessary for the experimental results to
acquire their full physical interpretation. At present, the world
average of the $B$-meson leptonic decays $BR(B \rightarrow \tau
\nu_{\tau})=(1.14\pm 0.22)10^{-4}$~\cite{Beringer:1900zz,Adachi:2012mm,Lees:2012ju},
which is potentially sensitive to NP effects already at
tree level, turns out to be in agreement with the SM
prediction $BR(B \rightarrow \tau
\nu_{\tau})_\mathrm{SM}=(0.81\pm 0.07)10^{-4}$~\cite{Bona:2009cj,utfitwebpage},
and also the recent measurements of
the $B_{s}^0 \rightarrow \mu^{+} \mu^{-}$ decay~\cite{Aaij:2012nna,
  Chatrchyan:2013bka, LHCb-PAPER-2013-046,
  CMS-LHCb-combination-July2013} have given a first, remarkable
evidence of SM consistency~\cite{Buras:2012ru,Buras:2013uqa,utfitwebpage}.
The neutral $B$-meson mixings, which can only occur at the loop level in
the SM, could be a privileged candidate process for detecting
amplified NP effects, and indeed they play a crucial role in the
Unitarity Triangle (UT) analysis (for recent results, see for example~\cite{Laiho:2012ss,
DescotesGenon:2012rq,Derkach:2013da}).
While in the SM
the frequency of the oscillations, $\Delta M_{B_{(s)}}$, receives the
contribution from a single four-quark operator, the knowledge of the
bag-parameters of the full four-fermion operator basis is required to
get predictions in general NP extensions of the SM. Two of these
B-parameters also enter the SM prediction of the lifetime difference
$\Delta \Gamma_s$ of neutral $B_s$ mesons, which has been recently
measured rather precisely by LHCb~\cite{Aaij:2013oba}.

In this paper we use gauge configurations with $N_f=2$ dynamical quarks at four values of the lattice spacing, generated by European 
Twisted Mass Collaboration (ETMC), to obtain the continuum limit results for a number of physical quantities that are 
relevant for $B$--Physics. These are the $b$-quark mass $m_b$, the pseudoscalar decay constants $f_{B}$ and $f_{B_s}$, 
and the bag-parameters of the full basis of $\Delta B=2$ four-fermion operators. In our previous paper~\cite{Dimopoulos:2011gx}, 
we provided a determination of the $b$-quark mass and of the $B$-meson decay constants obtained by studying the heavy 
quark on the lattice with the so called ratio method, proposed in Ref.~\cite{Blossier:2009hg}. The same strategy is 
also applied in the present study, and the lattice calculation presented here is based on the same set of gauge 
configurations used in our previous study. Nevertheless, several new results and improvements are presented in 
this paper. The main novelties, with respect to Ref.~\cite{Dimopoulos:2011gx}, are the following:
\begin{itemize}
\item[-] We have computed the full set of $B$-parameters for the $\Delta B=2$ four-fermion operators, 
which are relevant for $B$-meson mixings, within and beyond the SM, and for the theoretical predictions of the 
neutral $B$-meson lifetime differences $\Delta \Gamma_{(s)}$. For the full set of $B$-parameters, this is the 
first lattice calculation which takes into account the effect of dynamical quarks (preliminary results obtained 
with $N_f=2+1$ dynamical quarks have been presented in~\cite{Bouchard:2011xj}).
We have used these results to provide a refined Unitarity Triangle analysis improving the bounds coming from 
$B_{(s)}$-meson mixing constraints on NP.
\item[-] We have computed 2- and 3-point correlation functions by employing optimized smearing techniques. 
Given the temporal extensions of our lattices, while the use of smearing interpolating operators is mandatory 
to extract a signal for the $B$-parameters from the 3-point correlation functions, the suppression of the 
excited states contribution helps in improving also the determination of the $b$-quark mass and the decay constants. 
For this reason, the results obtained in this paper for $m_b$, $f_{B}$ and $f_{B_s}$ should be considered 
as an update of those given in  Ref.~\cite{Dimopoulos:2011gx}.
\item[-] One of the main sources of uncertainty in the determination of the decay constant $f_{B}$ 
in Ref.~\cite{Dimopoulos:2011gx} was due to the chiral extrapolation. In this study, we reduce this uncertainty 
by making use of the observation that the double ratio of decay constants $(f_{B_s}/f_{B})/(f_{K}/f_{\pi})$ 
exhibits a much smoother chiral behavior with respect to $f_{B_s}/f_{B}$ itself, due to a large  numerical 
cancellation of the corresponding chiral logarithms~\cite{Becirevic:2002mh}. Therefore, also in this respect, 
the present determinations of $f_{B_s}/f_{B}$ and of $f_{B}$ represent an improvement over the results of Ref.~\cite{Dimopoulos:2011gx}. 
\end{itemize}

The plan of this paper is as follows. 
In Section~\ref{sec:pheno}, based on the results of this work for the $\Delta B=2$ bag parameters, 
we discuss the implications for NP of our updated Unitarity Triangle analysis.
In Section~\ref{sec:lat_simul} we give information about lattice 
simulation details and we describe the techniques that have been used in this work. 
In Sections~\ref{sec:mb_fB_fBs} and \ref{sec:B_xi} we present and apply our strategy, 
namely the ratio method, in order to get continuum limit determinations for the $b$-quark mass, pseudoscalar 
decay constants of the $B$ and $B_s$ mesons, and the bag parameters for the full four-fermion operator basis,
as well as  other interesting quantities like  $\xi$ and $f_{Bq}\sqrt{B_i^{(q)}}$ ($i=1, \ldots, 5$ and $q=d/s$). 
In Section~\ref{sec:summary_final_results} we summarise the final results and discuss our error budget. 
We also provide some comparison plots between our numbers and those obtained by other lattice collaborations.

For reader's convenience we immediately give here our final results. 
For each quantity the quoted error corresponds to the total uncertainty which is the sum in quadrature of the 
statistical and systematic error. 

The result for the $b$-quark mass is given in the $\overline{\rm{MS}}$ scheme at the scale of its own value, $m_b$. We perform the 
running up to $m_b$ using either $N_f$ = 2 or $N_f=$ 4, we take the average over the two results and we consider their half 
difference as a systematic uncertainty\footnote{See Section~\ref{sec:summary_final_results} and Eqs~(\ref{eq:mb_Nf2}) and 
(\ref{eq:mb_Nf4}).}. We get: 
\begin{eqnarray}\label{eq:intro_mb_Nf4}
 m_b(m_b, \overline{\rm{MS}}) &=& 4.29(12) \,\,\, {\rm GeV},
\end{eqnarray}

Our results for the pseudoscalar decay constants for the $B$ and $B_s$ mesons as well as for their ratio are:
\begin{eqnarray}
f_{Bs} &=& 228(8)\,\,\,\, {\rm MeV}  \label{eq:intro_fBs} \\
f_{B}  &=& 189(8)\,\,\,\, {\rm MeV} \label{eq:intro_fB} \\
\dfrac{f_{Bs}}{f_B} &=& 1.206(24) \label{eq:intro_fratio} 
\end{eqnarray}

As a by-product of our work we have computed the decay constants for the   $D_s$ and $D$ mesons as well as their ratio. They read:
\begin{equation}\label{eq:fD_Ds_ratio}
f_{Ds} = 250(7) ~{\rm MeV}, \,\,\,\, f_{D} = 208(7) ~{\rm MeV}, \,\,\,\, f_{Ds}/f_{D} = 1.201(21) .
\end{equation}

\noindent The most general form of the $\Delta F=2$ effective weak Hamiltonian is
\begin{equation}
{\cal H}_{\rm{eff}}^{\Delta F=2} =\frac{1}{4}\, \sum_{i=1}^{5} C_i {\cal O}_i 
+\frac{1}{4}\, \sum_{i=1}^{3} \tilde{C}_i \tilde{{\cal O}}_i, \, 
\label{eq:Heff}
\end{equation}
where in the so-called SUSY basis~(\cite{Gabbiani:1996hi},~\cite{Bagger:1997gg})
the four-fermion operators ${\cal O}_i$ and $\tilde{{\cal O}}_i$  read
\begin{eqnarray}\label{def_Oi}
{\cal O}_1 &=& [\overline{h}^\alpha \gamma_\mu (1-\gamma_5)q^\alpha][\overline{h}^\beta \gamma_\mu (1-\gamma_5)q^\beta], \nonumber \\
{\cal O}_2 &=& [\overline{h}^\alpha (1-\gamma_5)q^\alpha][\overline{h}^\beta  (1-\gamma_5)q^\beta], \nonumber \\
{\cal O}_3 &=& [\overline{h}^\alpha (1-\gamma_5)q^\beta][\overline{h}^\beta  (1-\gamma_5)q^\alpha],  \\
{\cal O}_4 &=& [\overline{h}^\alpha (1-\gamma_5)q^\alpha][\overline{h}^\beta  (1+\gamma_5)q^\beta], \nonumber \\
{\cal O}_5 &=& [\overline{h}^\alpha (1-\gamma_5)q^\beta][\overline{h}^\beta  (1+\gamma_5)q^\alpha], \nonumber
\end{eqnarray}
\begin{eqnarray}
\tilde{{\cal O}}_1 &=& [\overline{h}^\alpha \gamma_\mu (1+\gamma_5)q^\alpha][\overline{h}^\beta \gamma_\mu (1+\gamma_5)q^\beta], \nonumber \\
\tilde{{\cal O}}_2 &=& [\overline{h}^\alpha (1+\gamma_5)q^\alpha][\overline{h}^\beta  (1+\gamma_5)q^\beta],  \\
\tilde{{\cal O}}_3 &=& [\overline{h}^\alpha (1+\gamma_5)q^\beta][\overline{h}^\beta  (1+\gamma_5)q^\alpha] . \nonumber 
\end{eqnarray}
For the neutral $B$-meson system $h \equiv b$ and $q \equiv d\, {\rm or}\, s$ with $\alpha$ and $\beta$ denoting color indices. 
Spin indices are implicitly contracted within square brackets.
The Wilson coefficients $C_i$ and $\tilde{C}_i$  have an implicit renormalization  scale dependence which
is compensated by the  scale dependence of  the renormalization constants of the corresponding  operators.

Notice that the parity-even parts of the operators ${\cal O}_i$ and $\tilde{{\cal O}}_i$ are identical.
Due to parity conservation in strong interactions, for the study of $\overline B^{0}_{q}-B^{0}_{q}$ oscillations it is  sufficient
to consider only the matrix elements $\langle \overline{B}^{0}_{q} | O_i | B^{0}_{q} \rangle$, 
where by $O_i$ ($i=1, \ldots, 5$) we denote the parity-even components of the operators~(\ref{def_Oi}).
We recall that in the SM only the matrix element of the operator ${\cal O}_1$ is relevant.

The bag parameters, $B_i$ ($i=1, \ldots, 5$), provide the value of four-fermion matrix elements in units
of the magnitude of their vacuum saturation approximation. More explicitly, they are defined
by the equations~\cite{Allton:1998sm, Bpar:SPQR1}
\begin{eqnarray}
 \langle \overline{B}^{0}_{q} | O_1(\mu) | B^{0}_{q} \rangle &=& {\cal C}_1\, B_1^{(q)}(\mu) ~ 
 m_{B_{q}}^2 f_{B_{q}}^2  \label{B1} \\
 \langle \overline{B}^{0}_{q} | O_i(\mu) | B^{0}_{q} \rangle &=& {\cal C}_i\, B_i^{(q)}(\mu) ~ 
\LSB \dfrac{ m_{B_{q}}^{2} f_{B_{q}}}{ m_b (\mu) + m_{q}(\mu)} \RSB^2  ~~~ {\rm for} ~~ i=2,\ldots, 5, 
\label{Bi} 
\end{eqnarray}
with ${\cal C}_i=(8/3,\, -5/3,\, 1/3,\, 2,\, 2/3)$.

In Table~\ref{tab:intro_bag_res_rimom} we collect our results for the bag parameters (see Eqs~(\ref{B1}) and
and (\ref{Bi})) of the full operator basis (i.e.\ the parity even componenents
of the operators in Eq.~(\ref{def_Oi})) in the RI/MOM scheme at the scale of 
the $b$-quark mass (Eq.~(\ref{eq:intro_mb_Nf4})).
In Table~\ref{tab:intro_bag_res} we gather results for the bag parameters 
expressed in the $\overline{\rm{MS}}$ scheme of Ref.~\cite{mu:4ferm-nlo} at the scale of the $b$-quark mass.  
Also, in Table~\ref{tab:intro_bag_res_beneke} we  give results for $B_i^{(d/s)}$ with $i=2, 3$,  
expressed in the scheme of Ref.~\cite{Beneke:1998sy} at the scale of the $b$-quark mass.
\begin{table}[!h]
\begin{center}
\begin{tabular}{|c|c|c|c|c|}
\hline
\multicolumn{5}{|c|}{(RI/MOM, $m_b$)}\tabularnewline
\hline
\hline
$B_{1}^{(d)}$ & $B_{2}^{(d)}$ & $B_{3}^{(d)}$ & $B_{4}^{(d)}$ & $B_{5}^{(d)}$\tabularnewline
\hline
\hline
0.84(4) & 0.88(4) & 1.10(18) & 1.12(7) & 1.89(16)\tabularnewline
\hline
\hline
$B_{1}^{(s)}$ & $B_{2}^{(s)}$ & $B_{3}^{(s)}$ & $B_{4}^{(s)}$ & $B_{5}^{(s)}$\tabularnewline
\hline
\hline
0.85(3) & 0.91(4) & 1.12(16) & 1.10(5) & 2.02(15)\tabularnewline
\hline
\end{tabular}
\caption{Continuum limit results of $B_i^{(d)}$ and $B_i^{(s)}$ ($i=1, \ldots, 5$), 
renormalized in the RI/MOM scheme at the scale of the $b$-quark mass. 
} 
\label{tab:intro_bag_res_rimom}
\end{center}
\end{table}   

\begin{table}[!h]
\begin{center}
\begin{tabular}{|c|c|c|c|c|}
\hline
\multicolumn{5}{|c|}{($\overline {\rm{MS}}$--BMU, $m_b$)}\tabularnewline
\hline
\hline
$B_{1}^{(d)}$ & $B_{2}^{(d)}$ & $B_{3}^{(d)}$ & $B_{4}^{(d)}$ & $B_{5}^{(d)}$\tabularnewline
\hline
\hline
0.85(4) & 0.72(3) & 0.88(13) & 0.95(5) & 1.47(12)\tabularnewline
\hline
\hline
$B_{1}^{(s)}$ & $B_{2}^{(s)}$ & $B_{3}^{(s)}$ & $B_{4}^{(s)}$ & $B_{5}^{(s)}$\tabularnewline
\hline
\hline
0.86(3) & 0.73(3) & 0.89(12) & 0.93(4) & 1.57(11)\tabularnewline
\hline
\end{tabular}
\caption{Continuum limit results of $B_i^{(d)}$ and $B_i^{(s)}$ ($i=1, \ldots, 5$), 
renormalized in the $\overline{\rm{MS}}$ scheme of Ref.~\cite{mu:4ferm-nlo} at the scale of the $b$-quark mass. 
} 
\label{tab:intro_bag_res}
\end{center}
\end{table}

\begin{table}[!h]
\begin{center}
\begin{tabular}{|c|c|c|c|}
\hline
\multicolumn{4}{|c|}{($\overline {\rm{MS}}$--BBGLN, $m_b$)}\tabularnewline
\hline
\hline
$B_{2}^{(d)}$ & $B_{3}^{(d)}$ & $B_{2}^{(s)}$ & $B_{3}^{(s)}$ \tabularnewline
\hline
\hline
0.76(3) & 0.87(13) & 0.78(3) & 0.89(12) \tabularnewline
\hline
\end{tabular}
\caption{Continuum limit results of $B_i^{(d)}$ and $B_i^{(s)}$ ($i=2, 3$), 
renormalized in the $\overline{\rm{MS}}$ scheme of Ref.~\cite{Beneke:1998sy} at the scale of the $b$-quark mass. 
} 
\label{tab:intro_bag_res_beneke}
\end{center}
\end{table}   

Our results for the SU(3)-breaking ratios $B_1^{(s)} /B_1^{(d)}$ and
$\xi$ (see Eq.~(\ref{eq:xi_def}))  are
\begin{eqnarray}
\dfrac{B_1^{(s)}}{B_1^{(d)}} &=& 1.01(2) \label{eq:intro_B_rat} \\
\xi &=& 1.225(31) \label{eq:intro_xi} 
\end{eqnarray}

\begin{table}[!h]
\begin{center}
\begin{tabular}{|c|c|c|c|c|c|}
\hline
\multicolumn{6}{|c|}{($\overline {\rm{MS}}$, $m_b$) [MeV]}\tabularnewline
\hline
\hline
$i$ & 1 & 2 & 3 & 4 & 5 \tabularnewline
\hline
\hline
$f_{Bd} \sqrt{B_i^{(d)}}$ & 174(8)  & 160(8) & 177(17) & 185(9) & 229(14)\tabularnewline
\hline
\hline
\hline
$f_{Bs} \sqrt{B_i^{(s)}}$ & 211(8) & 195(7) & 215(17) & 220(9) & 285(14)\tabularnewline
\hline
\end{tabular}
\caption{Continuum limit results of $f_{Bd} \sqrt{B_i^{(d)}}$ and $f_{Bs} \sqrt{B_i^{(s)}}$ ($i=1, \ldots, 5$). 
Bag parameters are expressed in the $\overline{\rm{MS}}$ scheme of Ref.~\cite{mu:4ferm-nlo} at the scale of the $b$-quark mass.  } 
\label{tab:intro_fbag_res}
\end{center}
\end{table}

Finally, in Table~\ref{tab:intro_fbag_res} we collect our results for the quantities 
$f_{Bq} \sqrt{B_i^{(q)}}$ where $q=d, s$ and $i=1, \ldots, 5$.  Again the 
bag parameters are expressed in the $\overline{\rm{MS}}$ scheme of Ref.~\cite{mu:4ferm-nlo} at the scale of the $b$-quark mass.
For convenience we also give our results for the SM relevant quantities  
in which the bag parameters are expressed in the RGI scheme and we have employed in the running $N_f=5$ 
and $\Lambda_{QCD}^{(N_f=5)}$=213(9) MeV~\cite{Bethke:2009jm}.
We get
\begin{eqnarray}\label{eq:fBrgi}
f_{Bd}\sqrt{\hat{B}_{1}^{(d)}} &=& 216(10) ~~\rm{MeV} \\
f_{Bs}\sqrt{\hat{B}_{1}^{(s)}} &=& 262(10)  ~~\rm{MeV}
\end{eqnarray} 

The RGI values of the bag parameters corresponding to the SM four-fermion operators read
\begin{equation} \label{eq:Brgi}
\hat{B}_{1}^{(d)} = 1.30(6),  \,\,\,\,\, \hat{B}_{1}^{(s)} = 1.32(5)
\end{equation}


\section{Model-independent constraints on $\Delta B=2$ operators and NP scale from the UT analysis}
\label{sec:pheno}
The $N_f=2$ results obtained in this work for the bag parameters of the full basis of $\Delta B=2$ four-fermion operators represent the first unquenched determination of these quantities. Besides the lattice studies of $B_1$, which is relevant for $B^0-\bar B^0$ mixing in the SM, a lattice result for $B_2^{(s)}$ has been obtained with $N_f=2+1$ dynamical quarks  in~\cite{Dalgic:2006gp}, while preliminary results for the full basis with $N_f=2+1$ have been presented in~\cite{Bouchard:2011xj}.

$\Delta F=2$ processes provide some of the most stringent constraints on NP generalizations of the SM. 
Several phenomenological analyses of $\Delta F=2$ processes have been performed in the last years, both for 
specific models and in model-independent frameworks~\cite{Bona:2007vi,Ligeti:2010ia,Buras:2010pz,Lenz:2010gu,
Lunghi:2010gv,Adachi:2011cb,Calibbi:2012at,KerenZur:2012fr,Mescia:2012fg,Buras:2012dp,Bertone:2012cu}.
A generalization of the Unitarity Triangle (UT) analysis, which allows for NP effects by including the most significant
flavor constraints on NP available at the time was performed in Ref.~\cite{Bona:2007vi}. The result was
a simultaneous determination of the CKM parameters and the size of NP contributions to
$\Delta F = 2$ processes in the neutral kaon and $B_{(s)}$ meson sectors.
The NP generalization of the UT analysis consists in including in the theoretical parametrization
of the various observables the matrix elements of operators which, though absent
in the SM, may appear in some of its extensions.

In a previous paper~\cite{Bertone:2012cu} we have presented the first ($N_f=2$) unquenched, continuum limit, 
lattice QCD results for the matrix elements
of the operators describing neutral kaon oscillations in extensions of the SM. 
In the same paper we have updated the UT analysis allowing for possible NP effects, improving the 
bounds coming from $K^0-\bar K^0$ mixing constraints.

In a similar way, we present here the $N_f=2$ lattice QCD results for the bag parameters 
of the full basis of $\Delta B =2$ four-fermion operators and we use them in updating the UT analysis beyond the SM.
The new ingredients entering the analysis
are collected in Table~\ref{tab:intro_bag_res}. For all the other input data we use the numbers quoted in
Ref.~\cite{utfitwebpage} in the Winter 2013 analysis.

In the NP-oriented analysis, the relations among experimental observables and the CKM matrix elements 
are extended by taking into consideration the most general form
of the $\Delta F = 2$ effective weak Hamiltonian, given in Eq.~(\ref{eq:Heff}).
In the present analysis we focus on $\Delta B = 2$ processes.
The effective weak Hamiltonian is parameterized
by Wilson coefficients of the form
\begin{equation}
  C_i (\Lambda) = \frac{F_i L_i}{\Lambda^2}\, ,\qquad i=2,\ldots,5\, ,
  \label{eq:cgenstruct}
\end{equation}
where $F_i$ is the (generally complex) relevant NP flavor coupling,
$L_i$ is a (loop) factor which depends on the interactions that
generate $C_i(\Lambda)$, and $\Lambda$ is the scale of NP, i.e.\ the
typical mass of new particles mediating $\Delta B=2$ transitions. For
a generic strongly interacting theory with an unconstrained flavor
structure, one expects $F_i \sim L_i \sim 1$, so that the
phenomenologically allowed range for each of the Wilson coefficients
can be immediately translated into a lower bound on
$\Lambda$. Specific assumptions on the flavor structure of NP
correspond to special choices of the $F_i$ functions.

Following Ref.~\cite{Bona:2007vi}, in deriving the lower bounds on
the NP scale $\Lambda$, we assume $L_i = 1$, that corresponds to
strongly-interacting and/or tree-level coupled NP. Two other
interesting possibilities are given by loop-mediated NP contributions
proportional to either $\alpha_s^2$ or $\alpha_W^2$. The first case
corresponds for example to gluino exchange in the minimal
supersymmetric SM. The second case applies to all models with SM-like
loop-mediated weak interactions. To obtain the lower bound on
$\Lambda$ entailed by loop-mediated contributions, one simply has to
multiply the bounds we quote in the following by
$\alpha_s(\Lambda)\sim 0.1$ or $\alpha_W \sim 0.03$.

The results for the upper bounds on the $|C^{B_d}_i|$ and $|C^{B_s}_i|$ coefficients and
the corresponding lower bounds on the NP scale $\Lambda$ are collected
in Tables~\ref{tab:Bd} and~\ref{tab:Bs}, where they are compared to the previous results
of Ref.~\cite{Bona:2007vi}. The superscript $B_d$ or $B_s$ is to recall that we
are reporting the bounds coming from the $B_d$- and $B_s$-meson sectors we are here
analyzing.
The constraints on the Wilson coefficients of the
non-standard operators and, consequently, on the NP scale turn out to be significantly 
more stringent than in Ref.~\cite{Bona:2007vi}, in particular for the $B_s$ sector.
Both experimental and theoretical inputs have been updated with
respect to Ref.~\cite{Bona:2007vi} (see Ref.~\cite{utfitwebpage}).
We notice, in particular, that the input values used in Ref.~\cite{Bona:2007vi} for 
$B_i^{(d/s)}$ were obtained in Ref.~\cite{Bpar:SPQR1} in the quenched approximation, 
at rather large pion masses and
at only one lattice spacing ($a \sim 0.1$ fm).  

We observe that the
analysis is performed (as in~\cite{Bona:2007vi}) by switching on one
coefficient at the time in each sector, thus excluding the possibility
of having accidental cancellations among the contributions of
different operators.  Therefore, the reader should keep in mind that
the bounds may be weakened if, instead, some accidental cancellation
occurs.

\begin{figure}[htbp]
  \begin{minipage}[b]{0.5\linewidth}
    \centering
    \includegraphics[width=\linewidth]{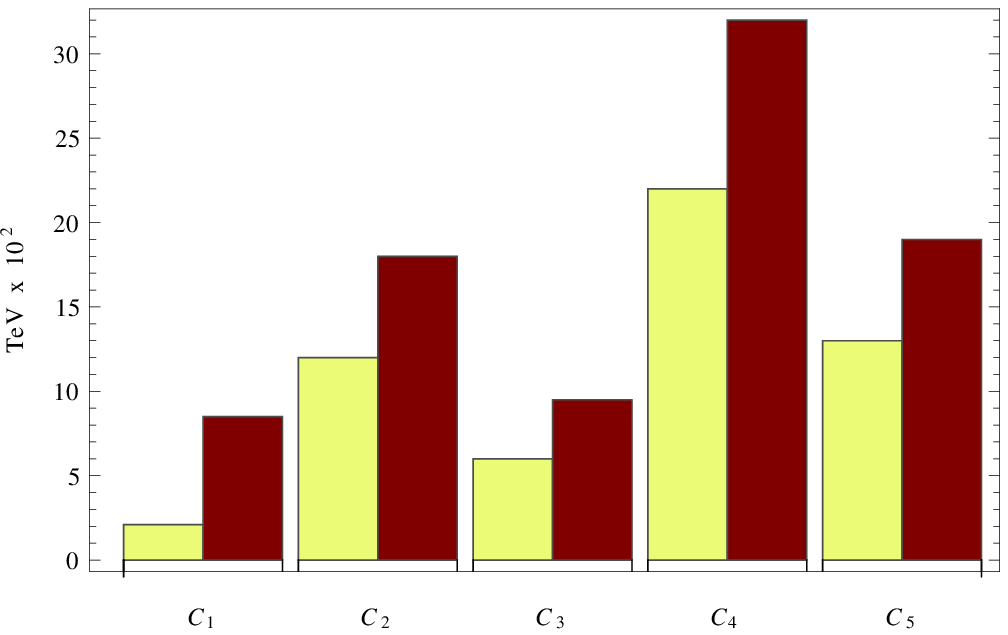}
    \caption{\sl The lower bounds on the NP scale, provided by the constraints on  
$|C^{B_d}_i|$ ($i=1,\ldots,5$) for generic NP flavor structure, are
shown as brown bars. For comparison, we plot  the bounds
of Ref.~\cite{Bona:2007vi} as yellow bars.}
    \label{fig:Bd}
  \end{minipage}
  \hspace{0.5cm}
  \begin{minipage}[b]{0.5\linewidth}
    \centering
    \includegraphics[width=\linewidth]{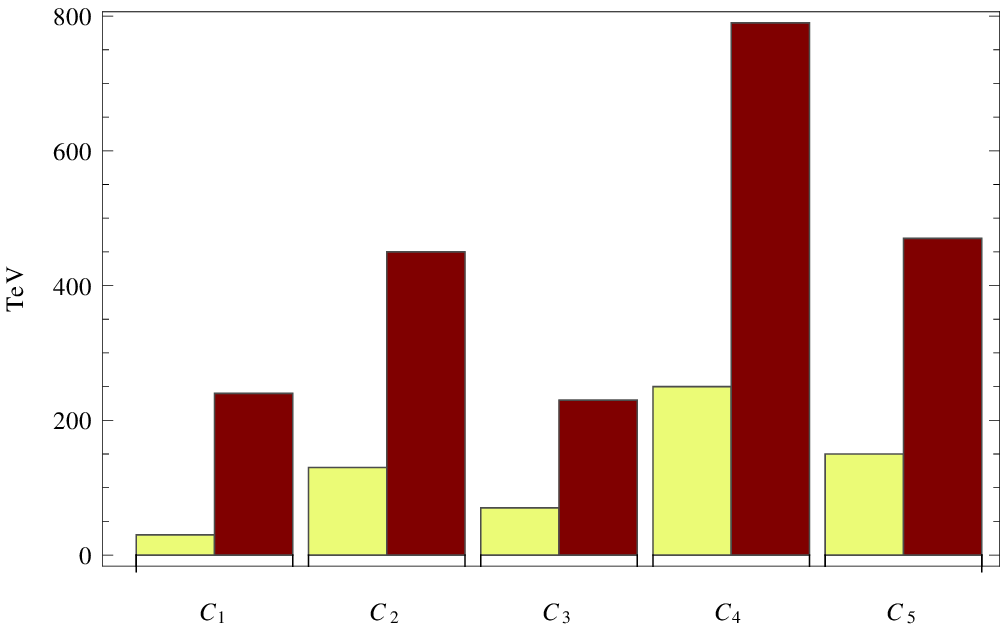}
    \caption{\sl The lower bounds on the NP scale, provided by the constraints on  
$|C^{B_s}_i|$ ($i=1,\ldots,5$) for generic NP flavor structure, are
shown as brown bars. For comparison, we plot  the bounds
of Ref.~\cite{Bona:2007vi} as yellow bars.}
    \label{fig:Bs}
  \end{minipage}
\end{figure}

In Figs.~\ref{fig:Bd} and~\ref{fig:Bs} we show the comparison between the lower bounds
on the NP scale obtained for the case of a generic strongly
interacting NP with generic flavor structure by the constraints on the
$|C^{B_d}_i|$ and $|C^{B_s}_i|$ coefficients coming from the present generalized UT
analysis, and the previous results of Ref.~\cite{Bona:2007vi}.

Comparing with the results of the UT--analysis in Ref~\cite{Bertone:2012cu}, we notice that
(at least for generic NP models with unconstrained flavour structure) the bounds
on the NP scale coming from $K^0$--$\bar K^0$ matrix elements turn out to be 
the most stringent ones.

\begin{table}
\parbox{.45\linewidth}{
\centering
\begin{tabular}{|@{}ccc|}
\hline\hline
 & $95\%$ upper limit  &
Lower limit on $\Lambda$ \\
&(GeV$^{-2}$) &
 (TeV)\\
\hline
\phantom{A} $|C^{B_d}_1|$ & $1.4 \cdot 10^{-12}$ & $8.5 \cdot 10^{2}$ \\
\phantom{A} $|C^{B_d}_2|$ & $3.0 \cdot 10^{-13}$ & $1.8 \cdot 10^{3}$  \\
\phantom{A} $|C^{B_d}_3|$ & $1.1 \cdot 10^{-12}$ & $9.5 \cdot 10^{2}$  \\
\phantom{A} $|C^{B_d}_4|$ & $9.5 \cdot 10^{-14}$ & $3.2 \cdot 10^{3}$  \\
\phantom{A} $|C^{B_d}_5|$ & $2.7 \cdot 10^{-13}$ & $1.9 \cdot 10^{3}$ \\
\hline
\hline
\phantom{A} $|C^{B_d}_1|$ & $2.3 \cdot 10^{-11}$ & $2.1 \cdot 10^{2}$ \\
\phantom{A} $|C^{B_d}_2|$ & $7.2 \cdot 10^{-13}$ & $1.2 \cdot 10^{3}$  \\
\phantom{A} $|C^{B_d}_3|$ & $2.8 \cdot 10^{-12}$ & $6.0 \cdot 10^{2}$ \\
\phantom{A} $|C^{B_d}_4|$ & $2.1 \cdot 10^{-13}$ & $2.2 \cdot 10^{3}$  \\
\phantom{A} $|C^{B_d}_5|$ & $6.0 \cdot 10^{-13}$ & $1.3 \cdot 10^{3}$  \\
\hline
\hline
\end{tabular}
\caption {$95\%$ upper bounds for the $|C^{B_d}_i|$ coefficients
  and the corresponding lower bounds on the NP scale, $\Lambda$, for
  a generic strongly interacting NP with generic flavor structure ($L_i=F_i=1)$. In the
  lower panel the results of~\cite{Bona:2007vi} are
  displayed for comparison.} 
\label{tab:Bd}
}
\hfill
\parbox{.45\linewidth}{
\centering
\begin{tabular}{|@{}ccc|}
\hline\hline
 & $95\%$ upper limit  &
Lower limit on $\Lambda$ \\
&(GeV$^{-2}$) &
 (TeV)\\
\hline
\phantom{A} $|C^{B_s}_1|$ & $1.8 \cdot 10^{-11}$ & $2.4 \cdot 10^{2}$ \\
\phantom{A} $|C^{B_s}_2|$ & $4.9 \cdot 10^{-12}$ & $4.5 \cdot 10^{2}$  \\
\phantom{A} $|C^{B_s}_3|$ & $1.8 \cdot 10^{-11}$ & $2.3 \cdot 10^{2}$  \\
\phantom{A} $|C^{B_s}_4|$ & $1.6 \cdot 10^{-12}$ & $7.9 \cdot 10^{2}$  \\
\phantom{A} $|C^{B_s}_5|$ & $4.5 \cdot 10^{-12}$ & $4.7 \cdot 10^{2}$ \\
\hline
\hline
\phantom{A} $|C^{B_s}_1|$ & $1.1 \cdot 10^{-9}$ &  $3.0 \cdot 10^{1}$ \\
\phantom{A} $|C^{B_s}_2|$ & $5.6 \cdot 10^{-11}$ & $1.3 \cdot 10^{2}$  \\
\phantom{A} $|C^{B_s}_3|$ & $2.1 \cdot 10^{-10}$ & $7.0 \cdot 10^{1}$ \\
\phantom{A} $|C^{B_s}_4|$ & $1.6 \cdot 10^{-11}$ & $2.5 \cdot 10^{2}$  \\
\phantom{A} $|C^{B_s}_5|$ & $4.5 \cdot 10^{-11}$ & $1.5 \cdot 10^{2}$  \\
\hline
\hline
\end{tabular}
\caption {$95\%$ upper bounds for the $|C^{B_s}_i|$ coefficients
  and the corresponding lower bounds on the NP scale, $\Lambda$, for
  a generic strongly interacting NP with generic flavor structure ($L_i=F_i=1)$. In the
  lower panel the results of~\cite{Bona:2007vi} are
  displayed for comparison.} 
\label{tab:Bs}
}
\end{table}

\clearpage

\section{Lattice setup and simulation details}
\label{sec:lat_simul}

The ETM Collaboration has generated $N_f=2$ gauge configuration ensembles at four values of the inverse bare gauge coupling,
 $\beta$, and at a number of light sea quark masses. 
The values of the simulated lattice spacings lie in the interval [0.05, 0.1] fm. 
Dynamical quark simulations have been performed using the tree-level improved Symanzik gauge action~\cite{Weisz:1982zw}
and the Wilson twisted mass action~\cite{Frezzotti:2000nk} tuned to maximal twist~\cite{FrezzoRoss1}.
Bare quark mass parameters, corresponding to a degenerate bare mass value of the $u/d$ quark, 
are chosen so as to have the light pseudoscalar mesons (``pions") in the range 
$280 \leq m_{\rm PS} \leq 500$~MeV. A list of the simulated charged pseudoscalar meson masses is given in~\cite{Blossier:2010cr}. 
Discussion about the computation of the neutral pseudoscalar meson mass using twisted mass fermions has been presented 
in~\cite{Boucaud:2008xu, Baron:2009wt}. 
More details on the action and our $N_f=2$ gauge ensembles can be found 
in Refs.~\cite{Boucaud:2008xu, Baron:2009wt, Boucaud:2007uk}. 
We stress that the use of maximally twisted fermionic action offers the advantage  of automatic O$(a)$  improvement for  
all the interesting physical observables computed on the lattice~\cite{FrezzoRoss1}. 
 
In the present work we treat the strange and the charm quarks as quenched. We have computed 2- and 3-point correlation functions
using valence quark masses whose range is extended from the light sea quark mass up to 2.5-3 times the charm quark mass. 
Simulation details are given in Table~\ref{tab:runs}, where  $\mu_{\ell}$, $\mu_{s}$ and $\mu_h$  indicate  the 
bare light, strange-like and heavy ({\it i.e.} charm-like and heavier) valence quark masses respectively. 

\begin{table}[!h]
\hspace*{0.1cm}
\begin{tabular}{||c||c||c||c||c||}
\hline
$\beta$  & $(L/a,~T/a)$    &  $a \mu_\ell$ &   $a \mu_s$    & $a \mu_h$ \\ \hline\hline
3.80 &    (24,~48)   &  0.0080, 0.0110   & 0.0175, 0.0194 & 0.1982, 0.2331, 0.2742, 0.3225, 0.3793,  \\
     &               &                   &   0.0213       & 0.4461, 0.5246, 0.6170, 0.7257, 0.8536 \\ \hline \hline
3.90 &    (32,~64)   &  0.0030, 0.0040   & 0.0159, 0.0177 & 0.1828, 0.2150, 0.2529, 0.2974, 0.3498,\\
     &               &                   &  0.0195        & 0.4114, 0.4839, 0.5691, 0.6694, 0.7873  \\ 
     &    (24,~48)   &  0.0040, 0.0064   &                &                         \\
     &               &  0.0085, 0.0100   &                &                         \\ \hline \hline
4.05 &    (32,~64)   &  0.0030, 0.0080   & 0.0139, 0.0154 & 0.1572, 0.1849, 0.2175, 0.2558, 0.3008,\\
     &               &                   & 0.0169         & 0.3538, 0.4162, 0.4895, 0.5757, 0.6771 \\ \hline \hline
4.20 &    (48,~96)   &  0.0020           & 0.0116, 0.0129 & 0.13315, 0.1566, 0.1842, 0.2166, 0.2548,  \\
     &               &                   & 0.0142          & 0.2997, 0.3525, 0.4145, 0.4876, 0.5734 \\ 
     &    (32,~64)   &  0.0065           &                &  \\
     &               &                   &                &  \\ \hline
\hline
\end{tabular}
\caption{Simulation details for correlator computation  at four values of the gauge coupling $\beta = 3.80, 3.90, 4.05 
~\rm{and}~ 4.20$. 
The quantities $a\mu_{\ell}$, $a\mu_{s}$ and $a\mu_h$ stand for  
light, strange-like and heavy ({\it i.e.} charm-like and heavier) bare valence quark mass values respectively, expressed in lattice units.}   
\label{tab:runs}
\end{table}

We have set light valence quark mass values equal 
to the light sea ones, $a\mu_{\ell} = a\mu_{sea}$. 
Renormalised quark masses, $\mu^R$, are obtained by the bare ones
using the renormalisation constant (RC)
$Z_{\mu} = Z_P^{-1}$,  $\mu^{R}=\mu / Z_P$~\cite{Frezzotti:2000nk,Frezzotti:2004wz}.
The values for $Z_P$ at the three coarsest lattice spacings have been computed in~\cite{Constantinou:2010gr} using RI-MOM 
techniques. Following the same method we have also computed $Z_P$ at the finest value 
of the lattice spacing corresponding to $\beta=4.20$. All $Z_P$ expressed in the $\overline{\rm{MS}}$ scheme at 2 GeV 
are gathered in Appendix C of~\cite{Bertone:2012cu}. Here we use the corresponding $Z_P$  values at the scale of 3 GeV, 
which is in the region of momenta  directly accessible in the RI-MOM calculation of Ref.~\cite{Constantinou:2010gr}. 
In Table~\ref{tab:ZP_a-lat} we collect the values of $Z_P$ at each value of $\beta$ as well as the 
corresponding   lattice spacing values.  
The latter have been computed employing
SU(2) (NLO) ChPT formulae to fit  in a combined way our data for the pion mass and decay constant by 
using as an input the experimental value of the pion decay constant~\cite{Blossier:2010cr}. 
\begin{table}[!h]
\begin{center}
\begin{tabular}{|l|c|c|c|c|}
\hline 
$\beta$ & 3.80 & 3.90 & 4.05 & 4.20\tabularnewline
\hline
\hline 
$Z_{P}$ & 0.447(13) & 0.473(8) & 0.516(5) & 0.539(5)\tabularnewline
\hline 
$a$ (fm)& 0.098(4) & 0.085(2) & 0.067(2) & 0.054(1)\tabularnewline
\hline
\end{tabular}
\caption{$Z_P$ in the $\overline{\rm{MS}}$ scheme at 3 GeV and lattice spacing 
values at the four values of the inverse gauge coupling 
$\beta$.} 
\label{tab:ZP_a-lat}
\end{center}
\end{table}  
Moreover in~\cite{Blossier:2010cr}, we have  computed the values for the light, strange and charm quark mass. In the 
$\overline{\rm{MS}}$ scheme at 3 GeV they read\footnote{Throughout this paper when we  use the ``overline" notation to the masses 
we mean renormalised quark masses in the $\overline{\rm{MS}}$ scheme at the scale of 3 GeV, 
unless a different renormalisation scale is explicitly indicated.}: 
$\overline{m}_{u/d}=3.3(2)$ MeV, $\overline{m}_{s}=88(5)$ MeV and 
$\overline{m}_{c}=1.05(3)$ GeV. 

We have computed 2- and 3- point correlation  functions by employing smearing techniques on a set of 100-240 independent gauge 
configurations for each ensemble and evaluated statistical errors using the bootstrap method.
Smeared interpolating operators become
mandatory in the cases where relativistic heavy quarks are involved. 
Smearing proves to be beneficial in reducing the coupling of the
interpolating field with the excited states, thus increasing its projection 
onto the lowest energy eigenstate. The usual drawback, {\it i.e.} increase of
the gauge noise due to fluctuations of the links entering in the smeared
fields, is controlled by replacing thin gauge links with APE smeared ones~\cite{Albanese:1987ds}.   
With this technical improvement we can extract heavy-light meson masses
and matrix elements at relatively small temporal separations while keeping
noise-to-signal ratio under control. We employed  
Gaussian smearing~\cite{Gusken:1989qx, Jansen:2008si} for heavy-light meson
interpolating fields at the source and/or the sink. The smeared
field is of the form:
\begin{equation}
\Phi^{\rm{S}}  = (1+6\kappa_{{\rm G}})^{-N_{\rm{G}}} 
 (1 + \kappa_{\rm{G}} a^2 \nabla^2_{\rm{APE}})^{N_G} \Phi^{\rm{L}},
\end{equation}
where $\Phi^{\rm{L}}$ is a standard local source and $\nabla_{\rm{APE}}$ 
is the lattice covariant derivative with APE 
smeared gauge links characterised by the parameters  $\alpha_{\rm{APE}}=0.5$ and
$N_{\rm{APE}}=20$. We have taken $\kappa_{\rm{G}}=4$ and $N_{\rm{G}}=30$. 
We have noticed that in practice we get better overlap with the ground state 
when the source, rather than the sink,
is smeared. Thus 2-point Smeared-Local (SL) correlation functions yield more improved 
plateaux for the lowest energy mass state than Local-Smeared (LS) or Smeared-Smeared (SS) ones.

\begin{figure}[!ht]
\subfigure[]{\includegraphics[scale=0.65,angle=-0]{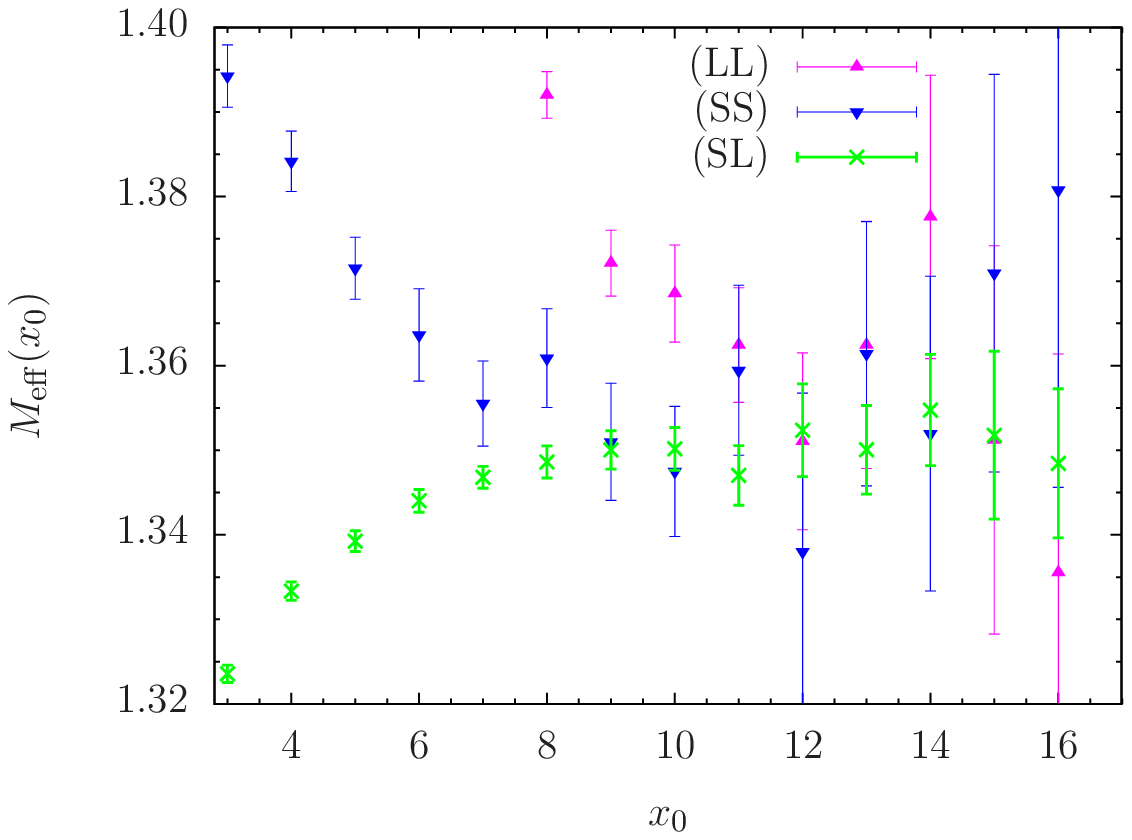}}
\hspace*{0.9cm}
\subfigure[]{\includegraphics[scale=0.65,angle=-0]{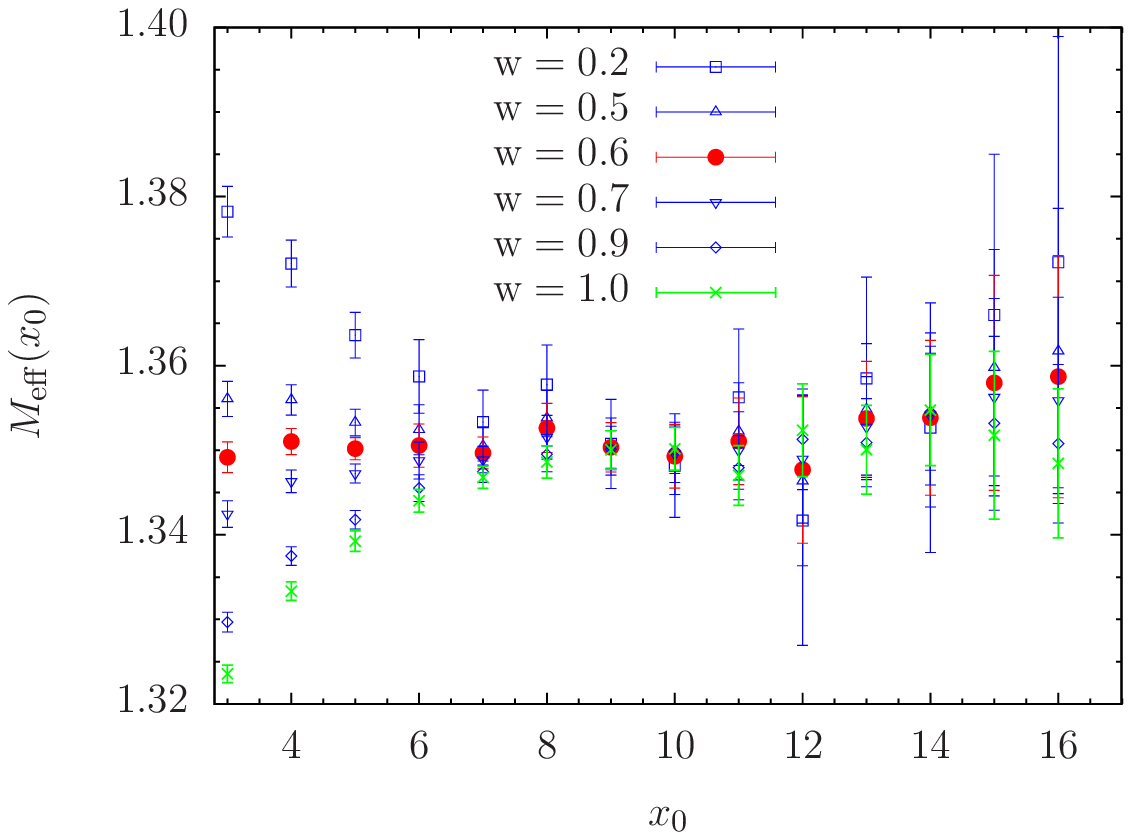}}
\begin{center}
\caption{\sl Example of the plateau for the effective pseudoscalar meson mass at $\beta=3.80$ and  $(a\mu_{\ell}, a\mu_h)=(0.0080, 0.5246)$: 
(a) improved plateau using a smeared field at the source and a local field at the sink (SL) -- green crosses -- compared to the correlators 
(LL) -- magenta upward triangles -- or (SS) -- blue downward triangles; 
(b) further improvement is obtained with optimised field at the source and local field at the sink after tuning the parameter w. 
Blue points represent the values of $M_{\rm{eff}}$ for different  values of the parameter w; 
red full circles correspond  to a case close to the optimal improvement; green crosses are for the w=1 case which 
corresponds  to the (SL) correlator case shown in panel~(a).
  }
\label{fig:mps_Phiopt}
\end{center}
\end{figure}

Even stronger overlap with the ground state is achieved with the use of an {\it optimised} source constructed as follows:
\begin{equation} \label{optsource}
\Phi^{\rm{opt}} \sim \, {\rm w}\,  \Phi^{\rm{S}} + (1-\rm{w})\, \Phi^{\rm{L}}, 
\end{equation} 
where we have introduced the tunable parameter $\rm{w}$. 
In practice, the use of these sources does not involve other inversions than those of the local and smeared sources.
We constructed correlators that have $\Phi^{\rm{opt}}$ as a source 
and  either $\Phi^{\rm{L}}$ or $\Phi^{\rm{S}}$ in the sink. We verified that in general 
the optimised correlators fulfil the expectations of providing an earlier Euclidean 
time projection on the ground state than the (SL) correlators. In Fig.~\ref{fig:mps_Phiopt}(a) we show an example of an 
improved ground state plateau using smeared source for the quark masses $(a\mu_{\ell}, a\mu_h)=(0.0080, 0.5246)$ at $\beta=3.80$. 
Fig.~\ref{fig:mps_Phiopt}(b) shows the improvement that we achieve when we use the optimised source of 
Eq.~(\ref{optsource}) with an appropriately tuned $\rm{w}$ parameter. In the figure  
the effect of tuning $\rm{w}$ at the level of the first decimal place is illustrated\footnote{Notice 
that the absolute values of $\rm{w}$ given in the figure and the number of decimal digits 
depend on the normalisation condition imposed in  Eq.~(\ref{optsource}).}. 
In our application however, we tried an even better tuning, {\it e.g.} up to the second digit. 
Our general conclusion is that employing 
the optimal source (\ref{optsource})   leads to significant improvement that results
in earlier time plateaux ({\it i.e.} at shorter time separations) 
for the effective pseudoscalar meson mass. Further details on the implementation of the method for 
computing pseudoscalar meson masses, decay constants and 
bag parameters are given in Appendix~\ref{APP_optimised_smeared}.

\section{Computation of the $b$-quark mass and decay constants $f_{B}$ and $f_{Bs}$ }
\label{sec:mb_fB_fBs}

The determination of the $b$-quark mass, the decay constants $f_{B}$ and $f_{Bs}$ as well as for their ratio, $f_{Bs}/f_B$, 
is carried out by adopting  
the so-called ratio method presented in Refs.~\cite{Blossier:2009hg, Dimopoulos:2011gx}. 
We refer the reader to these papers for a detailed presentation of the method.  
We will here discuss how our improved 2-point correlation functions lead to reduced 
systematic uncertainties. We start by recalling the main steps of the ratio method for the computation of the 
$b$-quark mass. The HQET suggests the asymptotic behavior, 
\begin{equation}
\lim_{\mu_h^{\rm{pole}}\to \infty}  \left(\frac{M_{h\ell}}{\mu_h^{\rm pole}}\right) = \rm{const.}, 
\label{eq:Mhl}
\end{equation}
where $M_{h\ell}$ is the heavy--light pseudoscalar meson mass and $\mu_h^{pole}$  is the heavy quark pole mass.
The key observation is that the static limit of appropriate ratios of the quantity in the (lhs) of Eq.~(\ref{eq:Mhl}) taken  
at nearby values of the quark pole mass is  equal to unity. This knowledge can be exploited in order to 
compute the $b$-quark mass by interpolating  between relativistic data obtained in the charm quark mass region 
and somewhat above it, and the infinite heavy quark mass limit. 
To this aim it is convenient to consider a sequence of heavy quark masses 
$(\overline\mu_h^{(1)}, \overline\mu_h^{(2)},  \cdots, \overline\mu_h^{(N)})$
which have a fixed ratio, $\lambda$, between any two successive values: $\overline\mu_h^{(n)} = \lambda \overline\mu_h^{(n-1)}$.
At each value of the lattice spacing we then build the following ratios,
\begin{eqnarray}
y(\overline\mu_h^{(n)},\lambda;\overline\mu_{\ell},a) & \equiv &
\frac{M_{h\ell}(\overline\mu_h^{(n)};\overline\mu_\ell,a)}{M_{h\ell}(\overline\mu_h^{(n-1)};\overline\mu_\ell,a)} \cdot
\frac{\overline\mu^{(n-1)}_h}{\overline\mu^{(n)}_h} \cdot
\frac{\rho( \overline\mu^{(n-1)}_h,\mu)}{\rho( \overline\mu^{(n)}_h,\mu)}= \nonumber \\
&=& \lambda^{-1} \frac{M_{h\ell}(\overline\mu_h^{(n)};\overline\mu_\ell,a)}{M_{h\ell}(\overline\mu^{(n)}_h/\lambda;\overline\mu_\ell,a)}
\cdot \frac{ \rho( \overline\mu^{(n)}_h/\lambda,\mu )}{\rho( \overline\mu^{(n)}_h,\mu)}\, ,\quad\quad n=2,\cdots,N \, .
\label{eq:yratio}
\end{eqnarray}
where the function $\rho( \overline\mu^{(n)}_h,\mu)$, that is known up to N$^3$LO in perturbation theory \cite{Chetyrkin:1999pq,
 Gray:1990yh, Broadhurst:1991fy, Chetyrkin:1999ys, Melnikov:2000qh}, 
relates the $\overline{\rm{MS}}$ renormalised quark mass (at the scale of $\mu=3$ GeV) 
to the pole mass: $\mu_h^{\rm{pole}} = \rho(\overline\mu_h, \mu) \, \overline\mu_h(\mu)$. 
By construction, ratios of pseudoscalar meson masses at successive values of the heavy quark mass 
are expected to show small discretisation errors even for rather large values of $\overline\mu_h$. 
\begin{figure}[!ht]
\subfigure[]{\includegraphics[scale=0.70,angle=-0]{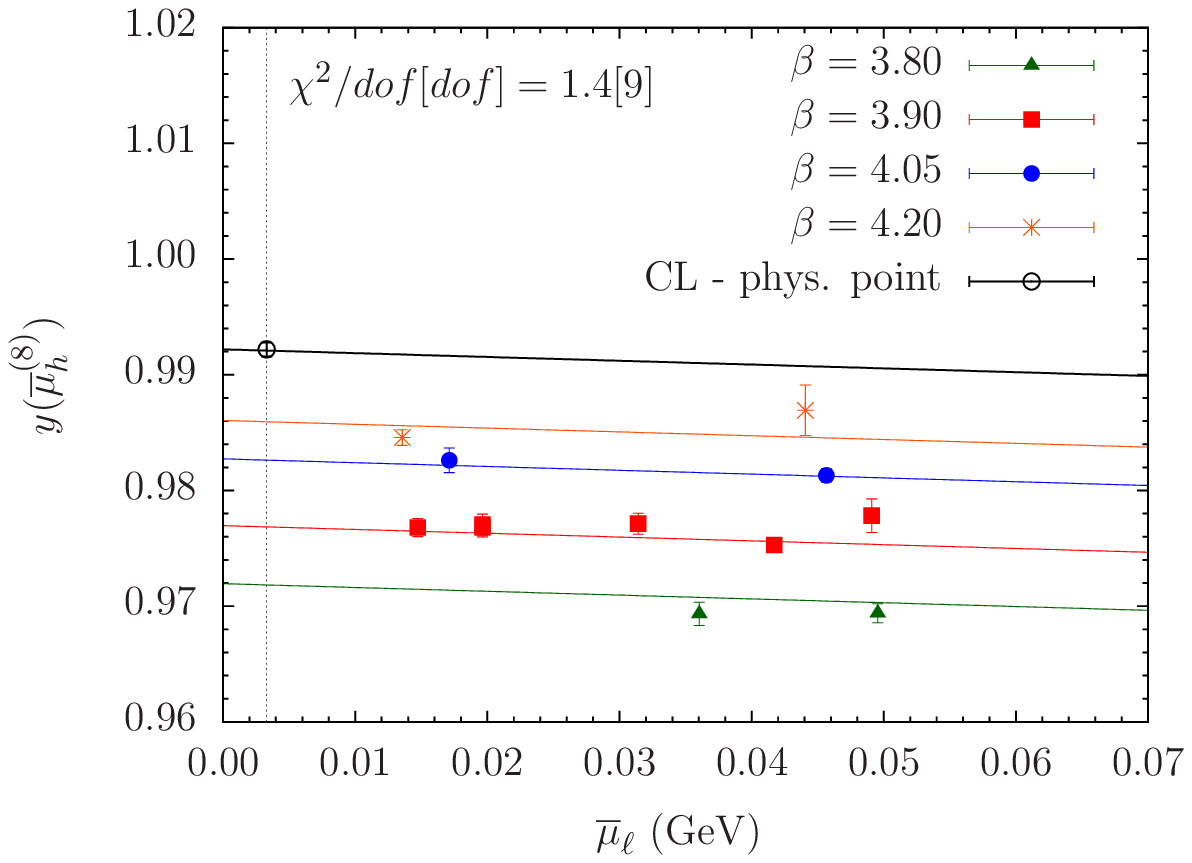}}
\subfigure[]{\includegraphics[scale=0.70,angle=-0]{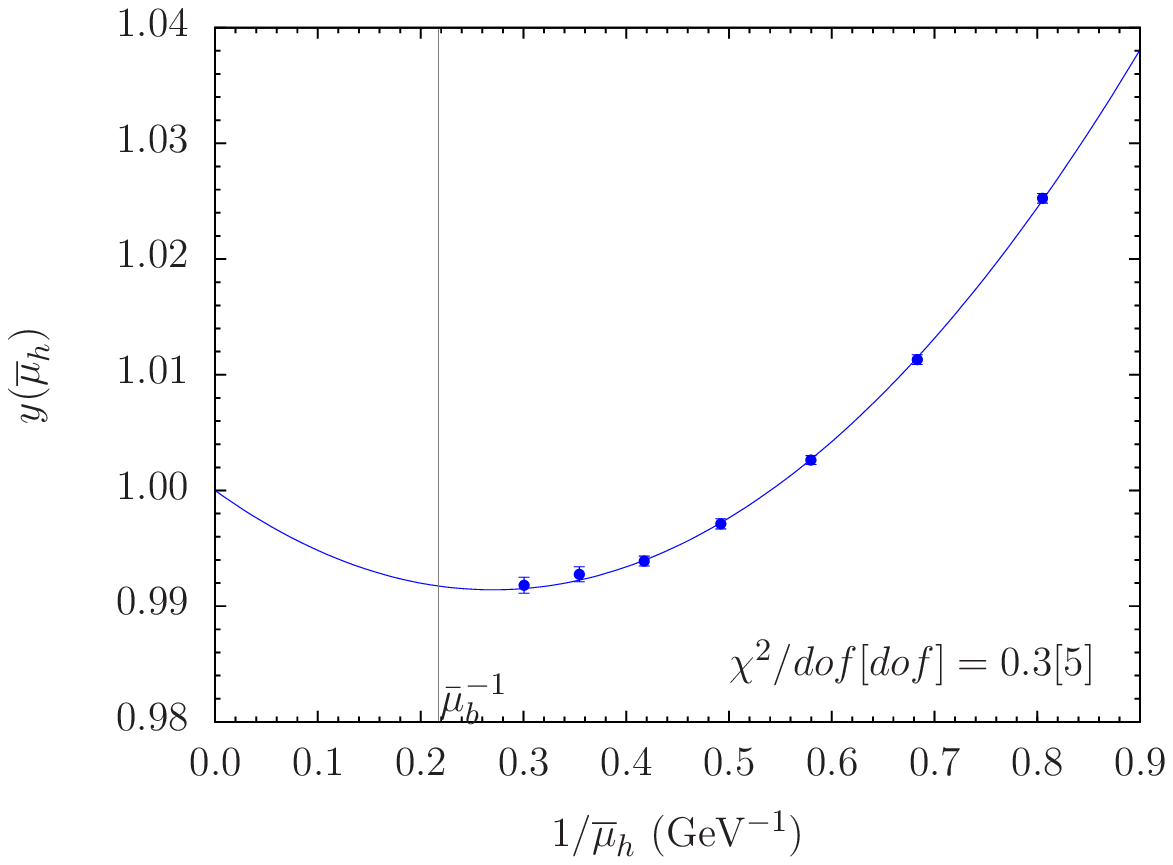}}
\begin{center}
\caption{\sl  (a) combined chiral 
and continuum fit of the ratio defined in Eq.~(\ref{eq:yratio}) against the renormalised light quark mass $\overline\mu_{\ell}$ 
for the largest value of the heavy quark mass (empty black circle is our result at the physical $u/d$ quark mass point 
in the continuum 
limit). (b) $y(\overline\mu_h)$ against $1/\overline\mu_h$ using the fit ansatz~(\ref{eq:y_ansatz}). 
We have used  for the scale in the running coupling that enters 
in the $\rho(\overline\mu_h, \mu)$ function, $\Lambda_{QCD}^{(N_f=2)} = 315(15)$ MeV  and $\lambda = 1.1784$.
The vertical black thin line marks the position of $1/\overline\mu_b$.  
  }
\label{fig:y}
\end{center}
\end{figure}
At each value of $\overline\mu_h^{(n)}$ we can thus  perform a well controlled combined chiral and continuum fit on the ratios of   
Eq.~(\ref{eq:yratio}) to extract the quantity  
$y(\overline\mu_h) \equiv y(\overline\mu_h, \lambda;\overline\mu_{u/d},a=0)$. As an example of the quality of the fit we 
report in Fig.~\ref{fig:y}(a) the linear fit in $\overline \mu_{\ell}$ of the data for $y(\overline\mu_h^{(n)})$
at the largest value of the heavy quark mass.

\begin{figure}[!h]
\subfigure[]{\includegraphics[scale=0.70,angle=-0]{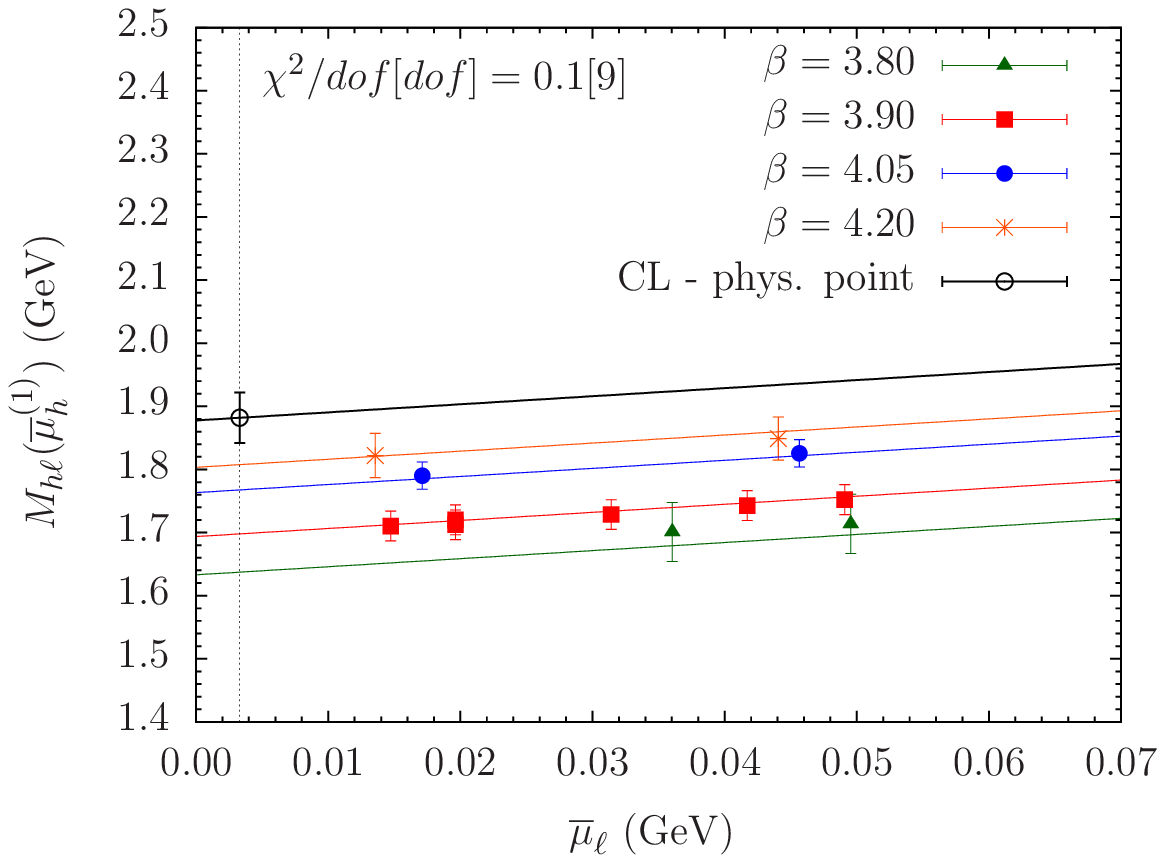}}
\subfigure[]{\includegraphics[scale=0.70,angle=-0]{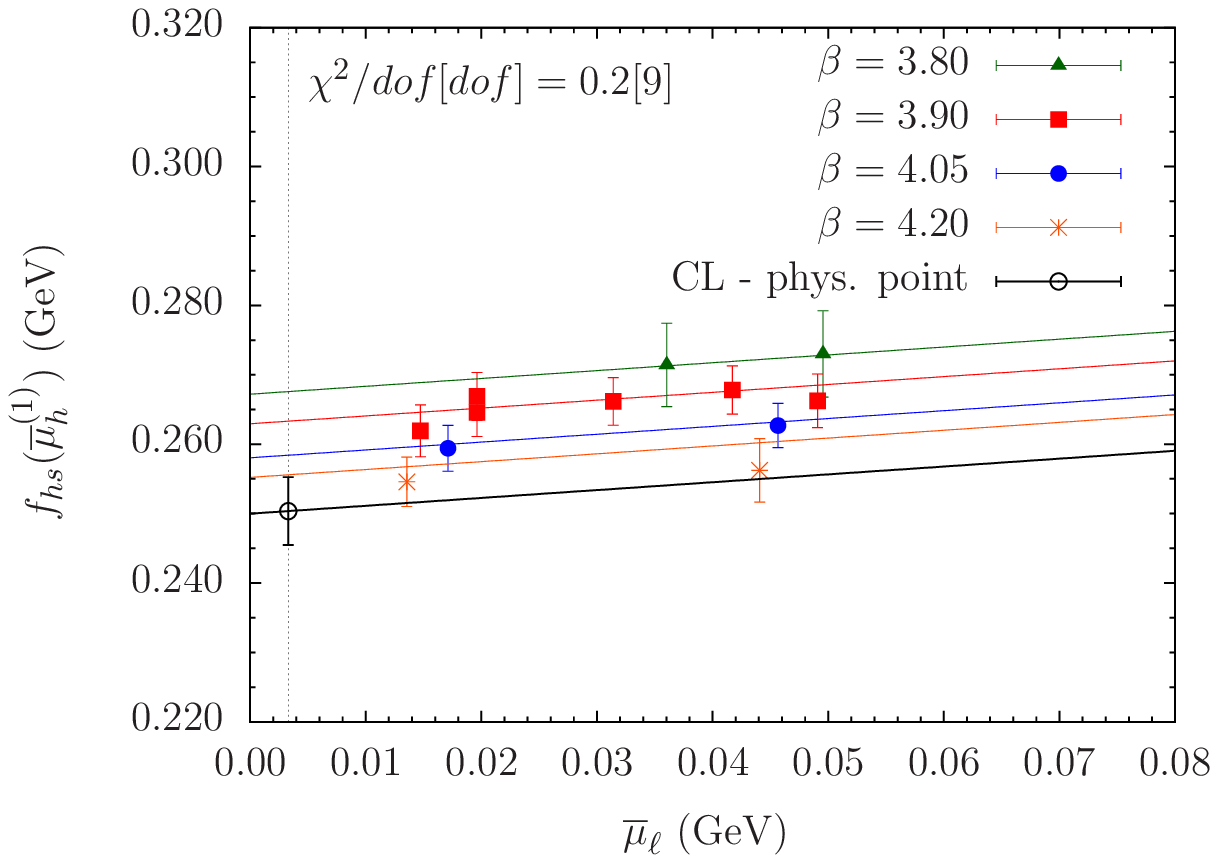}}
\begin{center}
\caption{\sl Combined chiral and continuum fit for the triggering point for (a) pseudoscalar meson mass 
$M_{h\ell}(\overline\mu_h^{(1)})$ and (b) pseudoscalar decay constant $f_{sh}(\overline\mu_h^{(1)})$  against the 
renormalised light quark mass $\overline\mu_{\ell}$. Both fit ans\"atze are linear in $\overline\mu_{\ell}$ 
and in $a^2$. Empty black circle is our result at the physical $u/d$ quark mass point in the continuum 
limit for both cases.
  }
\label{fig:trig}
\end{center}
\end{figure}

Relying on the well-known matching of heavy-light meson mass  evaluated in QCD onto HQET, we have defined the ratio $y(\overline\mu_h)$ in such
a way that its dependence on $\overline\mu_h$ can be described by the fit ansatz 
\begin{equation} \label{eq:y_ansatz}
y(\overline\mu_h) = 1 + \dfrac{\eta_1}{\overline\mu_h} + \dfrac{\eta_2}{\overline\mu_h^2},
\end{equation}   
that implements the constraint $\lim_{\overline{\mu}_h \rightarrow \infty} y(\overline\mu_h) = 1$. 
The fit parameters could be, in general, 
functions of $\log(\overline\mu_h)$. However in the range of the 
currently explored heavy quark mass values this logarithmic dependence can be safely neglected. 
Hence we approximate $\eta_{1,2}$ to constants whose value will be determined by the 
fit to the available  ratio data. 
Data (with $\rho$ from NLL order perturbative formulae) and fit are shown in Fig.~\ref{fig:y}(b). 
Finally the value for the $b$-quark mass can be computed from the 
{\it chain} equation
\begin{equation}\label{eq:y_chain}
   y(\overline\mu_h^{(2)})\, y(\overline\mu_h^{(3)})\,\ldots \, y(\overline\mu_h^{(K+1)})=\lambda^{-K} \,
\frac{M_{hu/d}(\overline\mu_h^{(K+1)})}{M_{hu/d}(\overline\mu_h^{(1)})} \cdot
\Big{[}\frac{\rho( \overline\mu_h^{(1)},\mu)}{\rho( \overline\mu_h^{(K+1)},\mu)}\Big{]}\,.
\end{equation}
Here on the one hand $\lambda$, $K$ and $\overline\mu_h^{(1)}$ are such
that $M_{hu/d}(\overline\mu_h^{(K+1)})$ coincides with 
the experimental value of the $B$-meson mass, $M_{B}=5.279$ GeV;
and on the other hand $M_{hu/d}(\overline\mu_h^{(1)})$ is the result of the combined chiral and continuum fit 
of  pseudoscalar meson mass values evaluated 
at the reference heavy quark mass, $\mu_h^{(1)}$.  Because of its role to Eq.~(\ref{eq:y_chain})
we will call it the {\it triggering point}. 
The quality of the linear fit in $\overline\mu_{\ell}$  
is shown in Fig.~\ref{fig:trig}(a).~\footnote{Following Ref.~\cite{Roessl:1999iu}, we fit pseudoscalar           
meson masses with  a charm-like and a light quark mass assuming that the dependence on the light quark mass is linear.}
For the present analysis we use 
$(\overline\mu_h^{(1)}, \lambda) = (1.05~\rm{GeV},~ 1.1784)$, for which Eq.~(\ref{eq:y_chain}) is satisfied for $K=K_b=9$.
Many other choices of $(\overline\mu_h^{(1)}, \lambda,\, K_b)$  could equally well be used.  
The $b$-quark mass result reads
\begin{equation}\label{eq:mub}
\overline\mu_b = \lambda^{K_b}\, \overline\mu_h^{(1)} =4.60(13)\,\, \rm{GeV} .
\end{equation} 
In section~\ref{sec:summary_final_results} we discuss the error budget attached to this result. 
We have also verified that using $M_{hs}$ instead of $M_{h\ell}$ leads to fully compatible results for the $b$-quark mass, 
the difference being at the per mille
level.

\begin{figure}[!ht]
\subfigure[]{\includegraphics[scale=0.70,angle=-0]{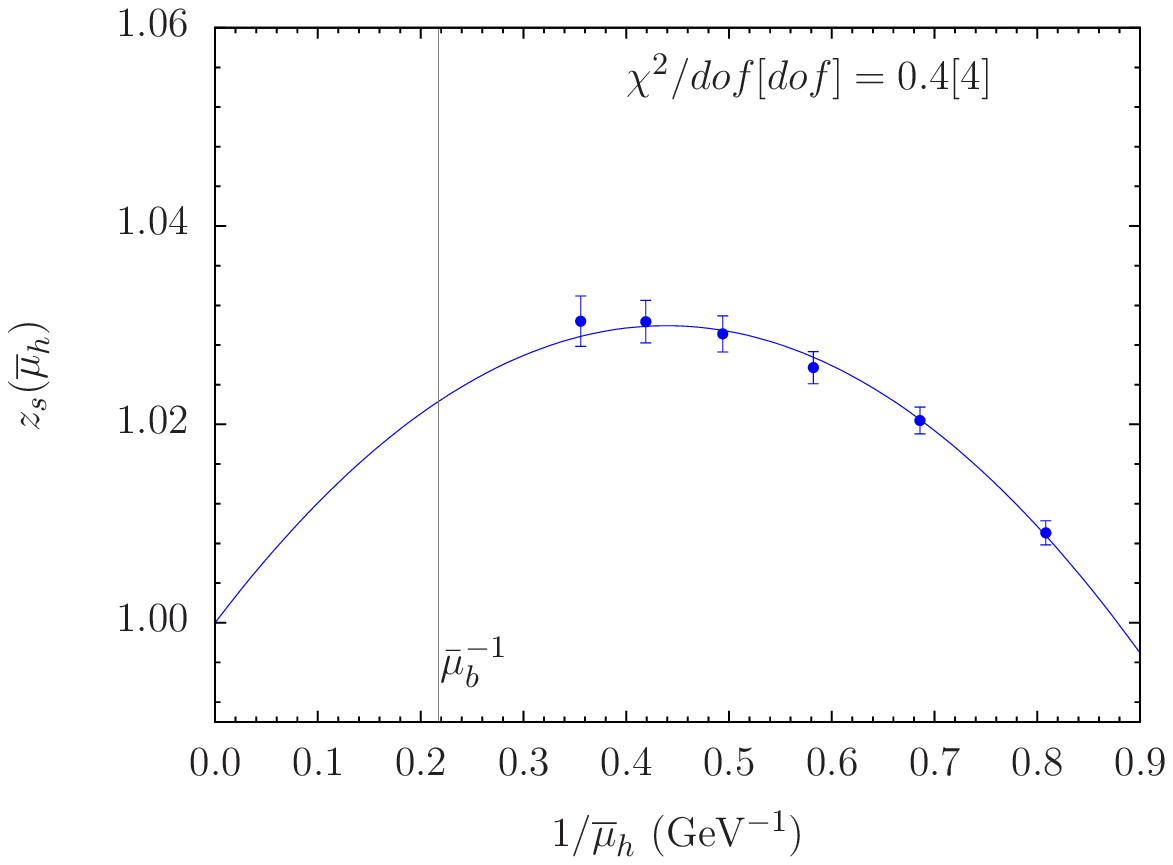}}
\hspace*{0.7cm}
\subfigure[]{\includegraphics[scale=0.70,angle=-0]{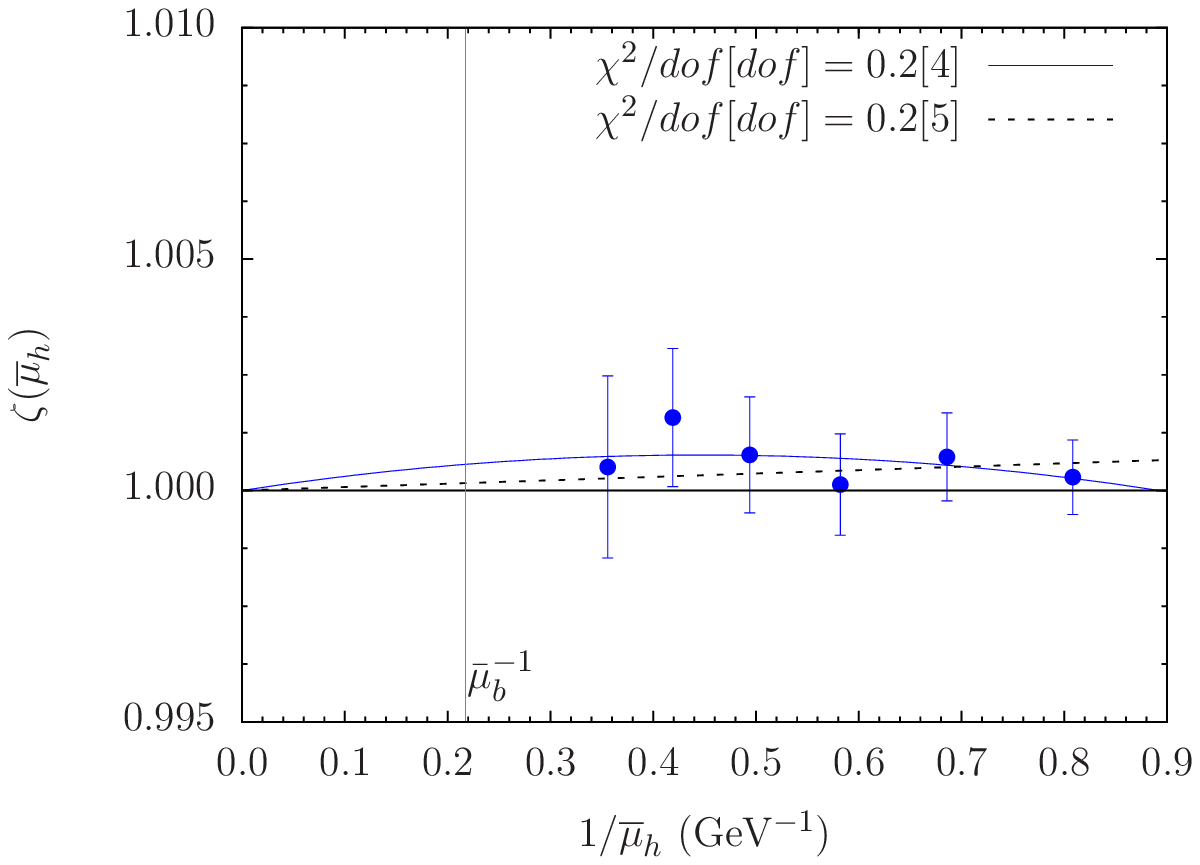}}
\begin{center}
\caption{\sl  $z_s(\overline\mu_h)$ and $\zeta(\overline\mu_h)$ against $1/\overline\mu_h$ in panels (a) and (b), respectively. 
We have used $\Lambda_{QCD}^{(N_f=2)} = 315(15)$ MeV and $\lambda = 1.1784$.
In both cases the fit function has a polynomial form of the type like the one of  Eq.~(\ref{eq:y_ansatz}) (blue curve). 
In panel (b) a fit of the form $\zeta(\overline\mu_h) = 1 + 1/\overline\mu_h$ has also 
been performed (black dashed straight line).  The vertical black thin line 
marks the position of $1/\overline\mu_b$.
  }
\label{fig:zs_zeta_vs_muh}
\end{center}
\end{figure}

We apply an analogous strategy to compute the pseudoscalar decay constant of the meson $B_{s}$, $f_{Bs}$ 
and the ratio of the decay constants $f_{Bs}/f_{B}$. The appropriate  HQET asymptotic 
conditions in the first two cases are
\begin{equation}  \label{eq:fhscondition} 
 \lim_{\mu_h^{\rm{pole}}\to \infty} f_{hs} \sqrt{\mu_h^{\rm{pole}}} =\mbox{constant}, 
 \end{equation}
 \begin{equation}
\lim_{\mu_h^{\rm{pole}}\to \infty} \Big(f_{hs}/f_{h\ell}\Big) = \mbox{constant}, 
\end{equation}   
where by `constant' we denote some finite non-zero value. 
Based on QCD to HQET matching of heavy-light meson decay constant ($f_{h\ell}$) and mass ($M_{h\ell}$) 
we define the ratios
\begin{eqnarray}
z_d(\overline\mu_h,\lambda;\overline\mu_\ell, a)&=&
\lambda^{1/2} \frac{f_{h\ell}(\overline\mu_h,\overline\mu_\ell, a)}{f_{h\ell}(\overline\mu_h/\lambda,\overline\mu_\ell, a)}
\cdot \frac{C^{stat}_A(\mu^*,\overline\mu_h/\lambda)}{C^{stat}_A(\mu^*, \overline\mu_h)}
\frac{[\rho( \overline\mu_h,\mu)]^{1/2}}{[\rho( \overline\mu_h/\lambda,\mu)]^{1/2}},   \label{z1} \\ 
z_s(\overline\mu_h,\lambda;\overline\mu_\ell,\overline\mu_s, a)&=&
\lambda^{1/2} \frac{f_{h s}(\overline\mu_h,\overline\mu_\ell,\overline\mu_s, a)}{f_{h s}(\overline\mu_h/\lambda,\overline\mu_\ell,\overline\mu_s, a)}
\cdot \frac{C^{stat}_A(\mu^*,\overline\mu_h/\lambda)}{C^{stat}_A(\mu^*,\overline\mu_h)}
\frac{[\rho( \overline\mu_h,\mu)]^{1/2}}{[\rho( \overline\mu_h/\lambda,\mu)]^{1/2}} \,. \label{z2}
\label{eq:z_zs_ratios}
\end{eqnarray}
\begin{figure}[!h]
\subfigure[]{\includegraphics[scale=0.70,angle=-0]{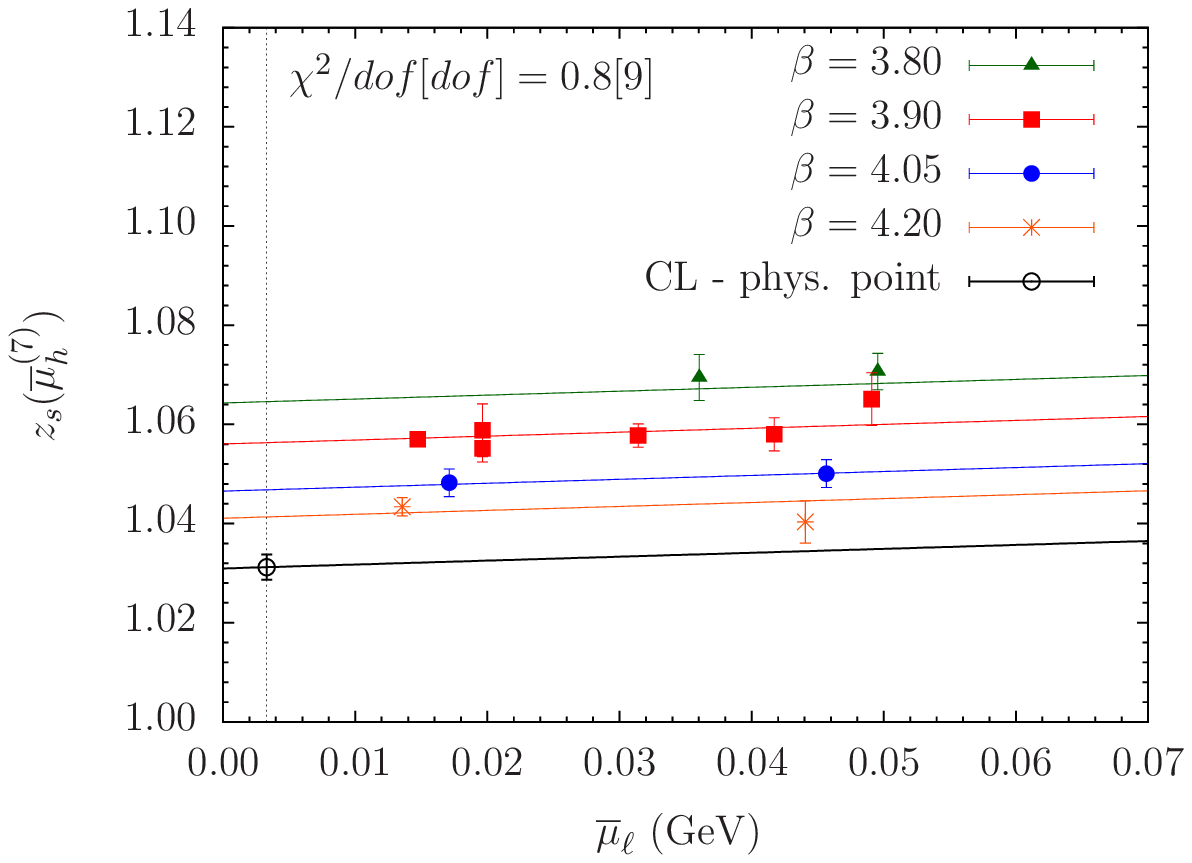}}
\subfigure[]{\includegraphics[scale=0.70,angle=-0]{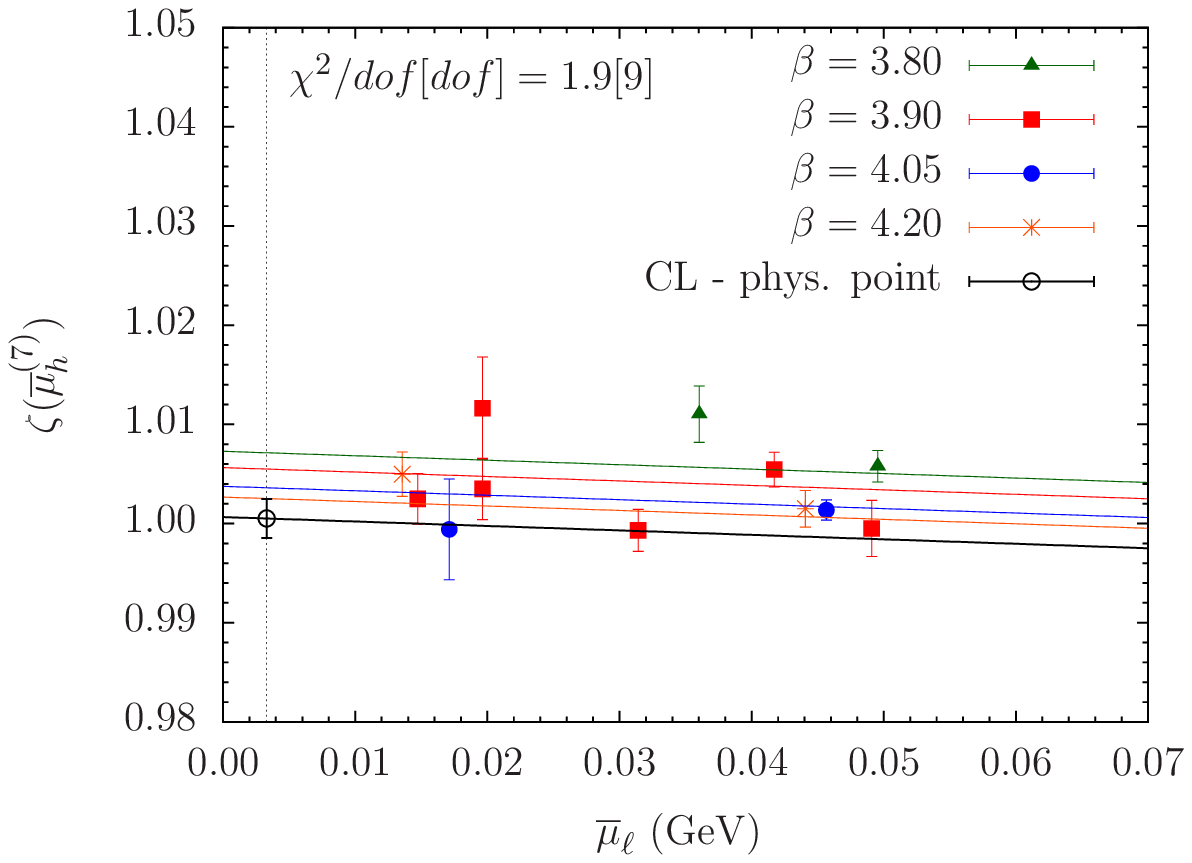}}
\begin{center}
\caption{\sl Combined chiral and continuum fits for the ratio $z_s(\overline\mu_h^{(n)})$ (Eq.~(\ref{eq:z_zs_ratios})) and 
the double ratio $\zeta(\overline\mu_h^{(n)})$ (Eq.~(\ref{eq:zeta})) against $\overline\mu_{\ell}$ are shown 
in panels (a) and (b) respectively.
In both cases ratios for the largest value of the heavy quark mass are reported ($n=7$). 
Empty black circle is our result at the physical $u/d$ quark mass point in the continuum limit.
  }
\label{fig:ratios_zs_zeta}
\end{center}
\end{figure}
The factor $C^{stat}_A(\mu^*,\overline\mu_h)$, 
known up to N$^2$LO in PT~\cite{Chetyrkin:2003vi}, provides  
the matching between the decay constant in QCD
for a heavy quark mass $\overline \mu_h$ and its static-light
counterpart in HQET (the arbitrary renormalization scale $\mu^*$
of HQET cancels in the ratio above).
From Eqs~(\ref{z1}) and (\ref{z2}) we also form the double ratio
\begin{equation} \label{eq:zeta}
\zeta(\overline\mu_h,\lambda;\overline\mu_\ell, \overline\mu_s, a) =\dfrac{z_s(\overline\mu_h,\lambda;\overline\mu_\ell,\overline\mu_s, a)}
{z_d(\overline\mu_h,\lambda;\overline\mu_\ell, a)} .
\end{equation}
By construction the ratios $z_d$, $z_s$ and $\zeta$   
have an exactly known static limit equal to unity and show a smooth  
chiral and continuum combined behavior. 
As in the case of the $y$-ratios, this is a consequence of the fact that $z_d$, $z_s$ and $\zeta$ are simply ratios 
of quantities calculated at 
nearby values of the heavy quark mass for which much of the discretisation errors cancel. 
Figs~\ref{fig:ratios_zs_zeta}(a) and \ref{fig:ratios_zs_zeta}(b) are two examples illustrating the quality of the combined 
chiral and continuum fits for $z_s(\overline\mu_h^{(7)})$ and $\zeta_s(\overline\mu_h^{(7)})$ respectively, 
at the largest heavy quark mass values used in the decay constant analysis.\footnote{Note that at the largest value of the 
heavy quark mass, $\overline\mu_h^{(8)}$, which has been used in the $b$-quark mass analysis, 
our estimates of the pseudoscalar meson decay constants 
proved to be rather noisy. Hence, in the decay constants' analysis we decided to use data corresponding up to the next largest 
heavy quark mass value, $\overline\mu_h^{(7)}$.}~\footnote{The rather high $\chi^2/dof$ value in the fit
of $\zeta_s(\overline\mu_h^{(7)})$ data  ratio is not representative of the quality of the ratio fits performed in the present work.   
The fits performed in the present analysis are dominated by systematic uncertainties.  
Correlation matrices are not taken into account since they turn out to be affected by large uncertainties, 
and the $\chi^2$ definition incorporates the contribution of priors. 
Furthermore, we would like to stress that in order to control 
systematic effects in the ratio analysis for all the physical quantities studied inn this work, 
we have repeated the whole procedure excluding the heaviest quark mass. We have 
treated the difference  between the analyses as a systematic uncertainty in the final result. See also the discussion 
in Section~\ref{sec:summary_final_results}.}
In Figs~\ref{fig:zs_zeta_vs_muh}(a) and \ref{fig:zs_zeta_vs_muh}(b) we show the dependence of $z_s(\overline\mu_h)$ 
and $\zeta(\overline\mu_h)$ on the inverse heavy quark mass, respectively. The fit ans\"atze  we have 
used are polynomial fit functions in the inverse heavy quark mass analogous to the one specified in Eq.~(\ref{eq:y_ansatz}). 
For the case of the double ratio 
$\zeta(\overline\mu_h)$ we have also tried a linear fit in $1/\overline\mu_h$ 
always implementing the static condition $\lim_{\overline{\mu}_h \rightarrow \infty} \zeta(\overline\mu_h) = 1$.  

Determinations of $f_{Bs}$ and $f_{Bs}/f_{B}$ are obtained by means of the equations 
\begin{eqnarray} \label{eq:zs_chain}
 \hspace*{-0.5cm} z_s(\overline\mu_h^{(2)})\, z_s(\overline\mu_h^{(3)})\,\ldots \, z_s(\overline\mu_h^{(K+1)}) &=& \lambda^{K/2} \,
\frac{f_{hs}(\overline\mu_h^{(K+1)})}{f_{hs}(\overline\mu_h^{(1)})} \cdot
\Big[ \frac{C^{stat}_A(\mu^*, \overline\mu_h^{(1)})}{C^{stat}_A(\mu^*, \overline\mu_h^{(K+1)})}
\sqrt{\frac{\rho( \overline\mu_h^{(K+1)},\mu)}{\rho( \overline\mu_h^{(1)},\mu)}} \Big],  \\
 \hspace*{-0.5cm} \zeta(\overline\mu_h^{(2)})\, \zeta(\overline\mu_h^{(3)})\,\ldots \, \zeta(\overline\mu_h^{(K+1)}) &=& \lambda^{K/2} \,
\Big[ \frac{f_{hs}(\overline\mu_h^{(K+1)})/f_{hu/d}(\overline\mu_h^{(K+1)})}{f_{hs}(\overline\mu_h^{(1)})/f_{hu/d}(\overline\mu_h^{(1)})} 
\Big] . 
\end{eqnarray}
The (lhs) of the above equations are taken from  the  fits of Figs~\ref{fig:zs_zeta_vs_muh}(a) and (b) respectively. 
Setting $\overline\mu_h^{(K+1)}=\overline\mu_b$, ({\it cf.} Eq.~(\ref{eq:mub}))  and having determined the values of 
$f_{hs}(\overline\mu_h^{(1)})$ and $[f_{hs}(\overline\mu_h^{(1)})/f_{hu/d}(\overline\mu_h^{(1)})]$ from a 
combined chiral and continuum fit, we finally 
get our results for $f_{Bs}$ and $[f_{Bs}/f_{B}]$ respectively. 
As expected the combined chiral and continuum fit for the quantity $f_{hs}(\overline\mu_h^{(1)})$ 
is smooth and shows tolerably small cutoff
effects, as well as a very weak dependence on the light quark mass (see Fig.~\ref{fig:trig}(b)). 

It is important to emphasize that in determining the value of the $B_s$-meson 
decay constant one could adopt, instead of
Eq.~(\ref{eq:fhscondition}), the following condition 
\begin{equation}\label{eq:fhs_sqrtMhs}
\lim_{M_{hs}\to \infty} f_{hs} \sqrt{M_{hs}} = \mbox{constant} \, ,
\end{equation}
by means of which any use of the heavy quark pole mass is avoided. 
In analogy to Eq.~(\ref{z2}) we can define the ratio
\begin{equation}\label{eq:tilde_zs}
\tilde z_s(\overline\mu_h,\lambda;\overline\mu_\ell,\overline\mu_s, a)=
\frac{f_{h s}(\overline\mu_h,\overline\mu_\ell,\overline\mu_s, a)\, 
\sqrt{M_{hs}(\overline\mu_h,\overline\mu_\ell,\overline\mu_s, a)}}
{f_{h s}(\overline\mu_h/\lambda,\overline\mu_\ell,\overline\mu_s, a)\, 
\sqrt{M_{hs}(\overline\mu_h/\lambda,\overline\mu_\ell,\overline\mu_s, a)}}
\cdot \frac{C^{stat}_A(\mu^*,\overline\mu_h/\lambda)}{C^{stat}_A(\mu^*,\overline\mu_h)}.
\end{equation}
We then determine the value of $f_{Bs}$ by means of the equation 
\begin{equation}\label{eq:chain_tilde_zs}
\tilde z_s(\overline\mu_h^{(2)})\, \tilde z_s(\overline\mu_h^{(3)})\,\ldots \, \tilde z_s(\overline\mu_h^{(K+1)}) = 
\frac{f_{hs}(\overline\mu_h^{(K+1)})\, \sqrt{M_{hs}(\overline\mu_h^{(K+1)})}}
{f_{hs}(\overline\mu_h^{(1)})\, \sqrt{M_{hs}(\overline\mu_h^{(1)})}} \cdot
\Big[ \frac{C^{stat}_A(\mu^*, \overline\mu_h^{(1)})}{C^{stat}_A(\mu^*, \overline\mu_h^{(K+1)})}
 \Big] 
\end{equation}
where we  set $M_{hs}(\overline\mu_h^{(K+1)})$ equal to the experimental value of the $B_s$-meson mass, $M_{Bs} = 5366.7$ MeV.

\begin{figure}[!h]
\begin{center}
\subfigure[]{\includegraphics[scale=0.70,angle=-0]{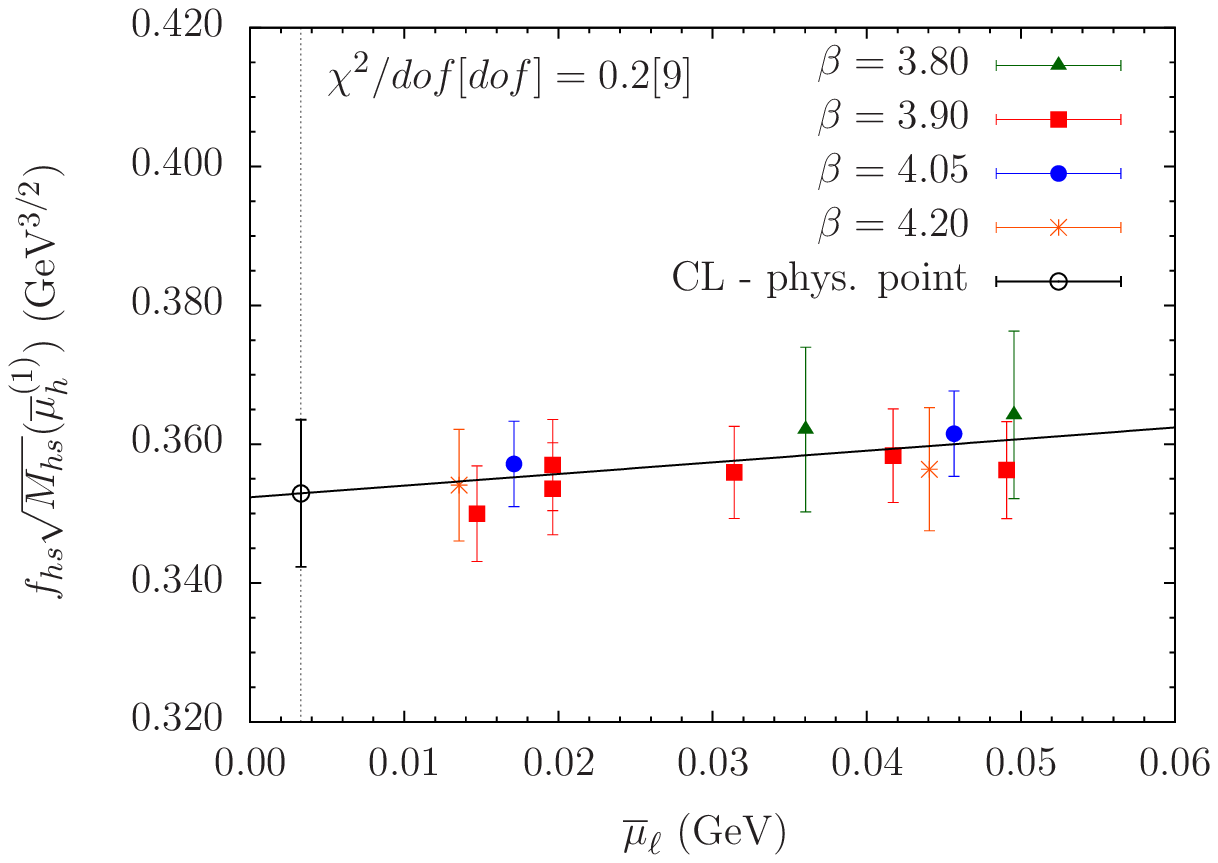}}
\subfigure[]{\includegraphics[scale=0.70,angle=-0]{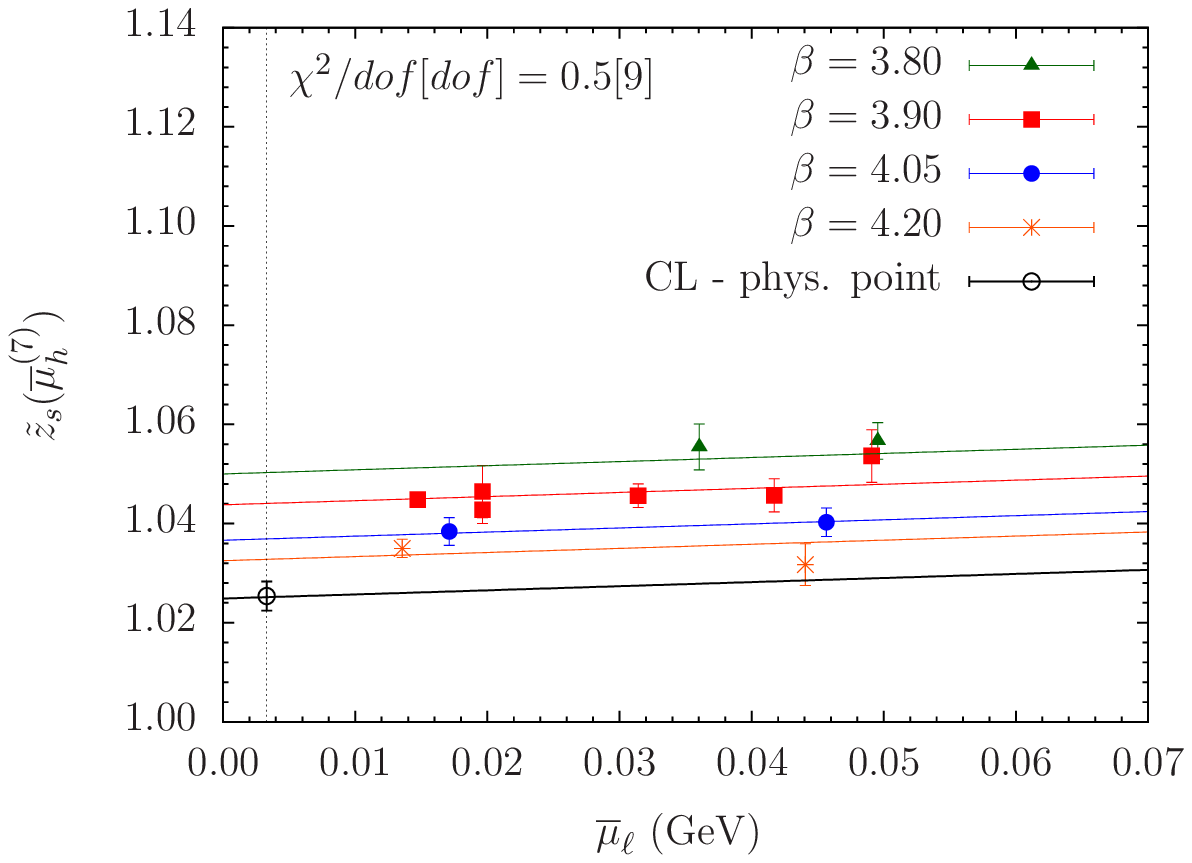}}
\subfigure[]{\includegraphics[scale=0.70,angle=-0]{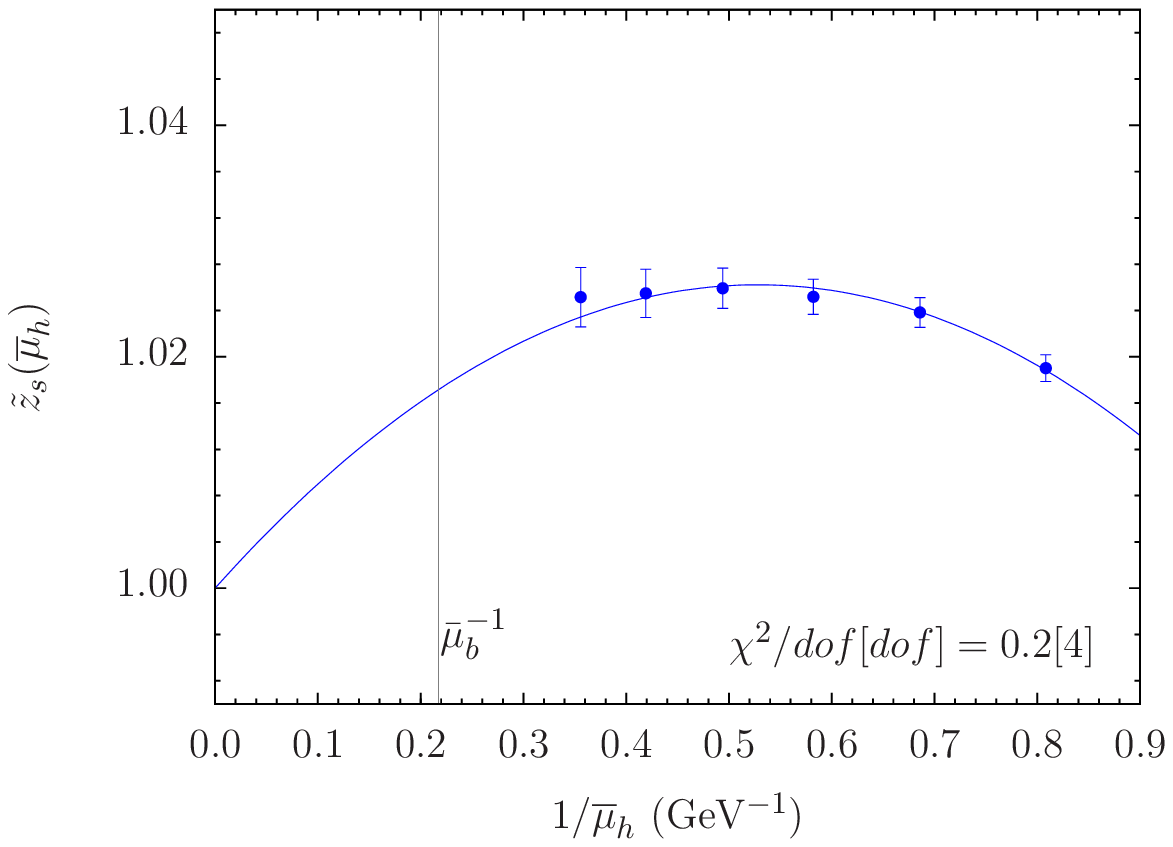}}
\caption{\sl (a) Combined chiral and continuum fit for the triggering point for the quantity 
$f_{hs}\sqrt{M_{hs}}$ against $\overline\mu_{\ell}$. 
(b) Combined chiral and continuum fit for the ratio
$\tilde z_{s}$, against  $\overline\mu_{\ell}$ calculated between the two largest heavy quark mass values used in this work ($n=7$). 
(c) $\tilde z_{s}(\overline \mu_h)$  against the inverse heavy quark mass. The static condition at unity is imposed
explicitly in the fit ansatz.}
\label{fig:zs_tilde}
\end{center}
\end{figure}

Figs~\ref{fig:zs_tilde}(a), (b) and (c) are the equivalent of Figs~\ref{fig:trig}(b), \ref{fig:ratios_zs_zeta}(a) 
and \ref{fig:zs_zeta_vs_muh}(a) when, in the computation of $f_{Bs}$ using the ratio method, we employ the condition presented in 
Eq.~(\ref{eq:fhs_sqrtMhs}) instead of the one given in the Eq.~(\ref{eq:fhscondition}). 
Note that, as it can be seen from Fig.~\ref{fig:zs_tilde}(a), 
the triggering point calculation of the quantity 
$f_{hs}\sqrt{M_{hs}}$ presents very small discretisation effects, and, though it is an accidental fact, it contributes to an accurate 
computation of the continuum limit. The fit ansatz used in fitting the data of Fig.~\ref{fig:zs_tilde}(c) against the inverse heavy 
quark mass is of the same 
form as the one presented in Eq.~(\ref{eq:y_ansatz}). 
We anticipate here (see also Section~\ref{sec:summary_final_results}) our finding that 
determinations of $f_{Bs}$ computed either via $z_s$ or $\tilde z_s$ 
ratios are fully compatible differing by less than 0.5\%.
We also note that ratios defined in terms of $M_{hs}$ (or $M_{h\ell}$),
rather than the heavy quark pole mass, could be used for all the
matrix elements discussed in the present paper.
  
\begin{figure}[!t]
\begin{center}
{\includegraphics[scale=0.73,angle=-0]{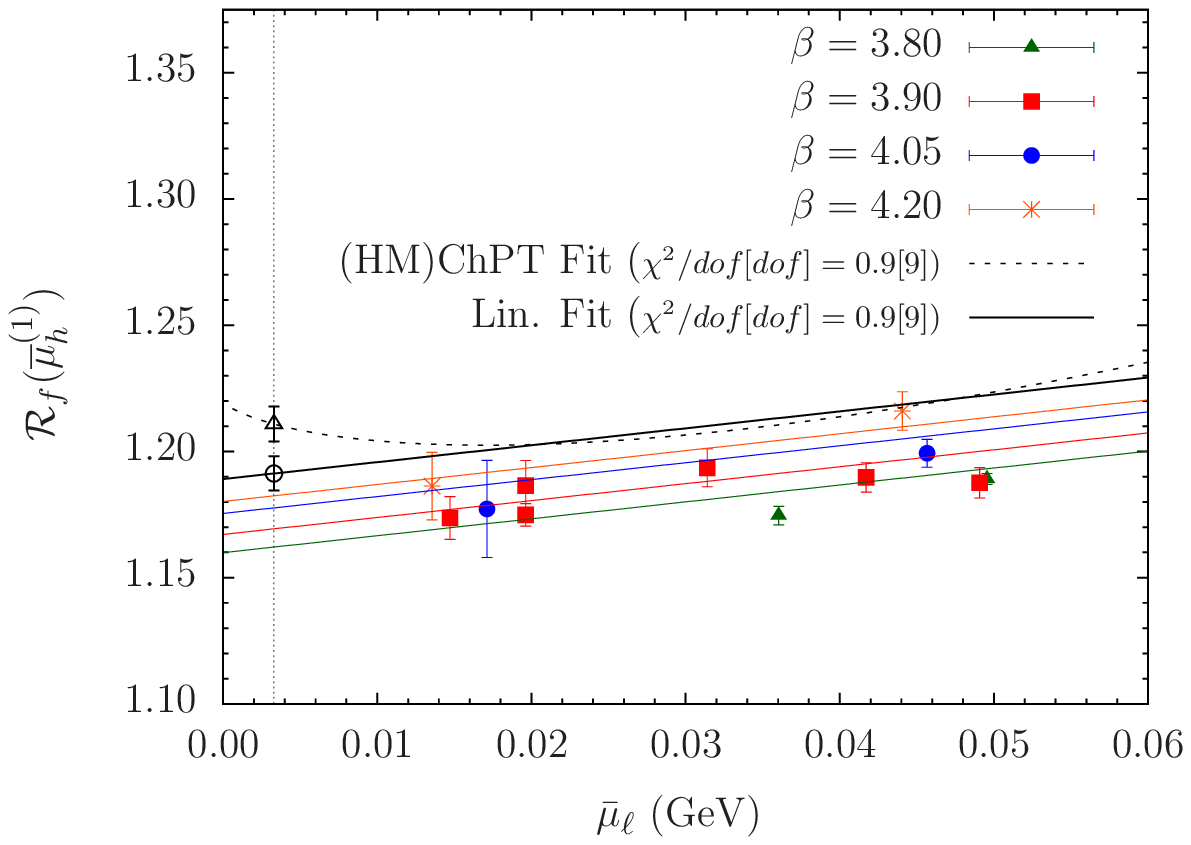}}
\caption{\sl Combined chiral and continuum fit at the triggering point for the quantity defined in Eq.~(\ref{eq:Rf}). 
We have used fit ans\"atze defined in Eqs~(\ref{eq:lin_ans}) and (\ref{eq:HMChPT_ratiof}) with $\hat{g}=0.61(7)$.
Colored lines correspond to the fit of  Eq.~(\ref{eq:lin_ans}). 
Empty black circle and empty black triangle are the results from the linear and the ChPT fit ansatz, respectively,  
at the physical $u/d$ quark mass point in the continuum limit.   
  }
\label{fig:zeta_trig}
\end{center}
\end{figure}

\vspace*{0.3 cm}
We also need to estimate the triggering point ratio
$[f_{hs}(\overline\mu_h^{(1)})/f_{hu/d}(\overline\mu_h^{(1)})]$ {\it i.e.} 
the value it takes after extrapolation to the continuum limit and
the physical light quark mass.
To this aim we make use of the useful observation~\cite{Becirevic:2002mh, Blossier:2009bx} 
that by forming the double ratio of the decay constants 
$[(f_{hs}/f_{h \ell}) / (f_{s \ell}/f_{\ell \ell})]$ one can exploit the possibility for a large cancellation of the 
chiral logarithmic terms. 
One can get the triggering point value by multiplying 
the above expression by an appropriate estimate of the
ratio of the $K$ to $\pi$ decay constants, $(f_K/f_{\pi})$. 
For notation simplicity we define the quantity  
\begin{equation} \label{eq:Rf}
{\cal R}_f = [(f_{hs}/f_{h \ell}) / (f_{s \ell}/f_{\ell \ell})] (f_K/f_{\pi})
\end{equation} 
and we plot it against $\overline\mu_{\ell}$, see Fig.~\ref{fig:zeta_trig}. 
We have used two fit ans\"atze. The first fit ansatz is linear in 
$\overline\mu_{\ell}$, the second one  is suggested by the combined use of the SU(2) ChPT and 
HMChPT. They read
\begin{eqnarray}\label{eq:zeta_trig}
(I) ~~~~~~~ {\cal R}_f &=& a_h^{(1)} + b_h^{(1)} \overline\mu_{\ell} + D_h^{(1)} a^2 \label{eq:lin_ans}\\
(II)~~~~~ {\cal R}_f  &=& a_h^{(2)}\Big[1 + b_h^{(2)} \overline\mu_{\ell} + \Big[ \frac{3(1+3\hat{g}^2)}{4} - \frac{5}{4} \Big]
\frac{2 B_0 \overline\mu_{\ell}}{(4 \pi f_0)^2} 
{\rm log}\Big( \frac{2 B_0 \overline\mu_{\ell}}{(4 \pi f_0)^2}  \Big)\Big]  + D_h^{(2)} a^2  \label{eq:HMChPT_ratiof} 
\end{eqnarray}
In the fit based on HMChPT, we take for the parameter $\hat g$ the value $\hat g=0.61(7)$~\cite{Nakamura:2010zzi} obtained from the experimental measurement of the $g_{D^*D\pi}$ coupling. We choose this value, instead of the HQET prediction $\hat g=0.44(8)$~\cite{Becirevic:2009yb}, 
because we fit data that are close to the charm mass region and in order to conservatively include in the average the maximum spread resulting from the different ways of performing the chiral extrapolation of our data.

As it can be noticed from Fig.~\ref{fig:zeta_trig} discretisation effects are small. The two estimates for the triggering point ratio 
at the physical light quark mass are compatible within two standard deviations. We take their average value as our final result and 
we consider their half difference as a systematic uncertainty. 
In this computation we have used the result $f_K/f_{\pi}=1.193(5)$ by FLAG~\cite{Colangelo:2010et} coming from an average over lattice 
determinations using $N_f=2+1$ dynamical quark simulations. This value is completely uncorrelated with relevant determinations 
by ETMC. The latest PDG result for the same ratio could be an alternative choice. 
This differs by one standard deviation from the  above one~\cite{Beringer:1900zz}.
In order to to get a (conservative) estimate of this particular systematic uncertainty we consider the spread 
between the above value, $f_K/f_{\pi}=1.193(5)$, and the one determined from $N_f=2$ dynamical quark 
simulations, $f_K/f_{\pi}=1.210(18)$~\cite{Colangelo:2010et}. 
Finally, we sum in quadrature the two systematic uncertainties, that is the one coming from the fit ansatz choice 
and the other from  $f_K/f_{\pi}$. Our result at the triggering point reads  
\begin{equation} \label{eq:fhsovfhl_trig}
\dfrac{f_{hs}}{f_{hu/d}}\Big|_{\overline\mu_h^{(1)}} = 1.201(7)(20), 
\end{equation}
where the first error is statistical and the second denotes the systematic uncertainty we have just discussed.  

\section{Computation of Bag parameters and $\xi$ }
\label{sec:B_xi}

For the evaluation of the four-fermion matrix elements on the lattice we use a mixed fermionic action setup
where we adopt different regularizations for sea and valence quarks as  proposed in Ref.~\cite{Frezzotti:2004wz}.
The Mtm-LQCD action of the  light quark flavor doublet that is used to generate 
unquenched gauge configurations can be written in
the so-called ``physical basis" in the form~\cite{FrezzoRoss1}
\begin{equation}
S^{\rm Mtm}_{sea} =  a^4 \sum_x \bar \psi (x) \Big \{ \frac{1}{2} \sum_\mu \gamma_\mu(\nabla_\mu 
+ \nabla^\ast_\mu ) - i \gamma_5 \tau^3 r_{\rm{sea}} \big [ M_{\rm cr} - 
\dfrac{a}{2} 
\sum_\mu \nabla^\ast_\mu \nabla_\mu \big ] + \mu_{sea} \Big \} \psi(x)\, .\label{SPHYS}
\end{equation}
The field $\psi$ denotes a mass degenerate up and down doublet with  bare (twisted) mass $\mu_{sea}$.
The parameter $M_{\rm cr}$ is the critical mass that one has to fix non-perturbatively at its optimal value
\cite{Boucaud:2007uk, Boucaud:2008xu} 
to guarantee the O($a$)-improvement of physical observables and get rid of all the unwanted
leading chirally enhanced cutoff effects. In the gauge sector the tree-level improved action
proposed in Ref.~\cite{Weisz:1982zw} has been used.

For valence quarks we use the OS regularization~\cite{Osterwalder:1977pc}.
The full valence action is given by the sum of the contributions of each {\it individual}
valence flavour $q_f$ and reads~\cite{Frezzotti:2004wz}
\begin{equation}
S^{\rm OS}_{val} =  a^4 \sum_{x,f} \bar q_f (x) \Big \{ \frac{1}{2} \sum_\mu \gamma_\mu(\nabla_\mu 
+ \nabla^\ast_\mu ) - i \gamma_5 r_f \big [ M_{\rm cr} - 
\frac{a}{2} \sum_\mu \nabla^\ast_\mu \nabla_\mu \big ] + \mu_f \Big \} q_f(x)\, ,
\label{OSvalact}
\end{equation}
where the index $f$ labels the valence flavors and $M_{\rm cr}$ is the same critical mass parameter
which appears in Eq.~(\ref{SPHYS}). We denote by $r_f$ and $\mu_f$ the values of the Wilson parameter
and the twisted quark mass of each valence flavor.

This particular setup offers the advantage that one can compute matrix elements that are at the same time 
O$(a)$-improved and free of wrong chirality mixing effects~\cite{Bochicchio:1985xa}. 
These two properties have already proved to be very beneficial in the
study of neutral Kaon meson oscillations~\cite{Bertone:2012cu, Constantinou:2010qv, Dimopoulos:2009es, Carrasco:2011gr}. 
For a detailed discussion  
about the choice of the action and its implementation for the calculation of the matrix elements we 
refer to Section 4 and Appendix A of Ref.~\cite{Bertone:2012cu}. 

The only significant difference concerning the four-fermion matrix elements computation of the present work with respect to 
Ref.~\cite{Bertone:2012cu} is that here we are dealing with heavy quarks ({\it i.e.} charm-like and heavier ones) 
rather than with the strange quark. 
With heavy quarks the signal to noise ratio deteriorates quickly as the time separation increases and moreover large time separations 
are necessary for the projection onto the ground state. 
We overcome both problems by employing smeared interpolating operators for the meson sources and in this way 
we are able to reduce the source time separation, 
$T_{\rm{sep}}$. The latter turns out to be less than half of the lattice time extension leading to safe plateau signals. 
In particular we have used $T_{\rm{sep}} = 16, ~18, ~22, ~~\rm{and}~~28$ for the 3-point correlation functions at 
$\beta=3.80, ~3.90, ~4.05, ~~\rm{and}~~ 4.20$, respectively.  Note that the rest of the simulation details are the same as those 
collected in Table~\ref{tab:runs}.
\begin{figure}[!t]
\subfigure[]{\includegraphics[scale=0.70,angle=-0]{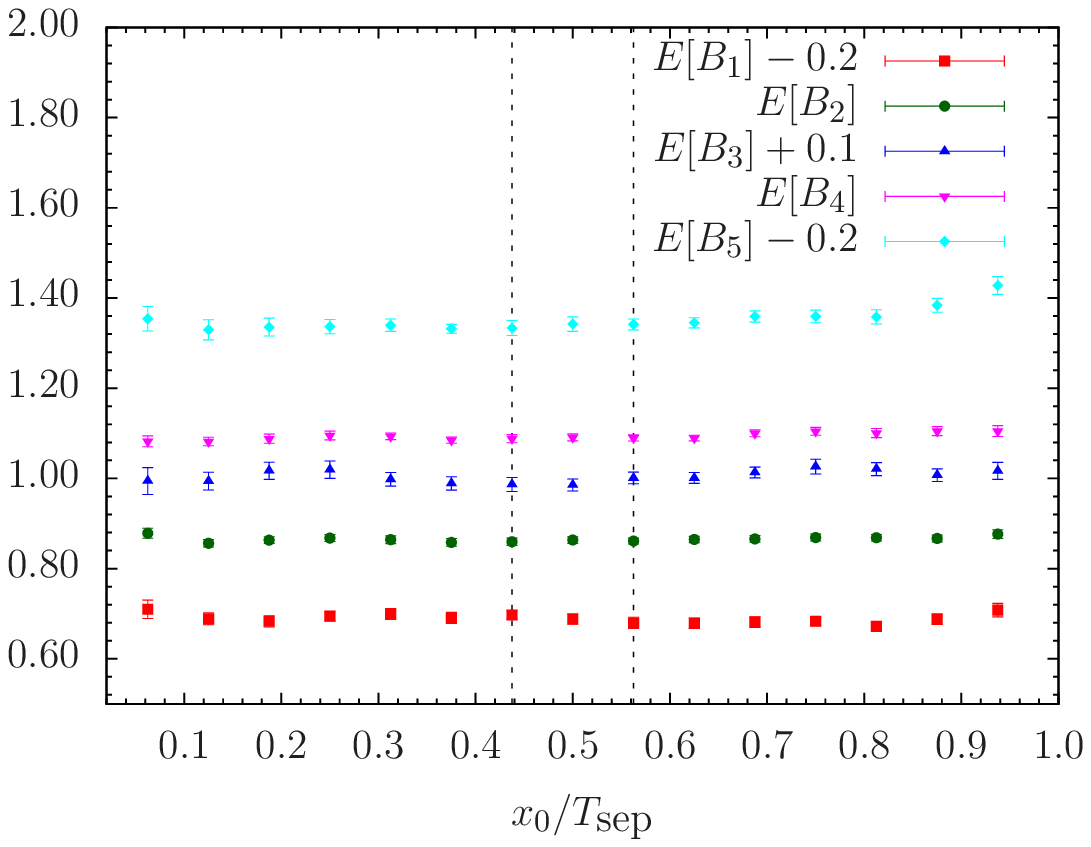}}
\subfigure[]{\includegraphics[scale=0.70,angle=-0]{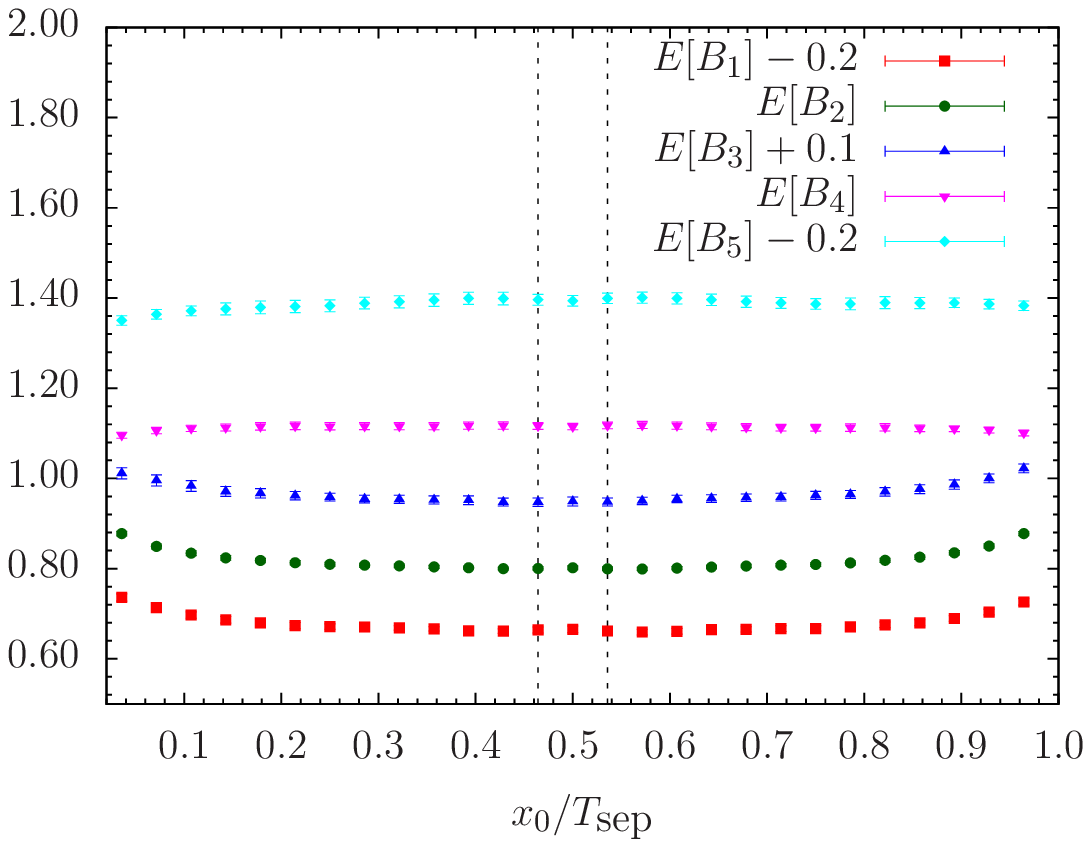}}
\begin{center}
\caption{\sl Plateau quality for the quantity $E[B_i]$ ($i=1, \ldots, 5$) defined in 
Eq.~(\ref{BBratio}) for a value of $a\mu_s$ close to the 
physical one: (a) $\beta=3.80$,  $(a\mu_{\ell}, a\mu_{h}) = (0.0080, 0.5246)$; 
(b) $\beta=4.20$,  $(a\mu_{\ell}, a\mu_{h}) = (0.0020, 0.3525)$. For clarity of presentation some ensembles of datapoints have been 
shifted as indicated in the label.
  }
\label{fig:BB_plateau}
\end{center}
\end{figure}

Bare $B_{i}$ ($i=1, \ldots, 5$) can be evaluated from ratios of 3-point, 
$C_{3;i}(x_0)$ and two 2-point, $C_2(x_0)$ and $C_2^{\prime}(x_0)$ correlation functions 
(for more details see the discussion that leads to Eqs.~(4.10)-(4.13) of Ref.~\cite{Bertone:2012cu}):
\begin{equation} 
E[B_{i}](x_0) \,= \, \dfrac{C_{3;i}(x_0)}{C_{2}(x_0) \,\, C_{2}^\prime(x_0)}, \quad  i=1, \ldots, 5\,.
\label{BBratio}
\end{equation}
For large time separation from the location of the two meson sources (defined at time positions $y_0$ and $y_0+T_{\rm{sep}}$),  
the plateau of the ratio~(\ref{BBratio}) provides an estimate of the (bare) $B_i$ ($i=1, \ldots, 5$) values, each 
multiplied by the corresponding factor 
$({\cal C}_1, {\cal C}_2, {\cal C}_3, {\cal C}_4, {\cal C}_5 )=(8/3,\, -5/3,\, 1/3,\, 2,\, 2/3)$. 
In Figs.~\ref{fig:BB_plateau}(a) and \ref{fig:BB_plateau}(b) we show two examples of the plateaux 
quality at $\beta=3.80$ and $\beta=4.20$.
    
RCs of the four-fermion operators have been calculated using the RI-MOM 
method~\cite{RIMOM}. For their computation and results we refer the reader to Appendices B and C of 
Ref.~\cite{Bertone:2012cu}.  
   
\subsection{Ratio method for the bag parameters and $\xi$}

We apply the ratio method to compute the bag parameters relevant to the neutral $B$-meson system. 
We introduce  ratios that approach unity in the static limit. 
In HQET, the static limit of each of the five bag
parameters is a constant. 
Hence, following a procedure analogous to that used for the $b$-quark mass and decay constants determination, we 
build ratios for each bag parameter computed at nearby values of the heavy quark mass that we define to be\footnote{Summation over 
repeated indices is assumed.}
\begin{eqnarray}
\omega^{(d)}_{i}(\overline\mu_h, \lambda; \overline\mu_{\ell}, a) &=& \dfrac{W_{ij}(\mu^{\star}, \overline\mu_h, \mu)}
{W_{ij}(\mu^{\star}, \overline\mu_h/\lambda, \mu)} \dfrac{B^{(d)}_j(\overline\mu_h, \overline\mu_{\ell}, a)}
{B^{(d)}_j(\overline\mu_h/\lambda, \overline\mu_{\ell}, a)} \,\,\,\,\,\, (i, j=1, \ldots, 5) \label{eq:omega_d}  \\
\omega^{(s)}_{i}(\overline\mu_h, \lambda; \overline\mu_{\ell},  \overline\mu_s, a) &=& 
\dfrac{W_{ij}(\mu^{\star}, \overline\mu_h, \mu)}
{W_{ij}(\mu^{\star}, \overline\mu_h/\lambda, \mu)} \dfrac{B^{(s)}_j(\overline\mu_h, \overline\mu_{\ell}, \overline\mu_s, a )}
{B^{(s)}_j(\overline\mu_h/\lambda, \overline\mu_{\ell},  \overline\mu_s, a)} \,\,\,\, (i, j=1, \ldots, 5) , \label{eq:omega_s} 
\end{eqnarray}    
where $B^{(d/s)}_i$ ($i=1, \ldots, 5$) denote the renormalised five bag parameters 
computed at the scale $\mu$ (in our case $\mu=3$ GeV). The indices $(d/s)$ correspond to the bag parameters controlling
($\overline B^{0}_{d}-B^{0}_{d}$) and ($\overline B^{0}_{s}-B^{0}_{s}$) mixing, respectively. 
The $5 \times 5$ matrix  $W$ converts (to a given order in RG-improved
perturbation theory) the QCD $B$-parameters into their 
counterparts in the HQET theory, thereby removing (to the same
perturbative order) the corrections logarithmic in $\overline\mu_h $ in
the O($(1/\overline\mu_h)^0$)--term of the $\omega_i^{(d,s)}$ ratios above.
It incorporates the QCD evolution from the scale $\mu$ to the reference scale identified by the 
heavy quark mass  and the  HQET evolution from the same heavy quark mass scale to some arbitrary
scale $\mu^{\star}$ (the dependence on which cancels in all ratios above),
as well as the matching between the two theories. At NLL order the $W$ matrix has a ($3 \times 3 \oplus 2 \times 2$) 
block diagonal form. Note that at TL or LL order the SM bag parameters $B^{(d/s)}_1$ 
evolve without mixing with $B^{(d/s)}_2$ and $B^{(d/s)}_3$. For more details on the 
QCD-HQET matching of the $B$-parameters see Appendix~\ref{APP_QCD-HQET}. 
Moreover in Appendix~\ref{APP_O123} we show that using the
ratio method (with QCD to HQET matching at TL order) one can numerically verify 
with good precision the relationship which connects, via the equations of motion, 
the operators $\mathcal{O}_{1}^{(q)}$,  $\mathcal{O}_{2}^{(q)}$ and $\mathcal{O}_{3}^{(q)}$ ($q = d, s$) in the static limit. 

Similarly we define the double ratio relevant to the SM bag parameters:
\begin{equation} \label{eq:zeta_omega}
\zeta_{\omega}(\overline\mu_h, \lambda; \overline\mu_{\ell},  \overline\mu_s, a) = 
\dfrac{\omega^{(s)}_{1}(\overline\mu_h, \lambda; \overline\mu_{\ell},  
\overline\mu_s, a)}{\omega^{(d)}_{1}(\overline\mu_h, \lambda; \overline\mu_{\ell}, a)} .
\end{equation}

It has been noticed that,  due to strong cancellations of systematic effects,
it is convenient for the unitarity triangle analysis, 
to compute the SU(3)-breaking parameter $\xi$, 
\begin{equation}\label{eq:xi_def} 
\xi = \dfrac{f_{Bs}}{f_{Bd}}\, \sqrt{\dfrac{B^{(s)}_1}{B^{(d)}_1}}, 
\end{equation}
whose knowledge fixes  the ratio of the CKM matrix elements, 
$|V_{td}|/|V_{ts}|$. 
According to the philosophy of this paper we define the ratio (see Eqs~(\ref{z1}), (\ref{z2}), 
(\ref{eq:omega_d}) and (\ref{eq:omega_s})) 
\begin{equation}\label{eq:zeta_xi}
\zeta_{\xi}(\overline\mu_h, \lambda; \overline\mu_{\ell},  \overline\mu_s, a) = 
\dfrac{z_s}{z_d} \sqrt{\dfrac{\omega^{(s)}_{1}}{\omega^{(d)}_{1}}} .
\end{equation} 
The static limit of the ratios defined in Eqs~(\ref{eq:omega_d}), (\ref{eq:omega_s}), (\ref{eq:zeta_omega}) and (\ref{eq:zeta_xi})
is 1.
At this point we consider chain equations analogous to the ones of Eqs.~(\ref{eq:y_chain}) and (\ref{eq:zs_chain}). 
For example we consider the chains 
\begin{eqnarray}
\omega_i^{(d)}(\overline\mu_h^{(2)})\, \omega_i^{(d)}(\overline\mu_h^{(3)})\,\ldots \, \omega_i^{(d)}(\overline\mu_h^{(K+1)}) &=& 
\dfrac{W_{ij}(\mu^{\star}, \overline\mu_h^{(K+1)}, \mu)\, B^{(d)}_j(\overline\mu_h^{(K+1)}, \overline\mu_{u/d}) }
{W_{ij}(\mu^{\star}, \overline\mu_h^{(1)}, \mu)\, B^{(d)}_j(\overline\mu_h^{(1)}, \overline\mu_{u/d})} 
\label{eq:omega_d_chain} \\
\omega_i^{(s)}(\overline\mu_h^{(2)})\, \omega_i^{(s)}(\overline\mu_h^{(3)})\,\ldots \, \omega_i^{(s)}(\overline\mu_h^{(K+1)}) &=& 
\dfrac{W_{ij}(\mu^{\star}, \overline\mu_h^{(K+1)}, \mu)\, B^{(s)}_j(\overline\mu_h^{(K+1)}, 
\overline\mu_{u/d}, \overline\mu_{s}) }
{W_{ij}(\mu^{\star}, \overline\mu_h^{(1)}, \mu)\, B^{(s)}_j(\overline\mu_h^{(1)}, \overline\mu_{u/d}, 
\overline\mu_{s})} .\label{eq:omega_s_chain}
\end{eqnarray}
\begin{figure}[!t]
\subfigure[]{\includegraphics[scale=0.70,angle=-0]{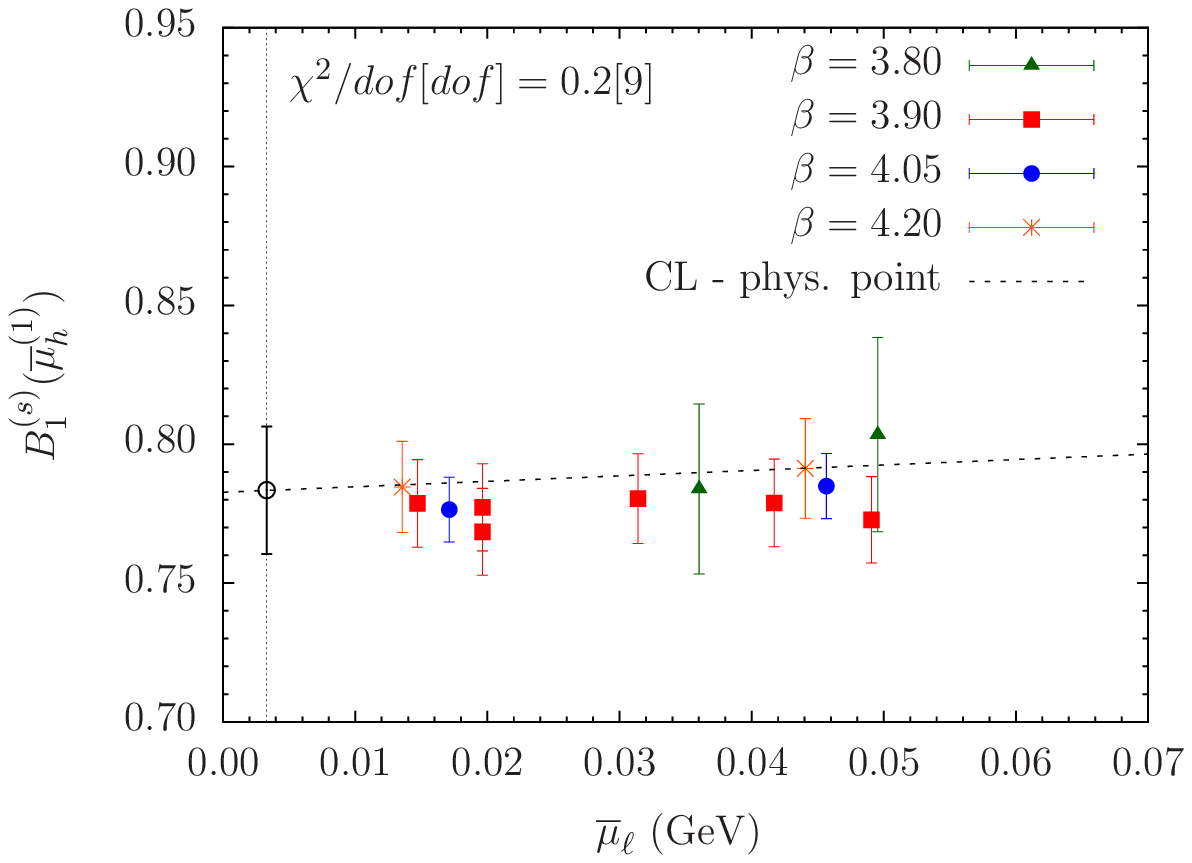}}
\subfigure[]{\includegraphics[scale=0.70,angle=-0]{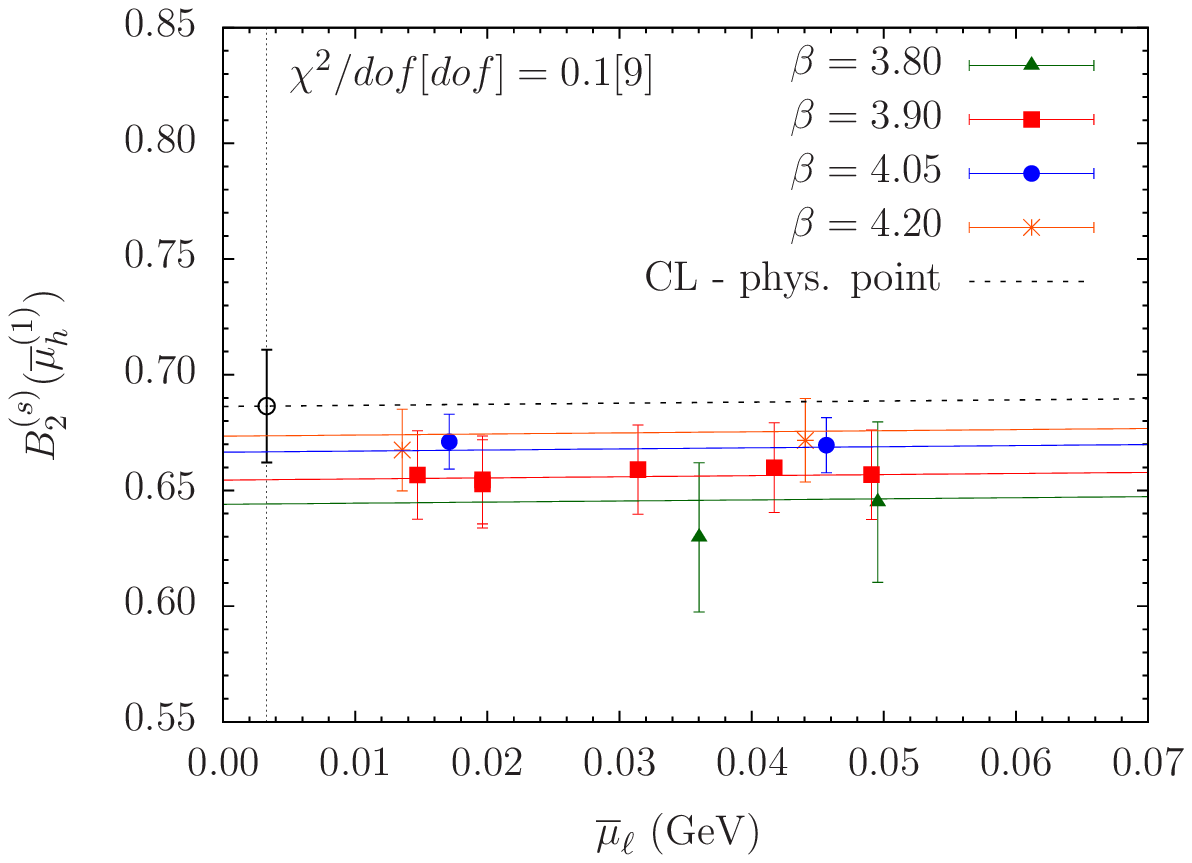}}
\begin{center}
\caption{\sl Combined chiral and continuum fit at the triggering point of the bag parameters $B_1^{(s)}$ and $B_2^{(s)}$ 
are shown in panels 
(a) and (b) respectively. Both fit ans\"atze are linear in $\overline\mu_{\ell}$ 
and in $a^2$. Empty black circle is our result at the physical $u/d$ quark mass point in the continuum 
limit for both cases. In panel (a) discretisation effects are rather small and to avoid cluttering the figure only the 
extrapolated line is shown.
  }
\label{fig:Bs12_trig}
\end{center}
\end{figure}
\begin{figure}[!t]
\subfigure[]{\includegraphics[scale=0.70,angle=-0]{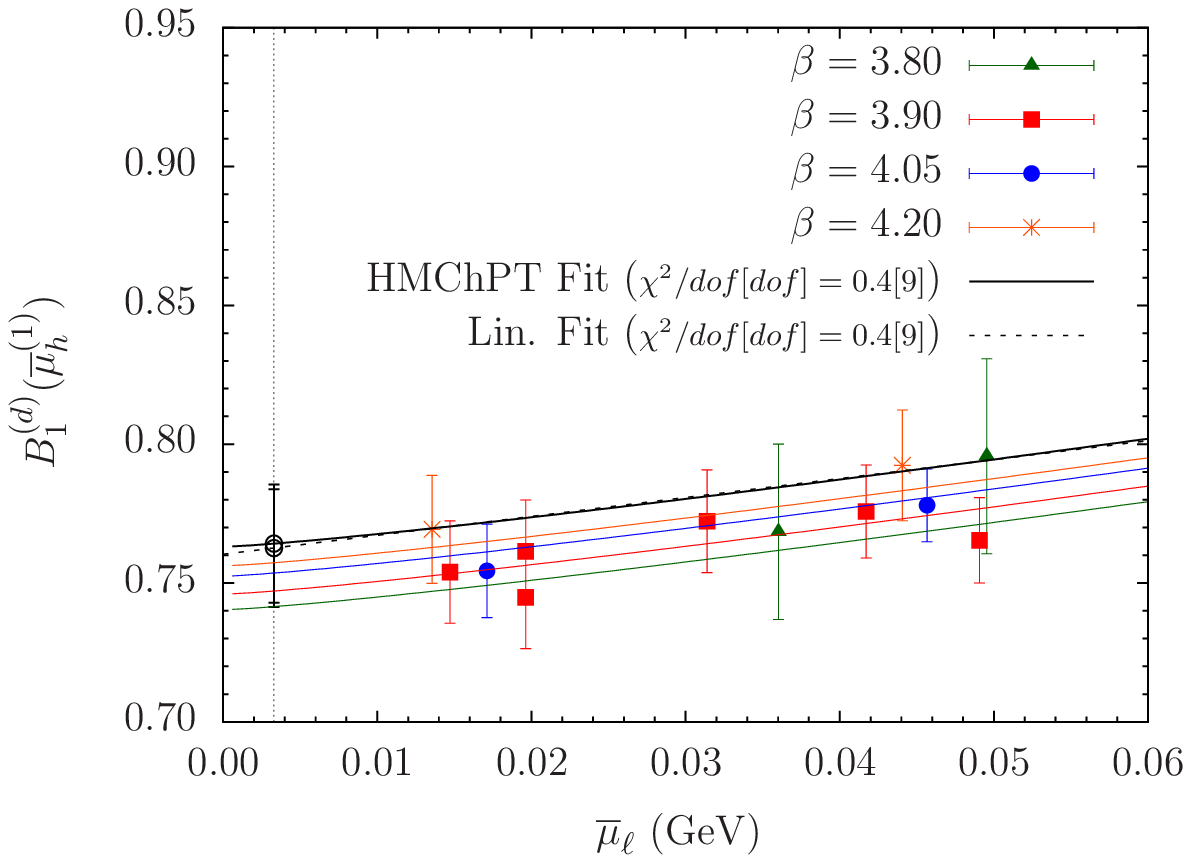}}
\subfigure[]{\includegraphics[scale=0.70,angle=-0]{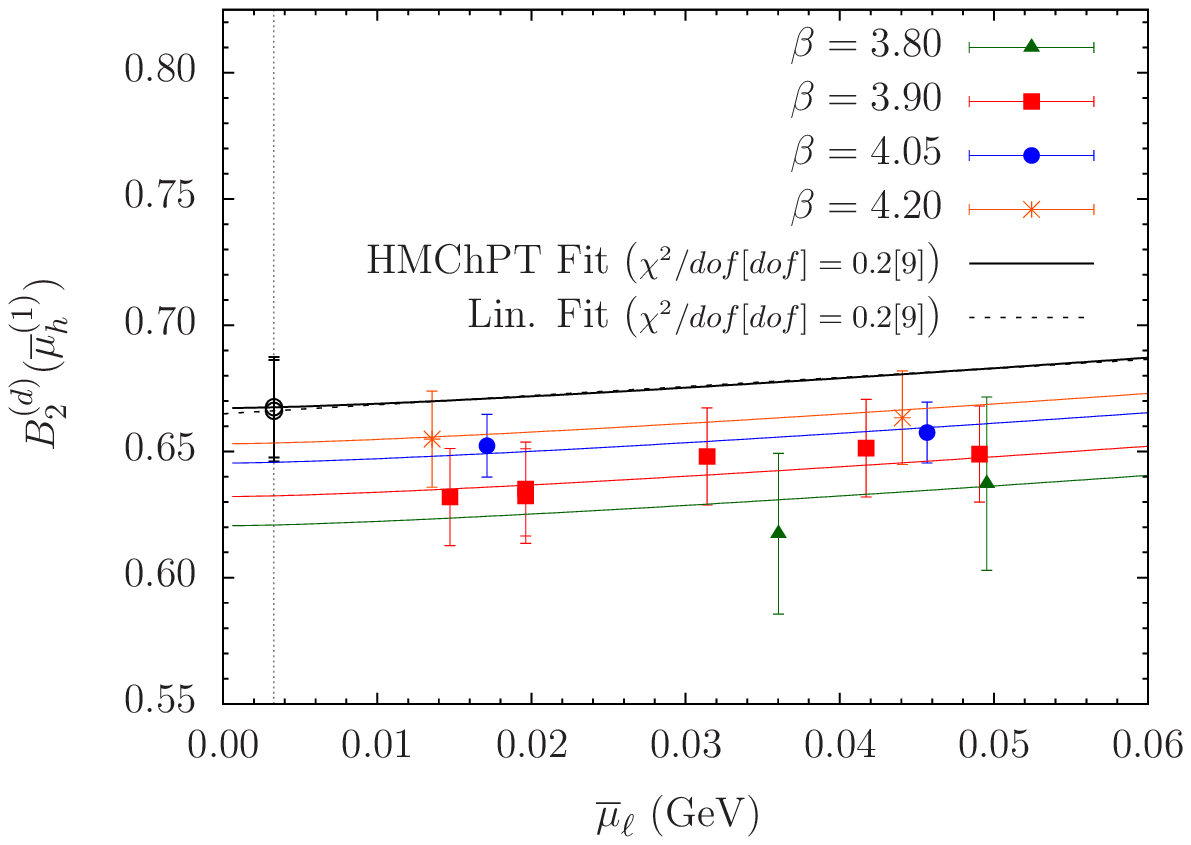}}
\begin{center}
\caption{\sl Combined chiral and continuum fit at the triggering point of the bag parameters $B_1^{(d)}$ and $B_2^{(d)}$ 
are shown in panels 
(a) and (b), respectively. For both quantities we have used a linear fit ansatz in 
$\overline\mu_{\ell}$ and the fit function proposed by 
HMChPT, (see Eq.~(\ref{eq:B1d_trig})). Colored lines correspond to the latter case. 
The results at the physical point  - empty black circle - obtained by the two fit function choices are 
indistinguishable.  
  }
\label{fig:Bd12_trig}
\end{center}
\end{figure}
Bag parameters defined at $\overline\mu_h^{(1)}$ are determined as usual by a combined chiral and continuum  
fit that shows small discretisation effects. In Figs~\ref{fig:Bs12_trig}(a) and \ref{fig:Bs12_trig}(b) we present the fit 
for the cases of $B^{(s)}_1$ and  $B^{(s)}_2$ respectively. We illustrate  the cases of  
$B^{(d)}_1$ and  $B^{(d)}_2$ in Figs~\ref{fig:Bd12_trig}(a) and ~\ref{fig:Bd12_trig}(b) respectively. 
A similar behaviour is observed for the rest of the bag parameters $B^{(s)}_i$ and $B^{(d)}_i$ with 
($i=3, 4, 5$). Note that for the $B_i^{(d)}$ cases HMChPT predicts a logarithmic chiral behaviour~\cite{Becirevic:2006me}. 
We have thus tried besides a linear fit ansatz, fit functions of the following type:
\begin{eqnarray}\label{eq:B1d_trig}
B_1^{(d)} &=& B_{1}^{\chi}\Big[1+ b_{1} \overline\mu_{\ell} - \frac{(1-3\hat{g}^2)}{2}\frac{2 B_0 \overline\mu_{\ell}}{(4 \pi f_0)^2} 
{\rm log}\Big( \frac{2 B_0 \overline\mu_{\ell}}{(4 \pi f_0)^2}  \Big) \Big]  + D_1 a^2  \\
B_i^{(d)} &=& B_{i}^{\chi}\Big[1+ b_{i} \overline\mu_{\ell} - \frac{(1-3\hat{g}^2\,Y)}{2}\frac{2 B_0 \overline\mu_{\ell}}{(4 \pi f_0)^2} 
{\rm log}\Big( \frac{2 B_0 \overline\mu_{\ell}}{(4 \pi f_0)^2}  \Big) \Big]  + D_i a^2,  \,\,\,\, i=2 \\
B_i^{(d)} &=& B_{i}^{\chi}\Big[1+ b_{i} \overline\mu_{\ell} + \frac{(1+3\hat{g}^2\,Y)}{2}\frac{2 B_0 \overline\mu_{\ell}}{(4 \pi f_0)^2} 
{\rm log}\Big( \frac{2 B_0 \overline\mu_{\ell}}{(4 \pi f_0)^2}  \Big) \Big]  + D_i a^2,  \,\,\,\, i=4, 5 
\end{eqnarray} 
where we use the HMChPT-based estimates $Y=1$~\cite{Becirevic:2006me} and $\hat{g}=0.61(7)$~\cite{Nakamura:2010zzi}. 
The bag parameter $B_3$ is related in HQET to the bag parameters $B_1$
and $B_2$ (see Appendix~\ref{APP_O123}). 
In the case $Y = 1$, which is the only one considered in this paper, the chiral expansion for $B_3$ 
is similar to the one of $B_2$ with the same chiral log.
The results at the physical point  -- empty black circle in Figs~\ref{fig:Bd12_trig}(a) and \ref{fig:Bd12_trig}(b) -- 
obtained by the two fit functional choices (HMChPT  and linear fit) are in practice indistinguishable.
\begin{figure}[!h]
\subfigure[]{\includegraphics[scale=0.70,angle=-0]{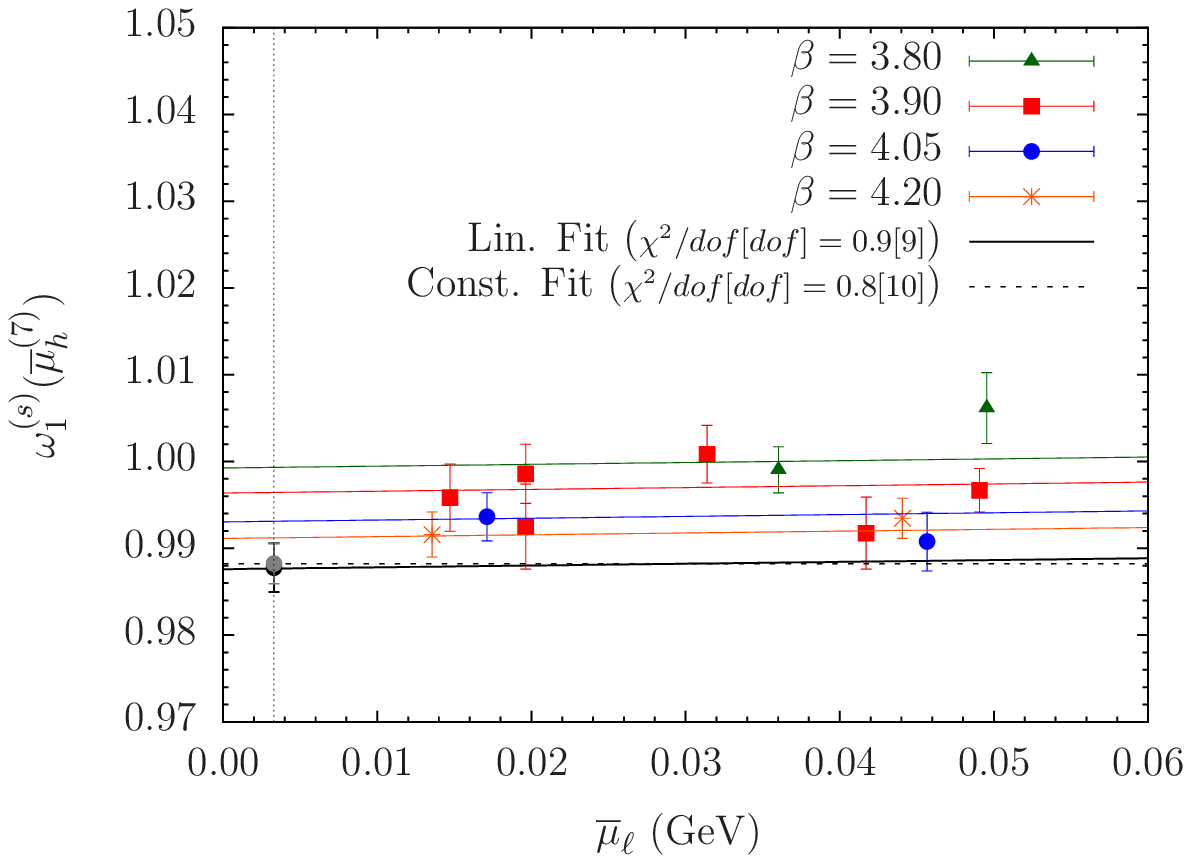}}
\subfigure[]{\includegraphics[scale=0.70,angle=-0]{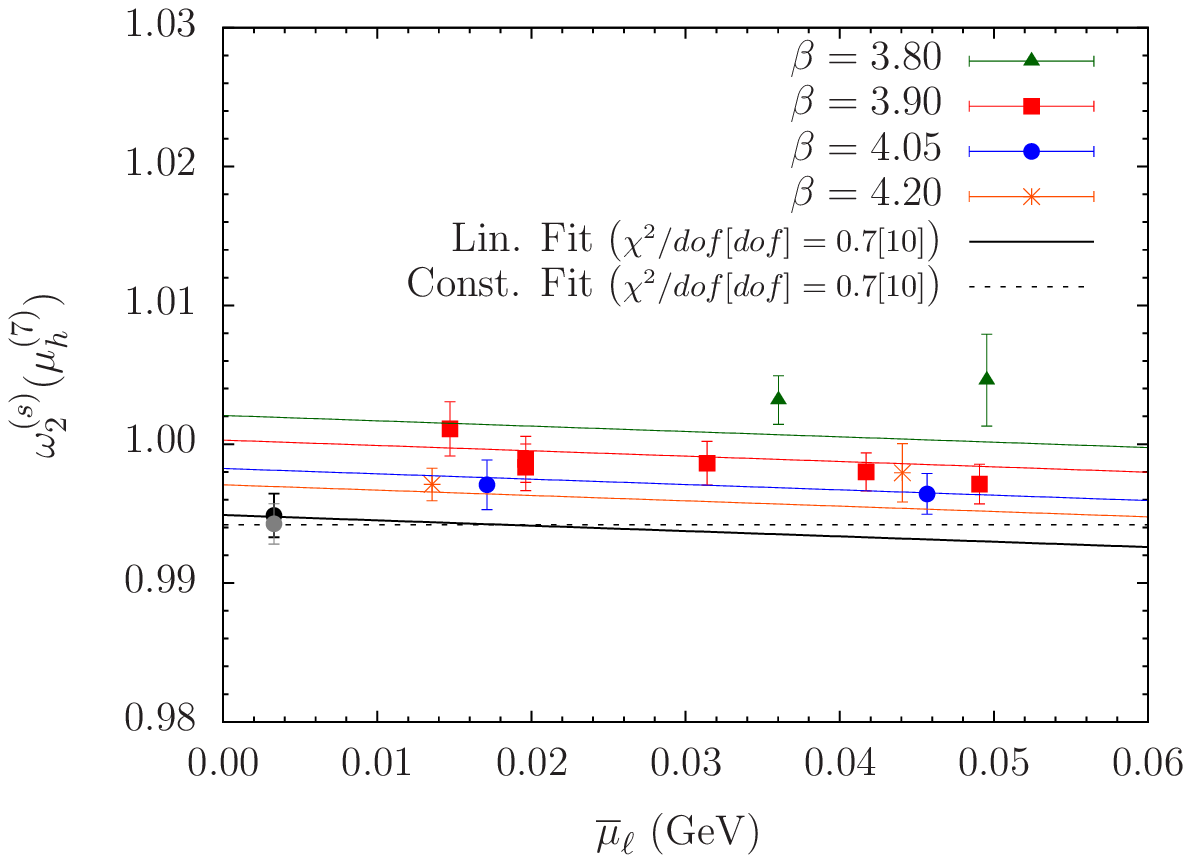}}
\begin{center}
\caption{\sl Combined chiral and continuum fits of the ratio $\omega_1^{(s)}(\overline\mu_h^{(n)})$  and 
 $\omega_2^{(s)}(\overline\mu_h^{(n)})$ (Eq.~(\ref{eq:omega_s_chain})) against $\overline\mu_{\ell}$ are shown 
 in the panels (a) and (b), respectively.
In both cases ratios for the largest value of the heavy quark mass are reported ($n=7$). 
Linear fit ansatz in $\overline\mu_{\ell}$ and 
fit to a constant are shown. Colored lines show the linear fit to the data. 
Empty black circle and full grey circle are the results, respectively, at the physical $u/d$ quark 
mass point in the continuum limit.
  }
\label{fig:ratios_ws}
\end{center}
\end{figure}
In Fig.~\ref{fig:ratios_ws}(a) and Fig.~\ref{fig:ratios_ws}(b) we illustrate two examples of the combined chiral and continuum fits 
for the ratios $\omega^{(s)}_1(\overline\mu_h^{(n)})$ and $\omega^{(s)}_2(\overline\mu_h^{(n)})$ at the largest value of heavy 
quark mass used in this part of the analysis ($n=7$). 

\begin{figure}[!hb]
\subfigure[]{\includegraphics[scale=0.60,angle=-0]{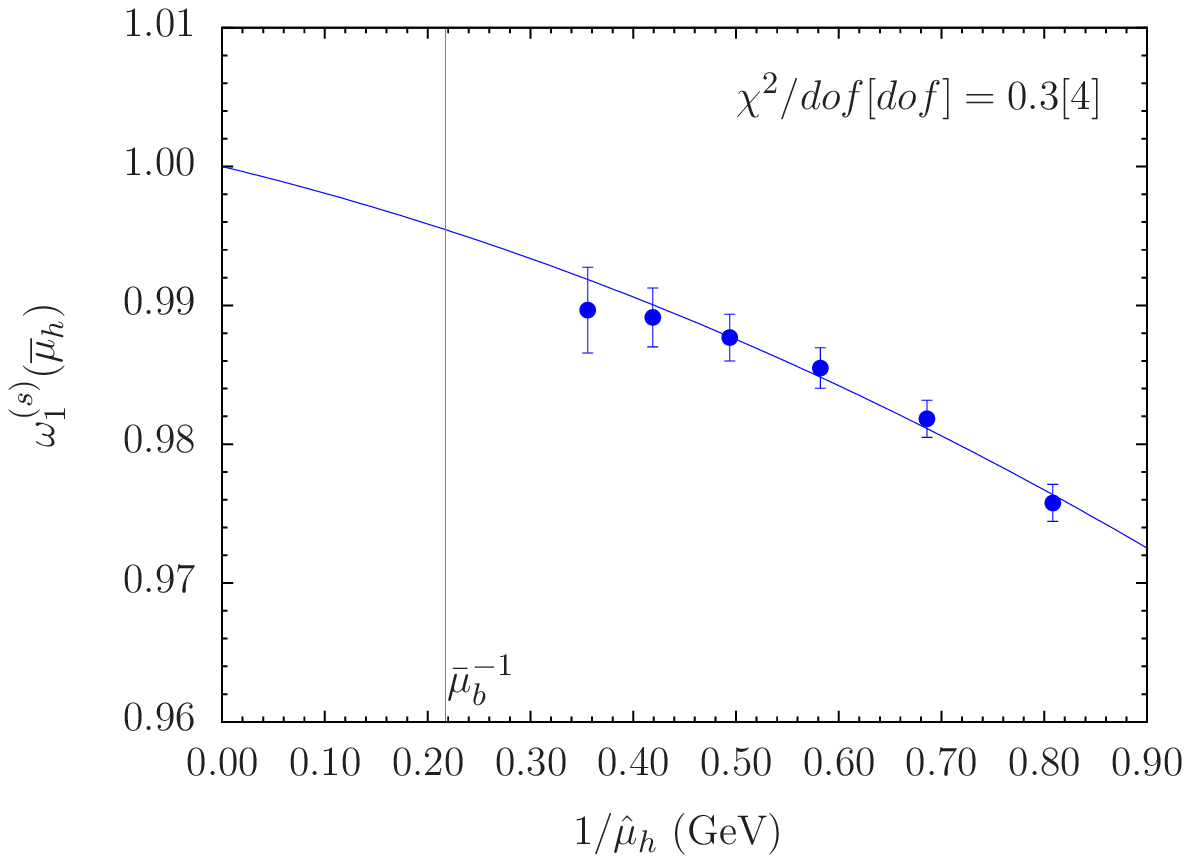}}
\subfigure[]{\includegraphics[scale=0.60,angle=-0]{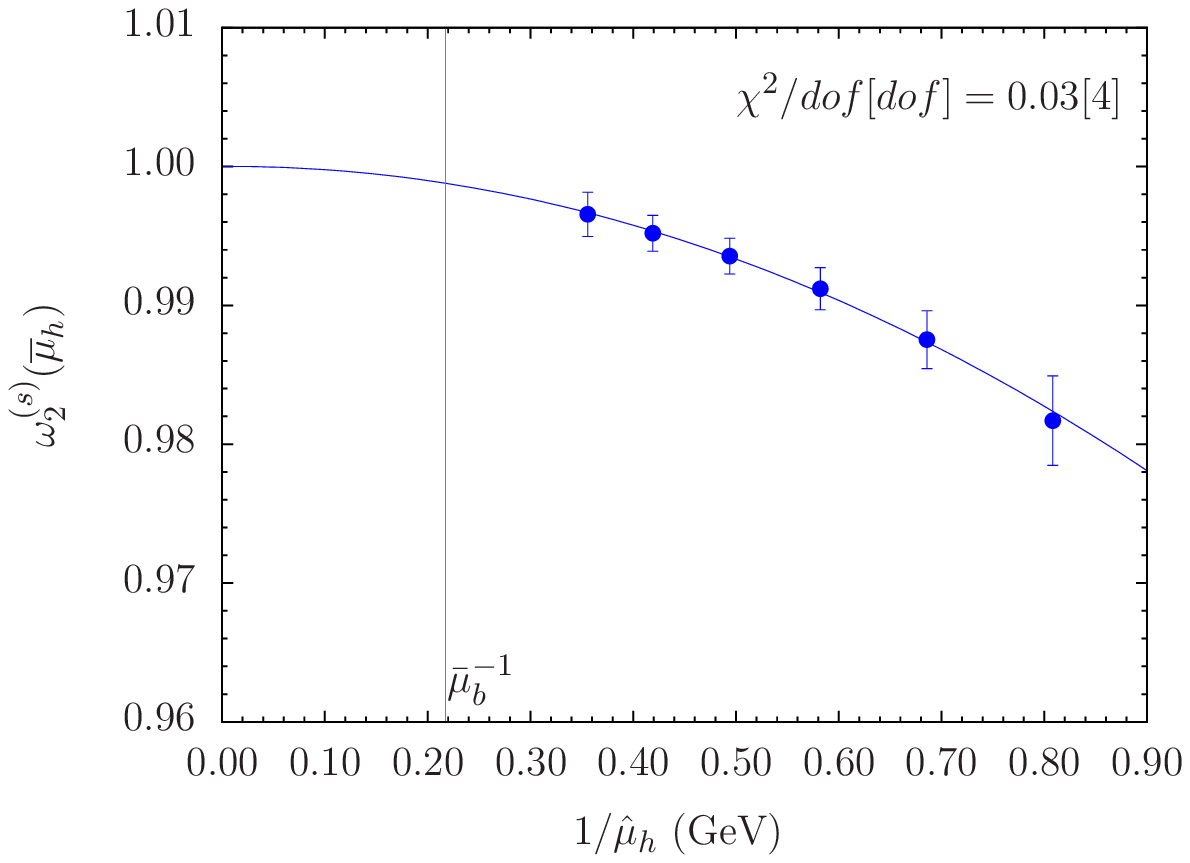}}
\subfigure[]{\includegraphics[scale=0.60,angle=-0]{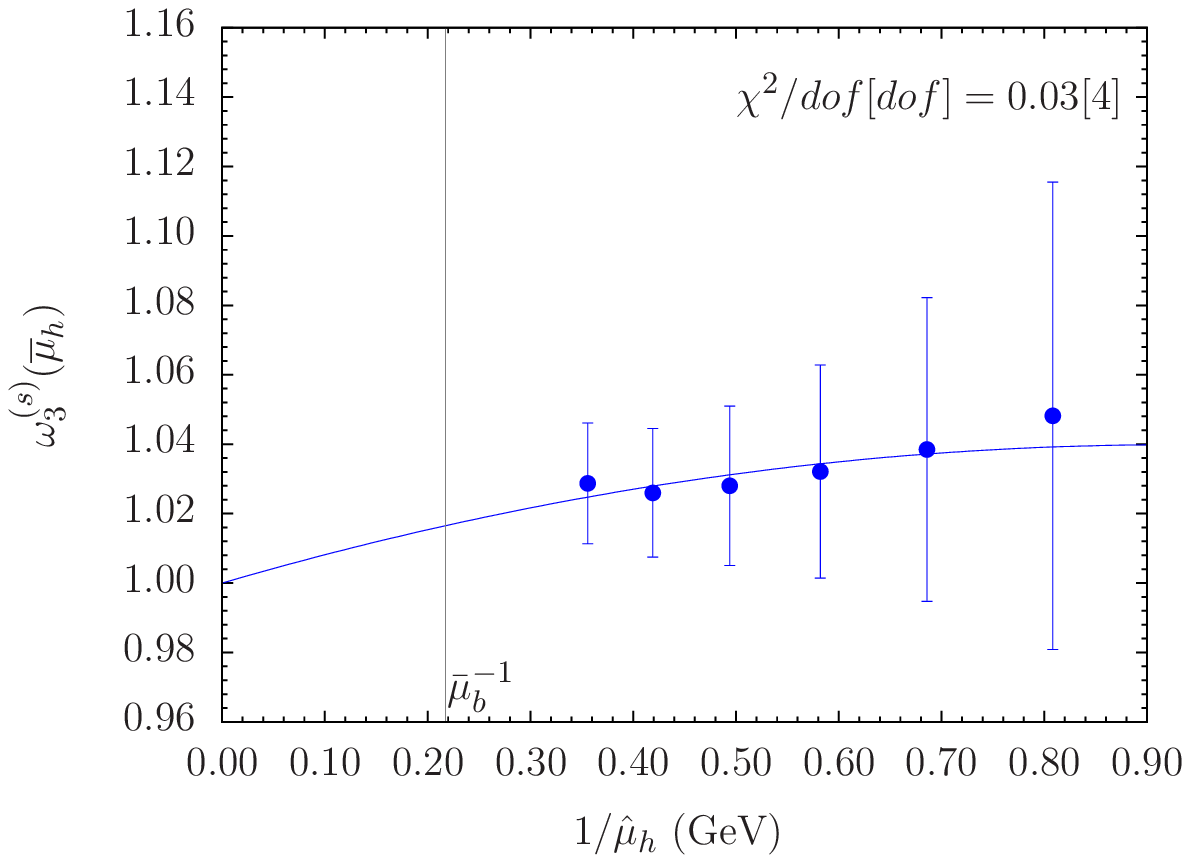}}
\subfigure[]{\includegraphics[scale=0.60,angle=-0]{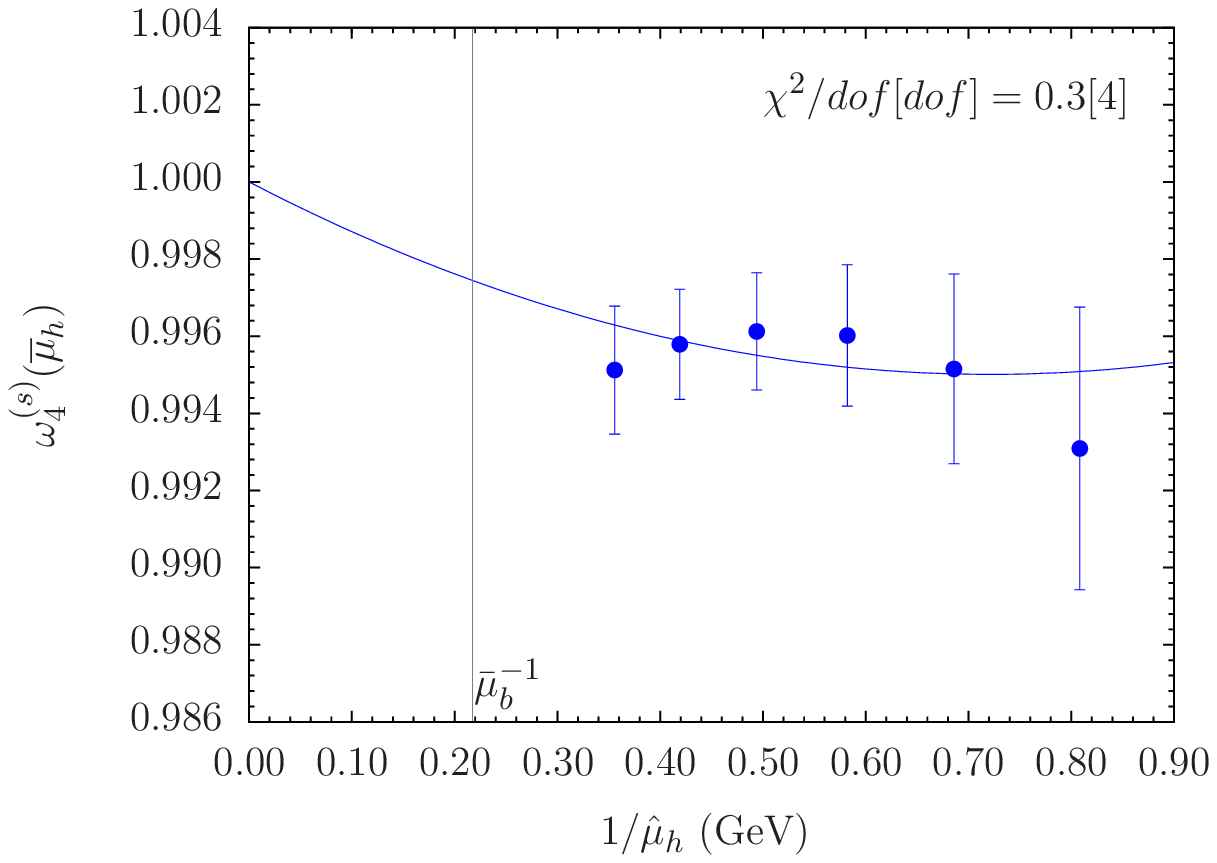}}
\subfigure[]{\includegraphics[scale=0.60,angle=-0]{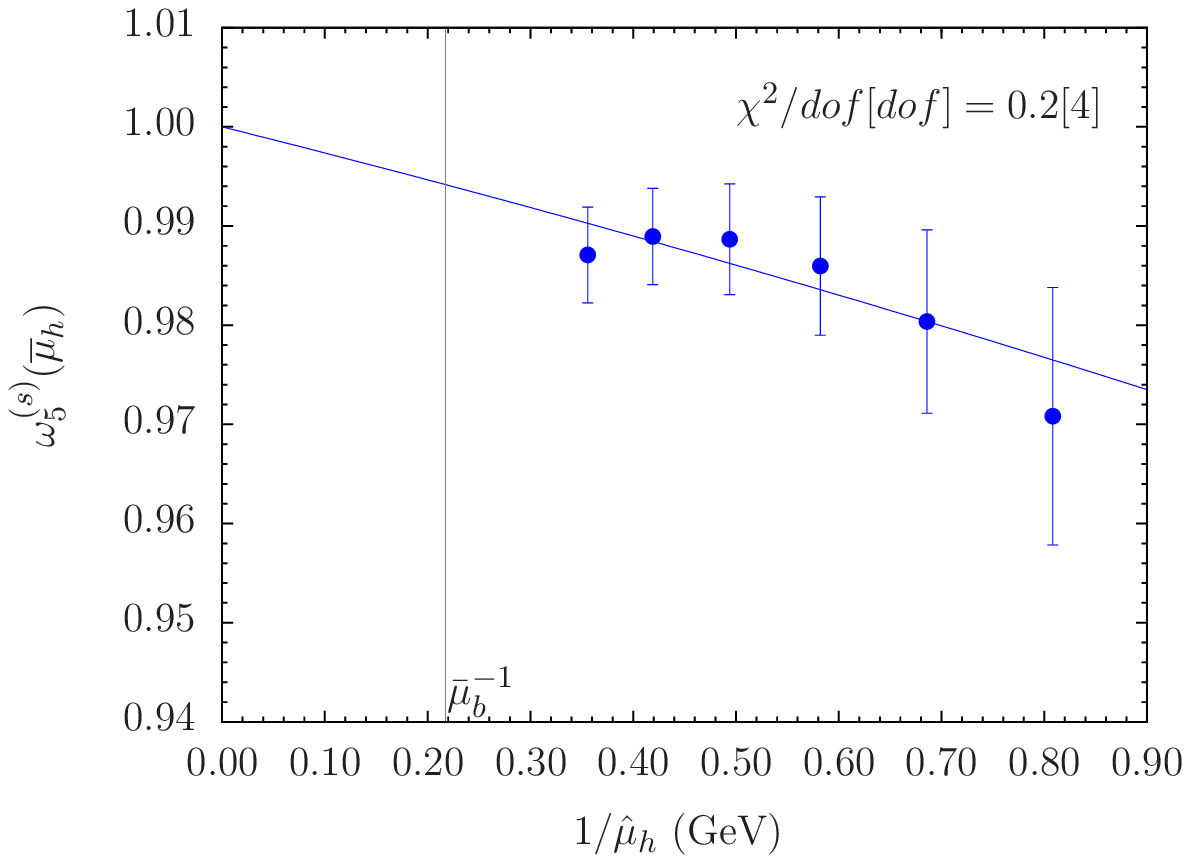}}
\begin{center}
\caption{\sl $\omega_i^{(s)}(\overline\mu_h)$  against $1/\overline\mu_h$ for $i=1, \ldots, 5$ are shown 
in panels (a),$\ldots$, (e),  
respectively. In all cases the fit function has a polynomial form like the one in Eq.~(\ref{eq:y_ansatz}). 
The vertical black thin line 
marks the position of $1/\overline\mu_b$.
  }
\label{fig:ws}
\end{center}
\end{figure}

We now pass to the discussion of the dependence on the inverse heavy quark mass of the $\omega$-ratios, 
evaluated in the continuum limit at 
the light quark physical point. 
As before we try polynomial fit ansatz of the  form,  
\begin{equation}\label{eq:omega_ansatz}
\omega_i^{(q)}(\overline\mu_h) = 1 + \dfrac{b_i^{(q)}}{\overline\mu_h} + \dfrac{c_i^{(q)}}{\overline\mu_h^2}~~~~~
 (q \equiv d, s;\,\, i=1, \ldots, 5),
\end{equation}  
using alternatively TL, LL, NLL order expressions for the matrix $W$.
The relevant fits (for the case of $W$ at NLL order) are 
illustrated in the five panels of 
Fig.~\ref{fig:ws} for  $B^{(s)}_i$ with $i=1, \ldots, 5$.
Finally, using Eqs.~(\ref{eq:omega_d_chain}) and 
(\ref{eq:omega_s_chain}) we obtain the values of the bag parameters at the $b$-quark mass 
by identifying $\overline\mu_h^{(K+1)}$ with $\overline\mu_b$ given in Eq.~(\ref{eq:mub}). 

\begin{figure}[!h]
\subfigure[]{\includegraphics[scale=0.70,angle=-0]{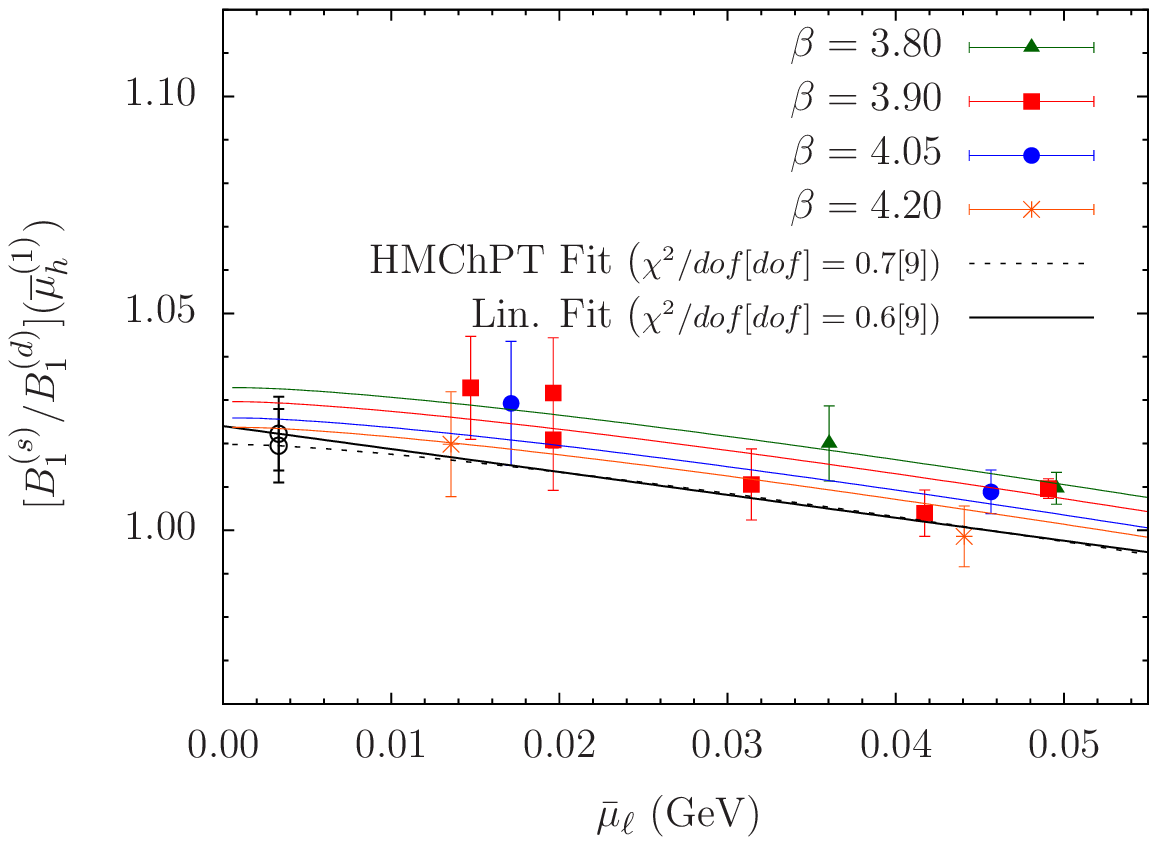}}
\hspace*{0.7cm}
\subfigure[]{\includegraphics[scale=0.70,angle=-0]{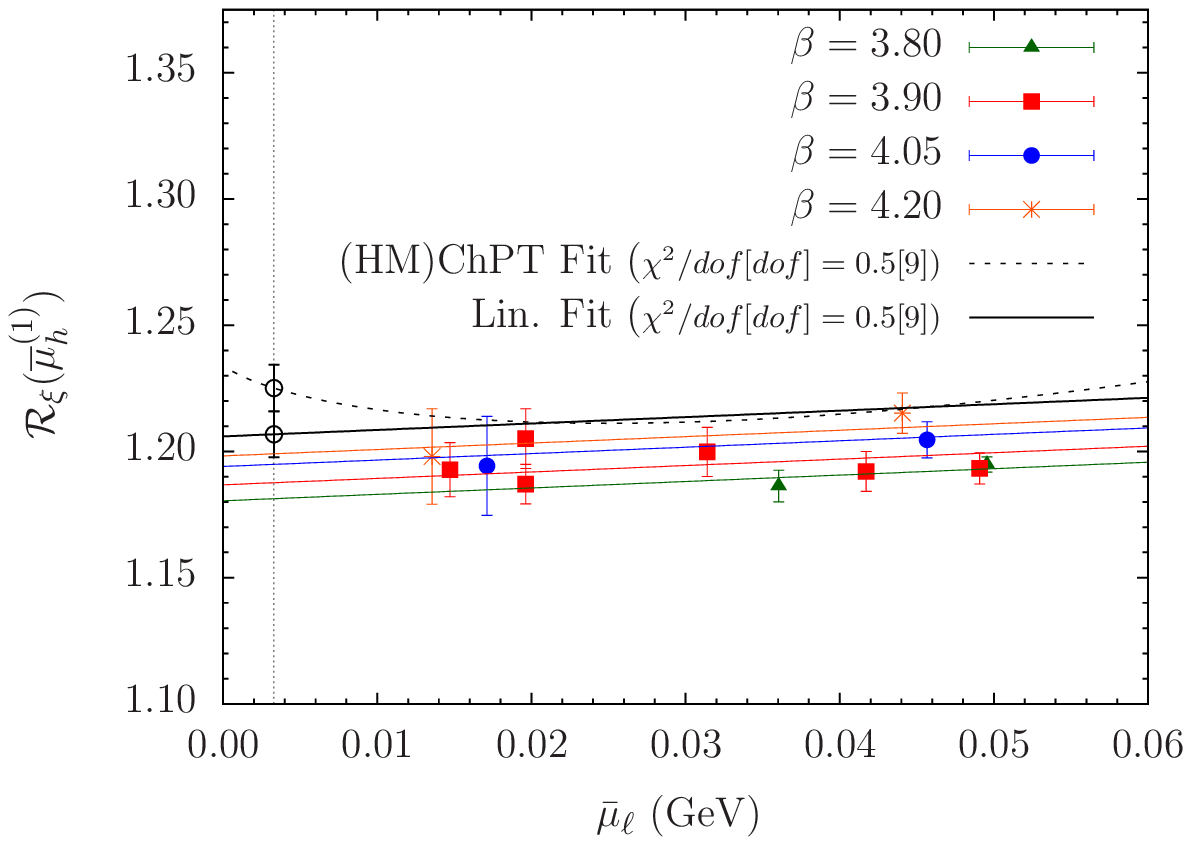}}
\begin{center}
\caption{\sl Combined chiral and continuum fits at the triggering point (a) of the double ratio 
$[B^{(s)}_1/B^{(d)}_1](\overline\mu_h^{(1)})$ and  
(b) of the quantity ${\cal R}_{\xi}(\overline\mu_h^{(1)})$  are shown. In both cases (HM)ChPT and linear fit functions have been used. 
In panel (a) colored lines correspond to the (HM)ChPT fit ansatz while in panel (b) to the linear fit one.  
In each panel empty black circle denotes the result at the physical $u/d$ quark mass point in the continuum 
limit. 
  }
\label{fig:BBsNNd_xi_trig}
\end{center}
\end{figure}

We follow the same procedure to determine the numerical value of the ratio $B^{(s)}_1 / B^{(d)}_1$ and the parameter $\xi$. 
Our final results will be presented in the next section. In 
Fig.~\ref{fig:BBsNNd_xi_trig}(a) we show 
the combined chiral and continuum fit for the bag parameter ratios at $\overline\mu_h^{(1)}$ for four values of the lattice spacing. 
We have made use of two  fit ans\"atze. The first one  suggested by the HMChPT reads
\begin{equation}\label{eq:HMChPT_ratioB} 
\Big[\dfrac{B^{(s)}_1}{  B^{(d)}_1}\Big](\overline\mu_h^{(1)}; \overline\mu_{\ell}, a) 
= b_h\Big[ 1+ c_h \overline\mu_{\ell} + \dfrac{1 - 3\hat{g}^2}{2}\dfrac{2B_0\overline\mu_{\ell}}{(4\pi f_0)^2}
{\rm log}\Big( \dfrac{2B_0\overline\mu_{\ell}}{(4\pi f_0)^2}\Big) + d_h a^2 \Big]
\end{equation}
with $\hat{g}=0.61(7)$ \cite{Nakamura:2010zzi}. The second is a linear fit with no logarithmic terms. 
It can be seen from Fig.~\ref{fig:BBsNNd_xi_trig}(a) that both fit ans\"atze lead to compatible results within half standard 
deviation.\footnote{Had we used $\hat{g}=0.53(4)$~\cite{Becirevic:2012pf} in the fit function (\ref{eq:HMChPT_ratioB})  the shift 
of our physical point result would be less than half a standard deviation {\it i.e.} a value similar to the shift due 
to the uncertainty of the estimate of $\hat{g}$ itself.} 
We average over the two results and we take their half difference as a systematic error:
\begin{equation}
\Big[\dfrac{B^{(s)}_1}{  B^{(d)}_1}\Big](\overline\mu_h^{(1)}) =  1.021(9)(2) .
\end{equation} 

We study the dependence of $\xi(\overline\mu_h^{(1)})$ on $\mu_{\ell}$ by making use of the quantity ${\cal R}_f$ defined in 
Eq.~(\ref{eq:Rf}) in order to better control and test the impact of the logarithmic terms from our data. 
The quantity that we fit against the light quark mass  
is defined by 
\begin{equation}
{\cal R}_{\xi} = {\cal R}_f \sqrt{B^{(s)}_1/B^{(d)}_1} .
\end{equation}
The fit is illustrated in Fig.~\ref{fig:BBsNNd_xi_trig}(b). 
We have tried a fit ansatz following SU(2) ChPT and HMChPT combining the formulae given in    
Eqs.~(\ref{eq:HMChPT_ratiof}) and (\ref{eq:HMChPT_ratioB}). 
We have also tried a fit assuming only a linear dependence 
on $\overline{\mu}_{\ell}$.   
Following the same reasoning as in the case of the ratio of the decay constants we take our final result as 
the average over the two values coming from the respective fit ans\"atze. We also consider similar arguments to get 
the estimate of our systematic uncertainty.    
Hence our result at the triggering point reads 
\begin{equation}\label{eq:xi_trig}
\xi(\overline\mu_h^{(1)}) = 1.215(9)(22), 
\end{equation}
where the first error is statistical while the second is systematic. 
\begin{figure}[!h]
\subfigure[]{\includegraphics[scale=0.70,angle=-0]{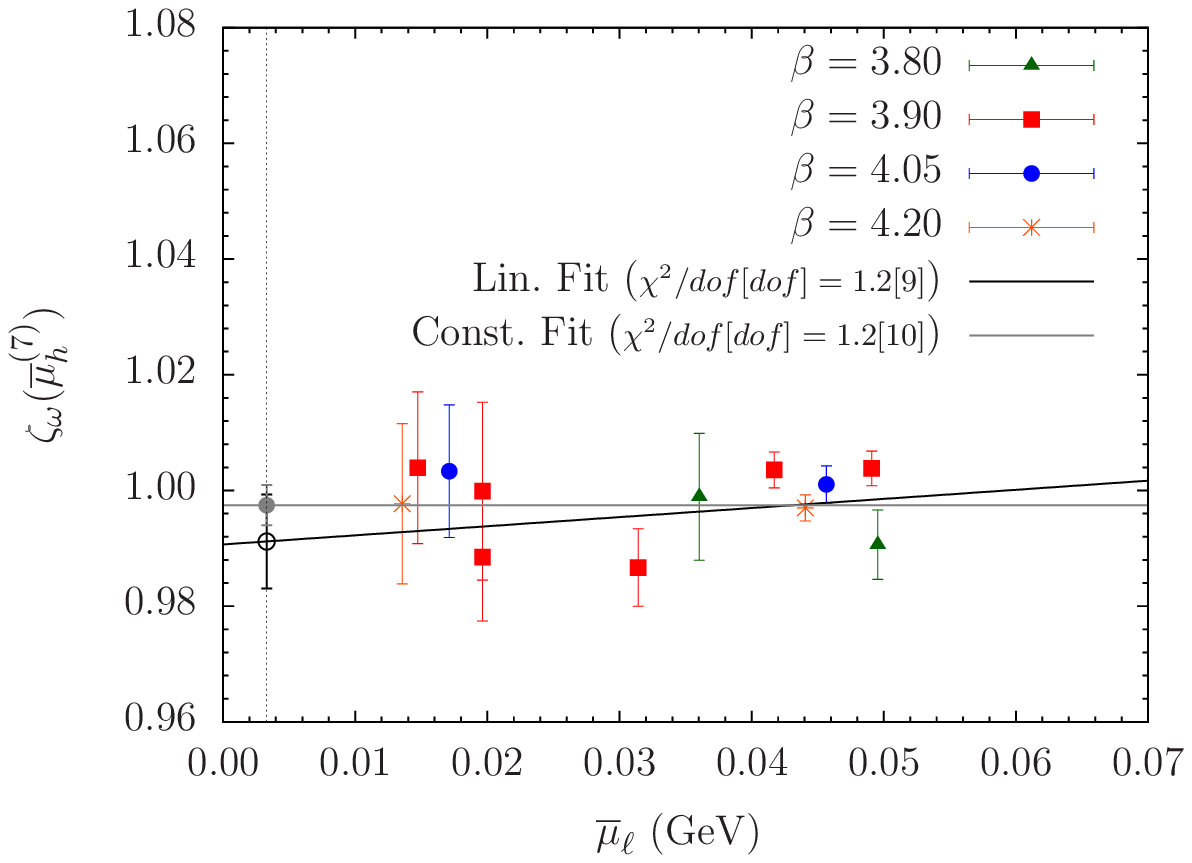}}
\subfigure[]{\includegraphics[scale=0.70,angle=-0]{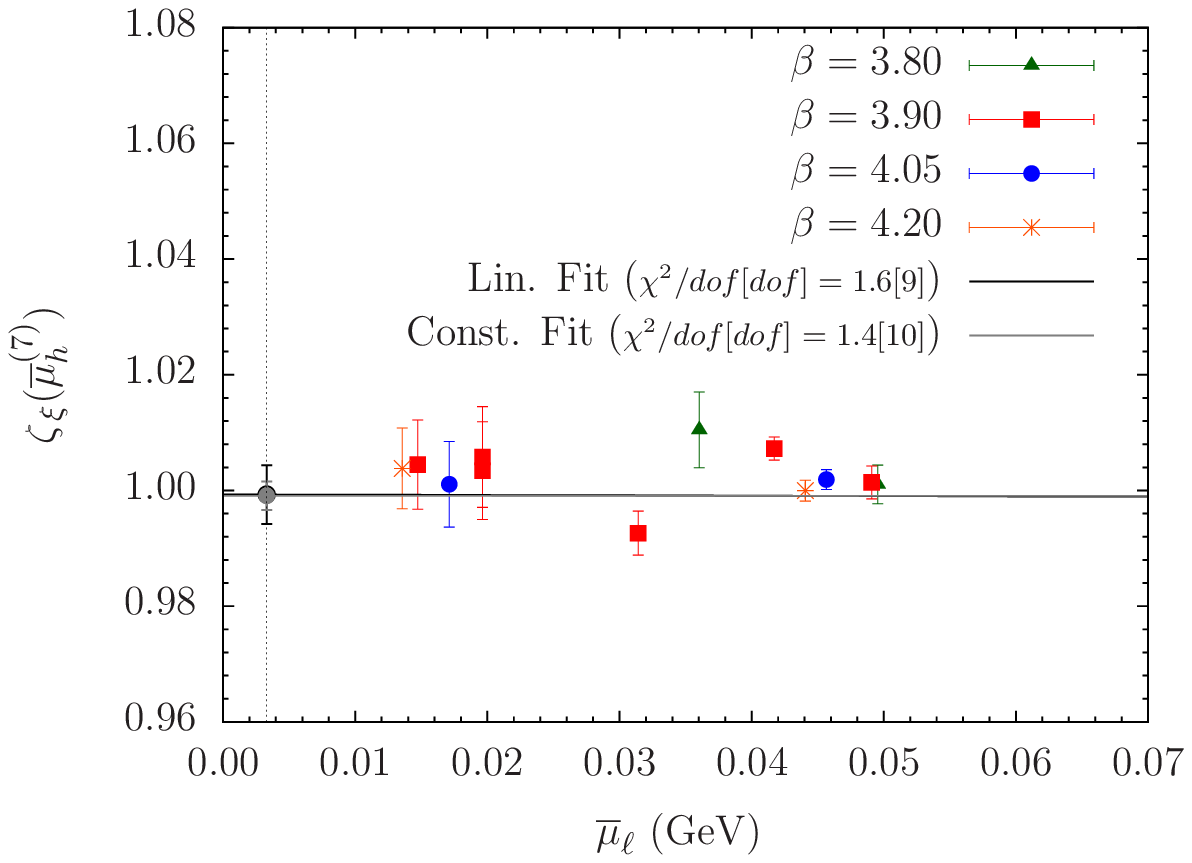}}
\begin{center}
\caption{\sl Combined chiral and continuum fits of the double ratios $\zeta_{\omega}(\overline\mu_h^{(n)})$  and 
$\zeta_{\xi}(\overline\mu_h^{(n)})$  against $\overline\mu_{\ell}$  are shown in panels (a) and (b), respectively.
Ratios for the largest value of the heavy quark mass are reported ($n=7$).    
Linear fit ansatz in $\overline\mu_{\ell}$ and 
fit to a constant are shown. Empty black circle and full grey circle are the results, respectively, at the physical $u/d$ quark 
mass point in the continuum limit.
  }
\label{fig:zeta_w_xi_vs_mul}
\end{center}
\end{figure}

\begin{figure}[!h]
\subfigure[]{\includegraphics[scale=0.70,angle=-0]{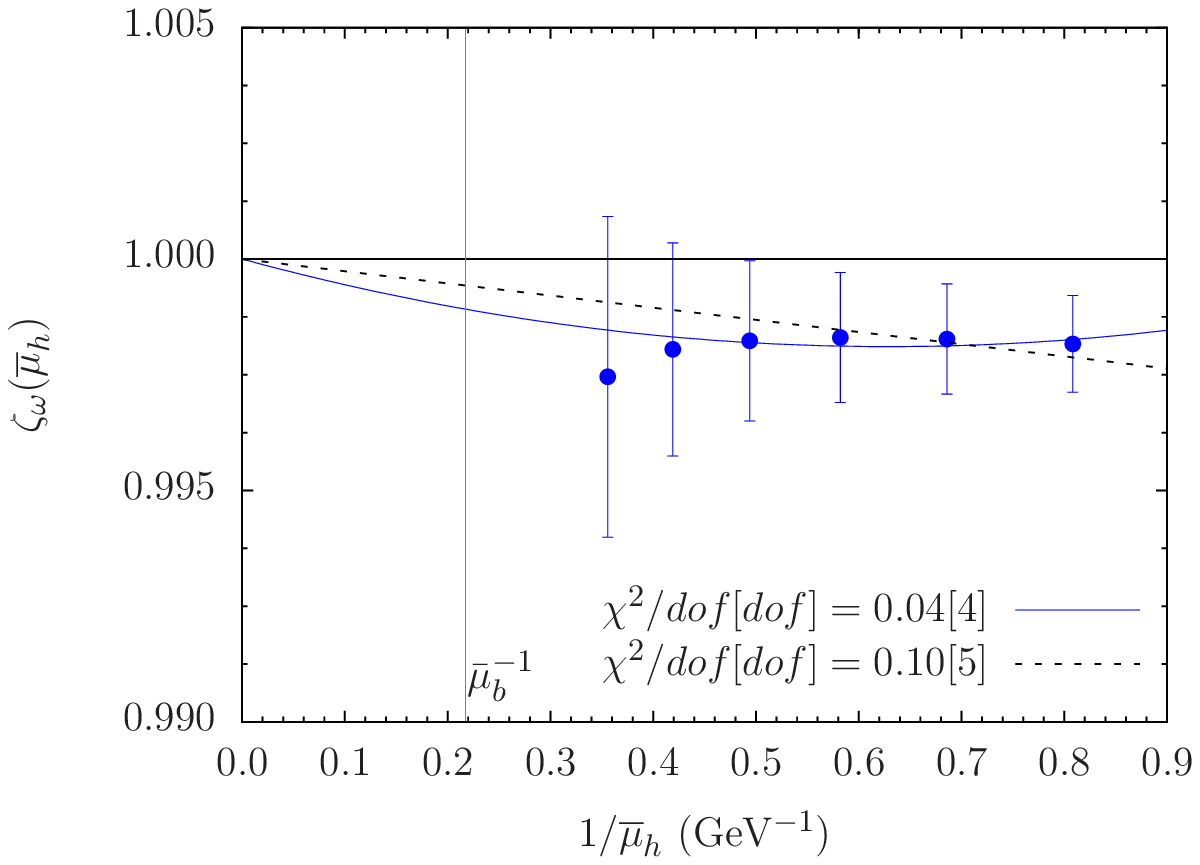}}
\subfigure[]{\includegraphics[scale=0.70,angle=-0]{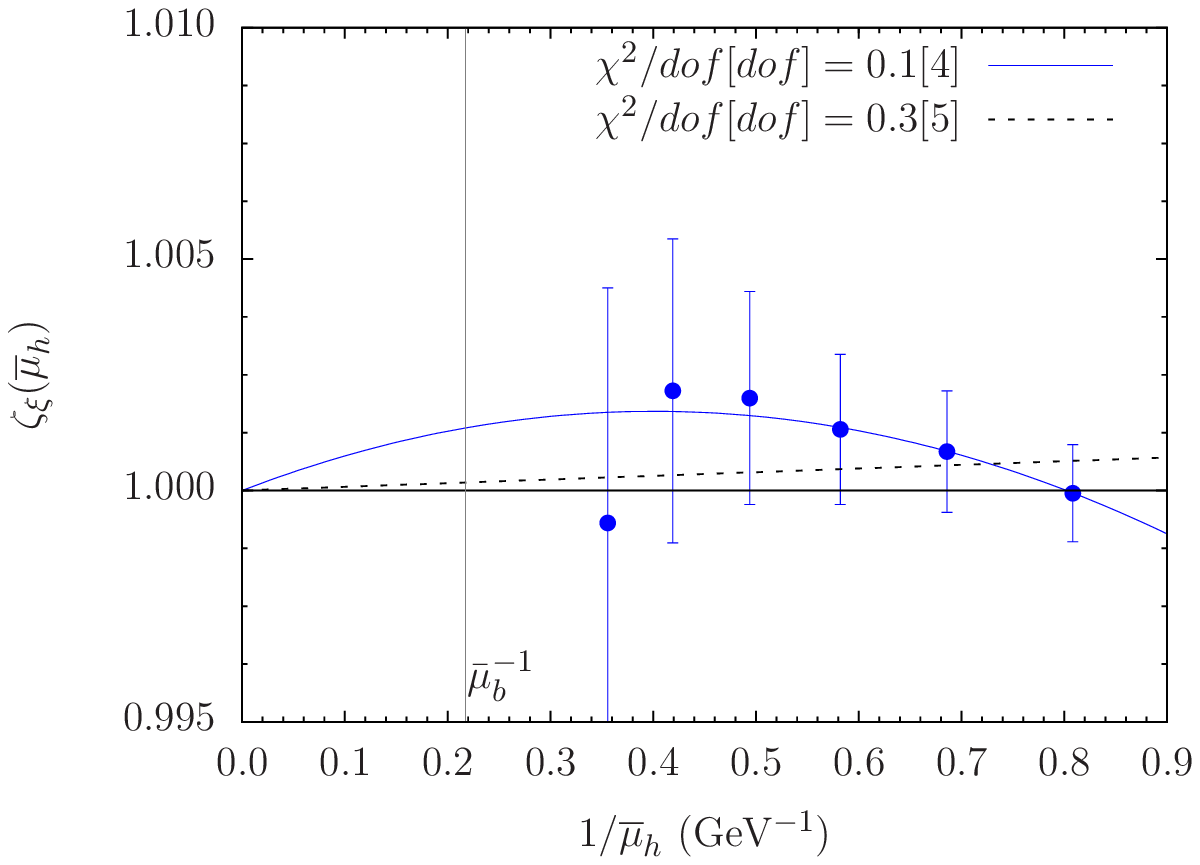}}
\begin{center}
\caption{\sl $\zeta_{\omega}(\overline\mu_h)$ and $\zeta_{\xi}(\overline\mu_h)$ against $1/\overline\mu_h$ 
are shown in panels (a) and (b), respectively. 
For both cases the fit function has a polynomial form of the type analogous to Eq.~(\ref{eq:y_ansatz}) (blue curve). 
Moreover, in panel (b) a fit of the 
form $\zeta(\overline\mu_h) = 1 + 1/\overline\mu_h$ has also been performed (black dashed straight line).  
The vertical black thin line 
marks the position of $1/\overline\mu_b$.
  }
\label{fig:zeta_w_xi_vs_muh}
\end{center}
\end{figure}

In Figs~\ref{fig:zeta_w_xi_vs_mul}(a) and \ref{fig:zeta_w_xi_vs_mul}(b) we illustrate  the 
dependence of $\zeta_{\omega}(\overline\mu_h^{(n)}, \overline\mu_{\ell}, 
\overline\mu_s, a)$ and $\zeta_{\xi}(\overline\mu_h^{(n)}, \overline\mu_{\ell}, \overline\mu_s, a)$ on $\overline\mu_{\ell}$ 
at the largest $\overline\mu_h$ value used in this part of the analysis ({\it i.e.} $n=7$). 
In Figs~\ref{fig:zeta_w_xi_vs_muh}(a) and  \ref{fig:zeta_w_xi_vs_muh}(b) we show  the dependence of 
$\zeta_{\omega}(\overline\mu_h^{(n)})$ and $\zeta_{\xi}(\overline\mu_h^{(n)})$ 
on the inverse heavy quark mass, $1/\overline\mu_h$. We have used for both of them fit function ans\"atze which display a linear and
quadratic dependence on $1/\overline\mu_h$ while the static condition to unity is explicitly imposed.

\clearpage

\section{Summary of results and discussion}
\label{sec:summary_final_results}
  
In this section we collect the results for the $b$-quark mass, the pseudoscalar decay constants of the $B$ and $B_s$ mesons, 
the bag parameters for the full four-fermion operator basis that quantify the QCD effects 
in the $\overline B^{0}_{d}-B^{0}_{d}$ and $\overline B^{0}_{s}-B^{0}_{s}$ 
oscillations, as well as the parameter $\xi$ and the quantities $f_{Bq}\sqrt{B_{Bq}}$ ($q=d$ or $s$). 
 
\vspace*{0.3cm}

\noindent {\bf b-quark mass}: from Eq.~(\ref{eq:mub}) given in  $\overline{\rm{MS}}$ scheme at the scale of 3 GeV 
we get the result for the $b$-quark mass at the scale of the $b$-quark mass itself. The running depends on the active flavour 
number and the value of $\Lambda_{QCD}$. We consider two cases. In the first case we consider the number of the simulated dynamical 
flavours, $N_f=2$, with $\Lambda_{QCD}^{(N_f=2)}=315(15)$ MeV.  
The $\Lambda_{QCD}$ value is the average over the values presented in Refs.~\cite{Blossier:2010ky, Jansen:2011vv, Fritzsch:2012wq}.
The second case consists in using $N_f=4$, ({\it i.e.} the number of active flavours between $\mu = 3$ GeV and 
$\mu = m_b$ in the physical case),  with $\Lambda_{QCD}^{(N_f=4)}=296(10)$ MeV~\cite{Bethke:2009jm}. We obtain
\begin{eqnarray}
 m_b(m_b, \overline{\rm{MS}}) &=& 4.31(9)(8) \,\,\, {\rm GeV}, \,\,\,\,\,\,\, (N_f = 2~~ \mbox{running }) \label{eq:mb_Nf2}\\
 m_b(m_b, \overline{\rm{MS}}) &=& 4.27(9)(8) \,\,\, {\rm GeV}, \,\,\,\,\,\,\, (N_f = 4~~~~~~  \mbox{''} ~~~~~~~) \label{eq:mb_Nf4}
\end{eqnarray}    
The first error (see the first entry for $m_b$ in the error budget Table~\ref{tab:error_budget_1})
has been computed using the bootstrap method. It includes the statistical uncertainties on the pseudoscalar meson mass at the 
triggering point (typically less than $0.5\%$) and on the ratios (less
than $0.1-0.2\%$), the uncertainty due to the quark mass renormalisation constant, 
$Z_P$, in employing renormalised light and heavy quark masses, 
as well as the statistical uncertainties of the lattice scale. 
Notice that the pure statistical error of the pseudoscalar mass
values computed from the relevant 2-point correlator functions lies typically at a sub-percent level. 
The second error  includes: 
(i) systematic uncertainties of the lattice scale; (ii) The  estimate of  discretisation effects that has been obtained by  repeating 
the whole analysis without including the data corresponding to the coarsest value of the lattice spacing; 
(iii) systematic uncertainties due to fit choices by (a) fitting the 
ratios $y$ against the inverse heavy quark mass adding an extra cubic dependence in the fit ansatz (see Eq.~(\ref{eq:y_ansatz})) 
and (b) excluding from the fit all the data from the ratio corresponding to the heaviest quark mass. 
Each of these checks in the fitting procedure leads to a small shift of 0.2-0.3\% from 
the central value. (iv) We consider he systematic error due to the  $\Lambda_{QCD}$ value as well as due to  
the half difference between results obtained  employing in the running either  
$N_f=2$ or $N_f=4$ dynamical  flavours. \\
In Table~\ref{tab:error_budget_1} we present a detailed description 
of the error budget giving the percentage error for each of the systematic sources of uncertainty.   
Our total error is computed by summing in quadrature 
all the above errors and it amounts to less than 3\%. 
By averaging the two results above and including their half-difference as an additional systematic uncertainty 
we finally obtain 
\begin{equation}\label{eq:mb_Nf2_Nf4}
 m_b(m_b, \overline{\rm{MS}}) = 4.29(9)(8)(2) \,\,\, {\rm GeV}
 \end{equation}
Our current result for the $b$-quark mass is fully
compatible with our previous determination in Ref.~\cite{Dimopoulos:2011gx}.       
\vspace*{0.3cm}

\noindent {\bf Pseudoscalar decay constants}: we have computed directly $f_{Bs}$ and the ratio $f_{Bs}/f_{Bs}$. From these we 
can get the value for $f_B=f_{Bs} / (f_{Bs}/f_{B})$. Our results read
\begin{eqnarray}
f_{Bs} &=& 228(5)(6)\,\,\,\, {\rm MeV}  \label{eq:fBs} \\
\dfrac{f_{Bs}}{f_B} &=& 1.206(10)(22) \label{eq:fratio} \\
f_{B} &=& 189(4)(5)(4) \,\, {\rm MeV} \label{eq:fB}
\end{eqnarray}
where the first error is due to statistical uncertainties estimated using the bootstrap method. 
The statistical error for the matrix element from which decay constants are computed is at a sub-percent level.
The statistical uncertainties at the triggering point and those of the ratios are about $0.5\%$  and $0.2\%$, 
respectively. It is also included the statistical uncertainty of the lattice scale.   
The second error refers to systematic uncertainties all added in quadrature. They are discussed in the following.  
(i)  For dimensionful quantities the systematic uncertainty of the lattice scale (about $2\%$) is taken into account.  
(ii) We estimate the systematic uncertainty due to residual discretisation effects by 
repeating the whole analysis without including data from the coarsest value of the lattice spacing.
(iii) The $f_{Bs}$ value has been obtained using the NLL approximation of the 
factors $C^{stat}_A$ and 
$\rho$, (see Eq.~(\ref{eq:z_zs_ratios})).\footnote{Their explicit form can be found in Ref.~\cite{Dimopoulos:2011gx}, 
Eqs.~(3.16) and (3.5).}  Had we used LL or TL approximation the total shift to the central value would be 
about 2 MeV which is well covered by the quoted errors. This is also included in the systematic error budget. 
A shift to the final value at the level of per mille is noticed if we exclude from the fitting procedure of ratios against
the inverse heavy quark mass data corresponding to the heaviest quark mass.
(iv) The second error in the decay constant ratio (\ref{eq:fratio}) as well as the third error
in $f_B$ is due to the systematic uncertainty arising from the chiral fit at the triggering point
-- see the discussion in Section~\ref{sec:mb_fB_fBs} and the result~(\ref{eq:fhsovfhl_trig}). 
In Table~\ref{tab:error_budget_1} we present in detail the full error budget.  
The alternative ratio method computation of $f_{Bs}$  based on the use of the HQET asymptotic 
behaviour expressed in Eq.~(\ref{eq:fhs_sqrtMhs}) leads to an estimate which differs from the above of Eq.~(\ref{eq:fBs}) by only
0.5\%, but its total error is a bit larger due to statistical and systematic uncertainties coming from the meson mass, $M_{hs}$. 

Our total uncertainty is given by the sum in quadrature of the above  errors. For $f_{Bs}$, 
$f_{Bs}/f_{B}$ and $f_{B}$ the total error is 3.4\%, 2.0\% and 4.0\%, respectively.
In Table~\ref{tab:error_budget_1} a detailed description of the error budget is presented. 
 We also notice that our present results are 
compatible within  one standard deviation with the older ones of Ref.~\cite{Dimopoulos:2011gx} which have been computed 
using local sources for the various propagator inversions. 

\begin{table}[!h]
\begin{center}
\begin{tabular}{|l|c|c|c|c|c|c|}
\hline 
source of uncertainty (in  \%) & $m_{b}$ & $f_{Bs}$ & $f_{Bs}/f_{B}$ & $f_{B}$ & $B_{1}^{(s)}/B_{1}^{(d)}$ & $\xi$\tabularnewline
\hline
\hline 
stat.  + fit  (CL and Chiral) & 2.1 & 2.2 & 0.8 & 2.1 & 1.5 & 1.3\tabularnewline
\hline 
lat. scale syst. error & 2.0 & 2.0 & - & 2.0 & - & -\tabularnewline
\hline 
syst. from discr. effects & 0.2 & 1.3 & 0.4 & 1.7 & 1.3 & 1.0\tabularnewline
\hline 
syst. from fit in $1/\mu_{h}$ & 0.4 & 1.0 & 0.1 & 1.1 & - & 0.5\tabularnewline
\hline 
syst. from trig. point & - & - & 1.7 & 1.7 & 0.1 & 1.8\tabularnewline
\hline 
syst. from $\Lambda_{QCD}$ and running & 0.5 & - & - & - & - & -\tabularnewline
\hline 
Total  & 3.0 & 3.4 & 2.0 & 4.0 & 2.0 & 2.5\tabularnewline
\hline
\end{tabular}
\caption{Error budget (in \%) for $m_b$, $f_{Bs}$, $f_{Bs}/f_{B}$, $f_{B}$, $B_{1}^{(s)}/B_{1}^{(d)}$ and $\xi$. Each of the numbers in the first row includes the uncertainties coming from the statistical errors of the correlators, the uncertainty of the quark mass renormalisation, the error of the combined chiral and continuum fits and the statistical uncertainty from the scale setting. In the second row we estimate the uncertainty due to the lattice scale systematics. In the third row we give an estimate of the systematic uncertainty related to discretisation effects. In the fourth row we display the systematic uncertainty connected to the fit of the ratios against $1/\mu_h$. In the fifth row the systematic uncertainty at the triggering point fit is shown. The sixth row is for the systematic error coming from the uncertainty in the value of $\Lambda_{QCD}$ and the half difference between results obtained employing either $N_f=2$ or $N_f=4$ in the evolution from $\mu =3$~Gev to the $b$-mass point $\sim 4.3$~GeV. The last row gives the total error.} 
\label{tab:error_budget_1}
\end{center}
\end{table}   

As a by-product of our work, since the heavy quark mass value at the triggering point has been set equal to the charm quark mass, 
we have computed the decay constants for the $D_s$ and $D$ mesons as well as their ratio. They read:
\begin{equation}\label{eq:fD_Ds_ratio}
f_{Ds} = 250(5)(5) ~{\rm MeV}, \,\,\,\, f_{D} = 208(4)(6) ~{\rm MeV}, \,\,\,\, f_{Ds}/f_{D} = 1.201(7)(20),
\end{equation} 
where the first error is statistical and the second one is systematic.

\begin{figure}[!p]
\subfigure[]{\includegraphics[scale=0.90,angle=-0]{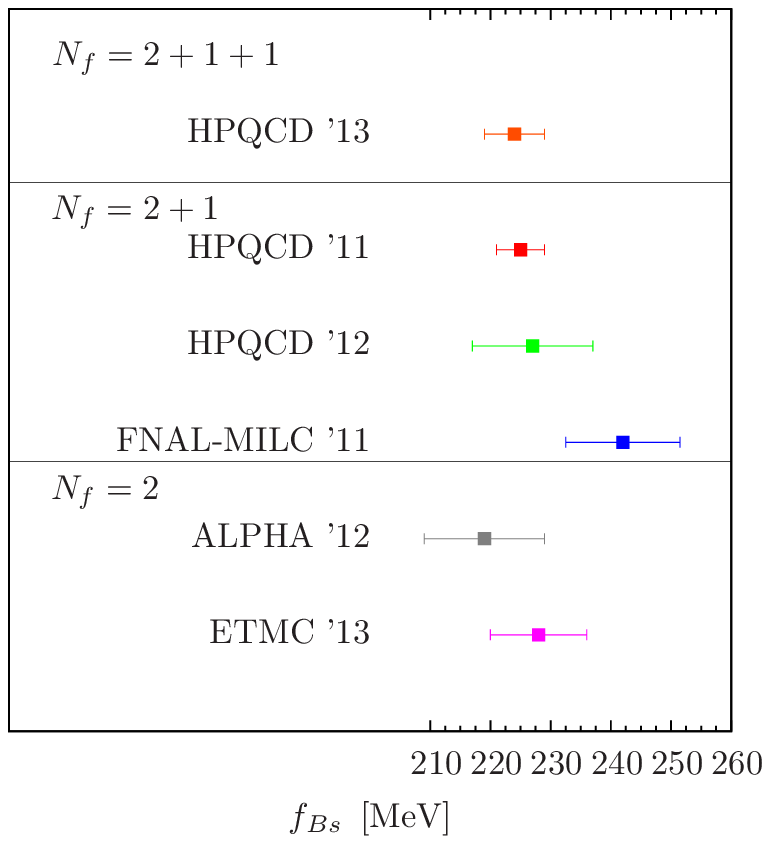}}
\subfigure[]{\includegraphics[scale=0.90,angle=-0]{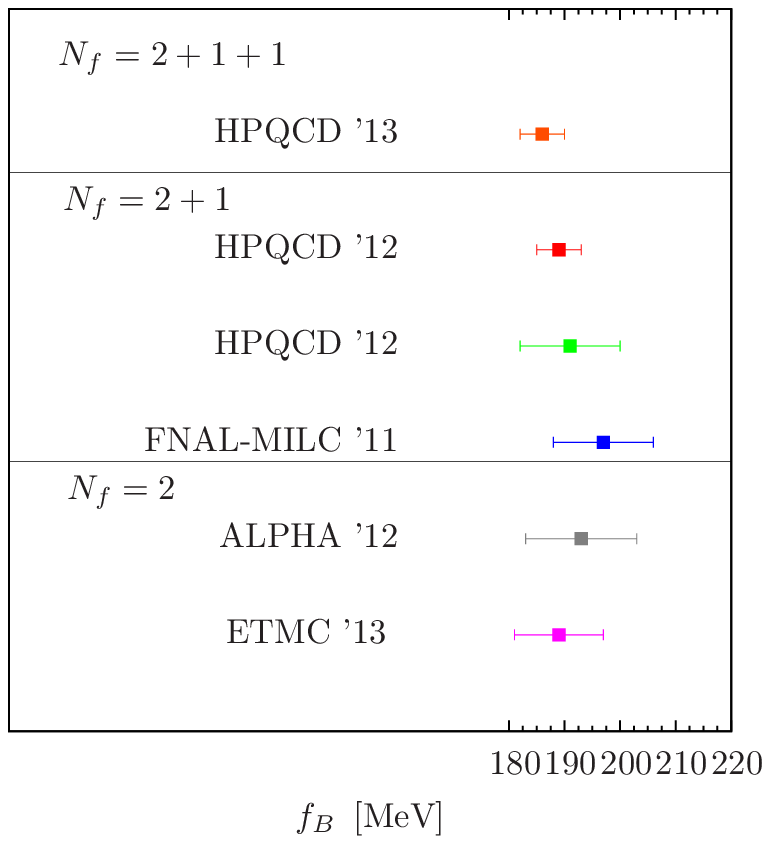}}
\subfigure[]{\includegraphics[scale=0.90,angle=-0]{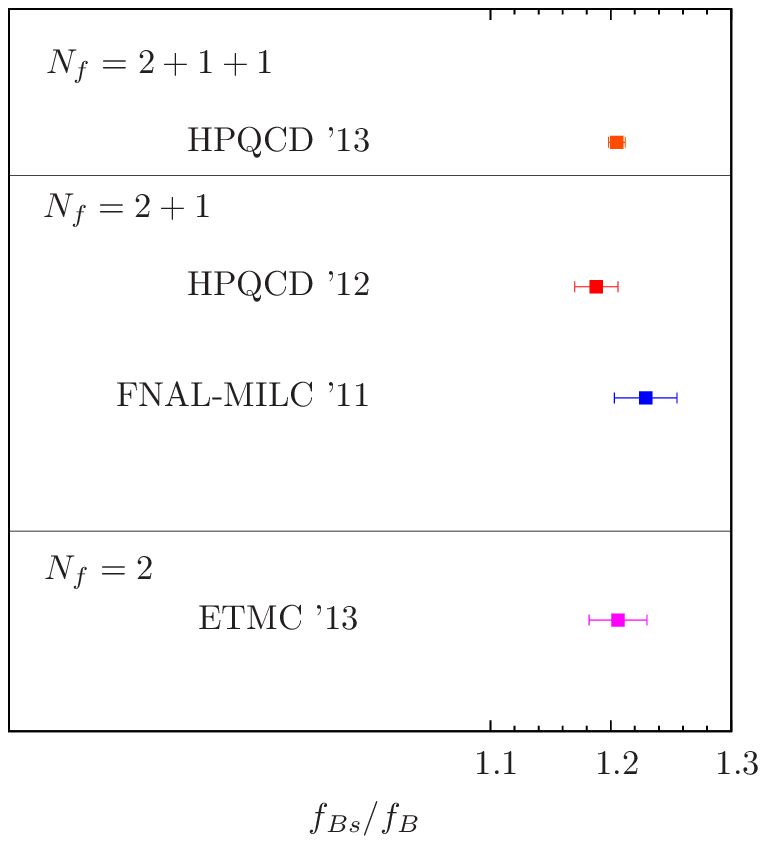}}
\hspace*{3cm}
\subfigure[]{\includegraphics[scale=0.90,angle=-0]{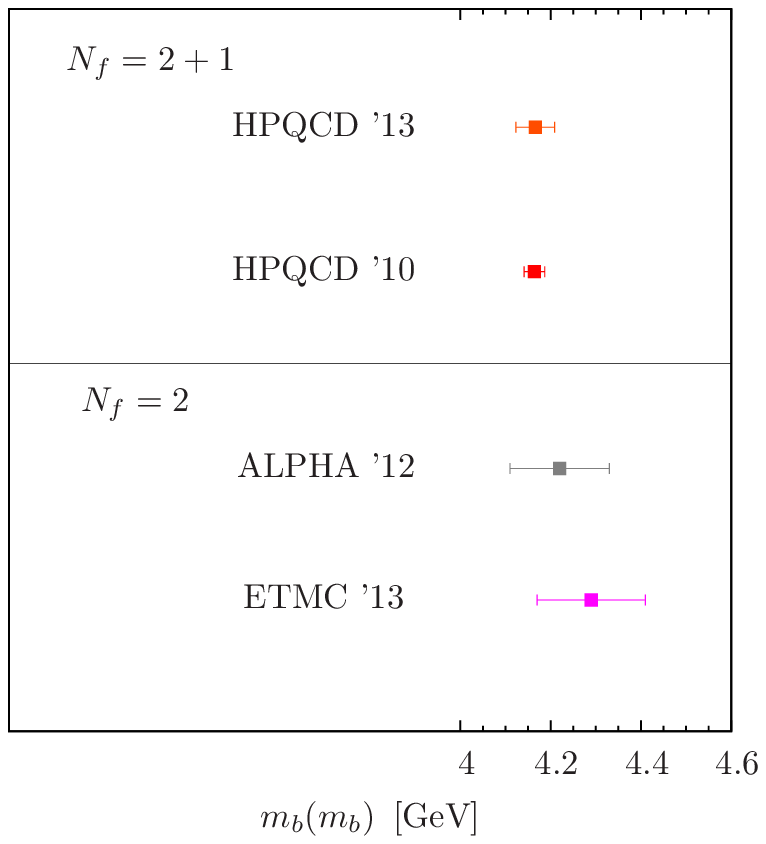}}
\begin{center}
\caption{\sl A comparison  of the available continuum extrapolated determinations of $f_{Bs}$ panel (a), 
 $f_B$ panel (b),  $f_{Bs}/f_B$ panel (c) and $m_b$ panel (d).    
The results of the present work have been labeled as ``ETMC '13". For the results of 
the other lattice groups we refer to  (from top to bottom):
(a) Refs.~\cite{Dowdall:2013tga, McNeile:2011ng, Na:2012kp, Bazavov:2011aa, Bernardoni:2012ti}; 
(b) Refs.~\cite{Dowdall:2013tga, Na:2012kp, Na:2012kp, Bazavov:2011aa, Bernardoni:2012ti}; 
(c) Refs.~\cite{Dowdall:2013tga, Na:2012kp, Bazavov:2011aa};
(d) Refs.~\cite{Lee:2013mla, McNeile:2010ji,  Bernardoni:2012ti}. (The results of Ref.~\cite{Bernardoni:2012ti} are still 
preliminary. 
Note also that  the results in panel (a) indicated as HPQCD'11 and HPQCD'12, 
as well as  those indicated as HPQCD'12 in panel (b), have both been obtained using $N_f=2+1$ (MILC) gauge  
ensembles but  employ different valence quark regularisations.)
  }
\label{fig:compr1}
\end{center}
\end{figure}

The results for the $B$-meson decay constants in Eqs.~(\ref{eq:fBs})-(\ref{eq:fB}) are obtained in the $N_f=2$ theory and do not account for the dynamical sea quark effects of the strange and the heavier quarks. In Figs.~\ref{fig:compr1}~(a) to (d) we compare these results and the one for the $b$-quark mass of Eq.~(\ref{eq:mb_Nf2_Nf4}) with those obtained by other lattice studies using either $N_f=2$, $N_f=2+1$ or $N_f=2+1+1$ simulations. The good agreement observed among the various results in the plots provides a first indication that the systematic effects due to the quenching of the strange and charm quarks are smaller than present uncertainties from other sources. 

More quantitatively, one can compare our results for the decay constants with the averages quoted by the FLAG working group from $N_f=2+1$ calculations~\cite{Aoki:2013ldr}:
\begin{equation}
\label{eq:fB-FLAG} 
f_{Bs} = 227.7(4.5)\,\,\,\, {\rm MeV}  \quad , \quad
f_{B} = 190.5(4.2) \,\, {\rm MeV} \quad , \quad
\dfrac{f_{Bs}}{f_B}= 1.202(22) \ .
\end{equation}
The differences in central values between our $N_f=2$ results of Eqs.~(\ref{eq:fBs})-(\ref{eq:fB}) and the $N_f=2+1$ FLAG averages are smaller than 1\%, showing that the quenching effect of the strange quark is smaller, at present, than other uncertainties. For the heavier charm quark, the quenching effect is expected to be even smaller in size and the comparison between the $N_f=2+1+1$ results of Ref.~\cite{Dowdall:2013tga} with the $N_f=2$ and $N_f=2+1$ determinations presented in Fig.~\ref{fig:compr1} does not show, indeed, any systematic deviation.

In the case of the charm quark, a parametric estimate of the quenching effect can be also derived using a perturbative argument. Since the sea quark contributions enter at the loop level, involving the exchange of at least two gluons, and are quadratically suppressed in the decoupling limit by the inverse quark mass, one expects them to be of order $\alpha_s(m_c)^2 (\Lambda_{QCD}/m_c)^2$ for the charm quark, corresponding to less than 1\%. Even though the accuracy of perturbation theory at the charm scale is limited, this estimate is of the same order of magnitude of the one indicated by the previous numerical comparison between lattice results. It supports the conclusion that for the decay constants this error is negligible with respect to other uncertainties. 

The estimate of the partial quenching effects in the determination of the quark masses, like the $b$-quark mass in the present study, is more subtle, since it also depends on the details of the renormalization procedure. In our calculation, we have renormalized the quark mass with the non-perturbative RI-MOM method at the scale $\mu=3$ GeV and we have chosen to quote our final result in the $\overline{\rm{MS}}$ scheme at the scale $m_b$. Thus, a dependence on the number of dynamical flavors also appears in the matching between the initial and the final scheme and scale. This effect is quantified by the difference between the results in Eq.~(\ref{eq:mb_Nf2}) and (\ref{eq:mb_Nf4}) and it is at the level of 1\%. In this case, being not fully negligible with respect to other uncertainties, it has been taken into account in the systematic error quoted in Eq.~(\ref{eq:mb_Nf2_Nf4}).


\vspace*{0.3cm}

\noindent {\bf Bag parameters}: In Table~\ref{tab:bag_res} we gather our results for the bag parameters 
of the full four-fermion operator basis.
\begin{table}[!h]
\begin{center}
\begin{tabular}{|c|c|c|c|c|}
\hline
\multicolumn{5}{|c|}{($\overline {\rm{MS}}$, $m_b$)}\tabularnewline
\hline
\hline
$B_{1}^{(d)}$ & $B_{2}^{(d)}$ & $B_{3}^{(d)}$ & $B_{4}^{(d)}$ & $B_{5}^{(d)}$\tabularnewline
\hline
\hline
0.85(3)(2) & 0.72(3)(1) & 0.88(12)(6) & 0.95(4)(3) & 1.47(8)(9)\tabularnewline
\hline
\hline
$B_{1}^{(s)}$ & $B_{2}^{(s)}$ & $B_{3}^{(s)}$ & $B_{4}^{(s)}$ & $B_{5}^{(s)}$\tabularnewline
\hline
\hline
0.86(3)(1) & 0.73(3)(1) & 0.89(10)(7) & 0.93(4)(1) & 1.57(7)(8)\tabularnewline
\hline
\end{tabular}
\caption{Continuum limit results for $B_i^{(d)}$ and $B_i^{(s)}$ ($i=1, \ldots, 5$), 
renormalized in the $\overline{\rm{MS}}$ scheme of Ref.~\cite{mu:4ferm-nlo} at the scale of the $b$-quark mass. 
See the text for discussion about the quoted errors.} 
\label{tab:bag_res}
\end{center}
\end{table}   
\vspace*{0.3cm}
Results are given in the $\overline{\rm{MS}}$ scheme of Ref.~\cite{mu:4ferm-nlo} at the scale of the $b$-quark mass $m_b$ 
of Eq.~(\ref{eq:mb_Nf2_Nf4}). We consider our final values to be the average over the results obtained by
using $N_f=2$ and $N_f=4$ in the evolution from the QCD renormalization scale
($\mu =3$~GeV) and the $b$-mass scale ($\sim 4.3$~GeV), while their difference, 
which is less than 1\% (in the worst case), is taken as an additional 
systematic error. 
In the results of Table~\ref{tab:bag_res} the first error corresponds to the sum of the statistical errors (typically about $1\%$) 
on the correlators, the RCs uncertainty (which is the largest among the others and amounts to $2-3\%$ depending on the case) 
and the fit error. 
The second error includes:  
 (i) the uncertainties due to the possible choices of 
fit ansatz in $1/\overline\mu_h$ and the maximum spread in the results induced by using the TL, LL or 
NLL order formulae for $W$ in the ratios of Eqs~(\ref{eq:omega_d}) and (\ref{eq:omega_s}).  The latter 
uncertainty is for  most of the cases rather small compared to the first error {\it i.e.} about $0.1\% - 1\%$,  
with the exception of $B_3^{(d)}$, $B_3^{(s)}$ and $B_5^{(s)}$ for which we notice errors of 7.3$\%$, 6.7$\%$ and 3.5$\%$, 
respectively;   
(ii) the uncertainty due to the fit ansatz (polynomial or HMChPT) at the triggering point (only for the case $B_i^{(d)}$); 
(iii)  the systematic uncertainty of residual discretisation effects, estimated by  repeating the whole analysis 
without including data 
from the coarsest value of the lattice spacing. 
our systematic uncertainty coming from the value  
(iv) the systematic error due to the  $\Lambda_{QCD}$ value as well as due to  
the half difference between results obtained  employing in the running either  
$N_f=2$ or $N_f=4$ dynamical  flavours.
In Tables~\ref{tab:error_budget_2} and \ref{tab:error_budget_3} we 
give the  detailed error budget for $B_i^{(d)}$ and $B_i^{(s)}$, respectively. 
\begin{table}[t]
\begin{center}
\begin{tabular}{|l|c|c|c|c|c|}
\hline 
source of uncertainty (in \%) & $B_{1}^{(d)}$ & $B_{2}^{(d)}$ & $B_{3}^{(d)}$ & $B_{4}^{(d)}$ & $B_{5}^{(d)}$\tabularnewline
\hline
\hline 
stat + fit + RCs & \multicolumn{1}{c|}{3.8} & 4.0 & 14 & 4.6 & 5.1\tabularnewline
\hline 
syst. from discr. effects & 1.7 & 0.2 & 0.7 & 2.0 & 3.8\tabularnewline
\hline 
syst. from fit in $1/\mu_{h}$  & 1.6 & 1.8 & 7.3 & 1.5 & 3.6 \tabularnewline 
\hline 
syst. from trig. point & 0.1 & 0.1 & 0.1 & 2.2 & 2.4 \tabularnewline
\hline
syst. due to $\Lambda_{QCD}$ and running & 0.2 & 0.5 & 0.2 & 0.1 & 0.9\tabularnewline
\hline 
Total & 4.5 & 4.4  & 16 & 5.7 & 7.8\tabularnewline
\hline
\end{tabular}
\caption{Error budget (in \% ) for $B_i^{d}$ ($i=1, \ldots, 5$). First row includes the statistical uncertainties of correlators, of 
the fits' extrapolation and interpolation as well as the uncertainties of the RCs. The second row includes the systematic 
uncertainties due to discretisation effects. In the third row we give the estimates of the 
systematic errors by employing different choices of fit ans\"atze in fitting the ratios as a function of $\mu_{\ell}$ and the 
systematics related to the fit of the continuum value of the ratios against $1/\mu_h$. We have also included the 
systematic uncertainty estimated by repeating our analysis without using data from the heaviest quark mass pair.
In the fourth we show the systematic uncertainty from the fitting procedure at the triggering point. 
In the fifth row we display our systematic uncertainty coming from the value  
of $\Lambda_{QCD}$ and the half difference between results obtained  employing in the running either  
$N_f=2$ or $N_f=4$ dynamical  flavours. The last row gives the total error.} 
\label{tab:error_budget_2}
\end{center}
\end{table}
\begin{table}[t]
\begin{center}
\begin{tabular}{|l|c|c|c|c|c|}
\hline 
source of uncertainty (in \%) & $B_{1}^{(s)}$ & $B_{2}^{(s)}$ & $B_{3}^{(s)}$ & $B_{4}^{(s)}$ & $B_{5}^{(s)}$\tabularnewline
\hline
\hline 
stat + fit + RCs & \multicolumn{1}{c|}{3.1} & 3.6 & 11.1 & 4.3 & 4.2\tabularnewline
\hline 
syst. from discr. effects & 0.5 & 0.5 & 2.9 & 1.2 & 0.6\tabularnewline
\hline 
syst.  from fit in $1/\mu_{h}$  & 1.3 & 1.7 & 6.7 & 0.2 & 5.0\tabularnewline
\hline 
syst. due to $\Lambda_{QCD}$ and running & 0.2 & 0.5 & 0.2 & 0.1 & 0.9\tabularnewline
\hline 
Total & 3.4 & 4.0 & 13.3 & 4.5 & 6.6\tabularnewline
\hline
\end{tabular}
\caption{Error budget (in \% ) for $B_i^{s}$ ($i=1, \ldots, 5$). For details see caption of Table~\ref{tab:error_budget_2}.  } 
\label{tab:error_budget_3}
\end{center}
\end{table}

We notice that the results for $B_i^{(d/s)}$ in Ref.~\cite{Bpar:SPQR1},  
obtained in the quenched approximation and using Wilson-clover quarks
are in the same ballpark as the present ones.
A detailed comparison, for example on the impact of quenching, is not very meaningful because 
older quenched results have been obtained at rather large pion masses and
at only one lattice spacing ($a \sim 0.1$ fm), while the unquenched ones of this work have been extrapolated  
to the continuum limit and to the physical pion mass.

\noindent {\bf Other quantities:} $B_{1}^{(s)}/B_1^{(d)}$, $\xi$, $f_{Bq} \sqrt{B_i^{(q)}}$.  
Following the analysis presented in Sections~\ref{sec:mb_fB_fBs} and \ref{sec:B_xi} we  obtain
\begin{eqnarray}
\dfrac{B_1^{(s)}}{B_1^{(d)}} &=& 1.007(15)(14) \label{eq:B_rat} \\
\xi &=& 1.225(16)(14)(22) \label{eq:xi} .
\end{eqnarray}
The first error in both results is of statistical nature and it is due to the fitting procedures employed for 
the triggering point, $\sim 1\%$ (in both quantities) 
and for the corresponding ratios (less than $1\%$ for each one of them). 
The second error is our estimated uncertainty coming from employing 
different types of ansatz in fitting  
the ratios (less than $\sim 0.5\%$) as a function of  the inverse heavy quark mass, 
as well as the uncertainty coming from using different orders for the QCD 
and HQET running and matching (less than $0.5\%$ in both cases). The third  error in $\xi$ corresponds to the 
uncertainty due to  possible fit ansatz choices at the triggering point discussed in Section~\ref{sec:B_xi}, (see  
Eq.~(\ref{eq:xi_trig})). Our final uncertainty is taken as the sum in quadrature of the various uncertainties and it amounts 
to 2.0\% for $B_{1}^{(s)}/B_1^{(d)}$ and 2.5\% for $\xi$. The error budget for both quantities is summarized
in Table~\ref{tab:error_budget_1}.

In Table~\ref{tab:fbag_res} we collect our results for the quantities $f_{Bq} \sqrt{B_i^{(q)}}$ 
with $q=d, s$ and $i=1, \ldots, 5$.  
Bag parameters are expressed in the $\overline{\rm{MS}}$ scheme of Ref.~\cite{mu:4ferm-nlo} at the scale of the $b$-quark mass.
The error for each of these quantities is determined by combining the errors previously discussed on decay constants and 
bag-parameters. 
For convenience we also give our results for the SM relevant quantities  
in which the bag parameters are expressed in the RGI scheme. 
In this case for the running of the coupling constant we take
$N_f=5$  and $\Lambda_{QCD}^{(N_f=5)}$=213(9) MeV~\cite{Bethke:2009jm}.
We find
\begin{eqnarray}
f_{Bd}\sqrt{\hat{B}_{1}^{(d)}} &=& 216(6)(8) ~~\rm{MeV}\\
f_{Bs}\sqrt{\hat{B}_{1}^{(s)}} &=& 262(6)(8) ~~\rm{MeV}, 
\end{eqnarray} 
where the first error is statistical and the second one is systematic.

The RGI values of the bag parameters corresponding to the SM four-fermion operators read
\begin{equation} 
\hat{B}_{1}^{(d)} = 1.30(5)(3),  \,\,\,\,\, \hat{B}_{1}^{(s)} = 1.32(5)(2), 
\end{equation}
where the first error is statistical and the second one is systematic.

\begin{table}[t]
\begin{center}
\begin{tabular}{|c|c|c|c|c|c|}
\hline
\multicolumn{6}{|c|}{($\overline {\rm{MS}}$, $m_b$) [MeV]}\tabularnewline
\hline
\hline
$i$ & 1 & 2 & 3 & 4 & 5 \tabularnewline
\hline
\hline
$f_{Bd} \sqrt{B_i^{(d)}}$ & 174(5)(6)  & 160(5)(6) & 177(14)(9) & 185(6)(7) & 229(8)(11)\tabularnewline
\hline
\hline
\hline
$f_{Bs} \sqrt{B_i^{(s)}}$ & 211(5)(6) & 195(5)(5) & 215(14)(9) & 220(7)(6) & 285(8)(12)\tabularnewline
\hline
\end{tabular}
\caption{Continuum limit results for $f_{Bd} \sqrt{B_i^{(d)}}$ and $f_{Bs} \sqrt{B_i^{(s)}}$ ($i=1, \ldots, 5$). 
Bag parameters are expressed in the $\overline{\rm{MS}}$ scheme of Ref.~\cite{mu:4ferm-nlo} at the scale of the $b$-quark mass.  } 
\label{tab:fbag_res}
\end{center}
\end{table}   

\begin{figure}[p]
\subfigure[]{\includegraphics[scale=0.90,angle=-0]{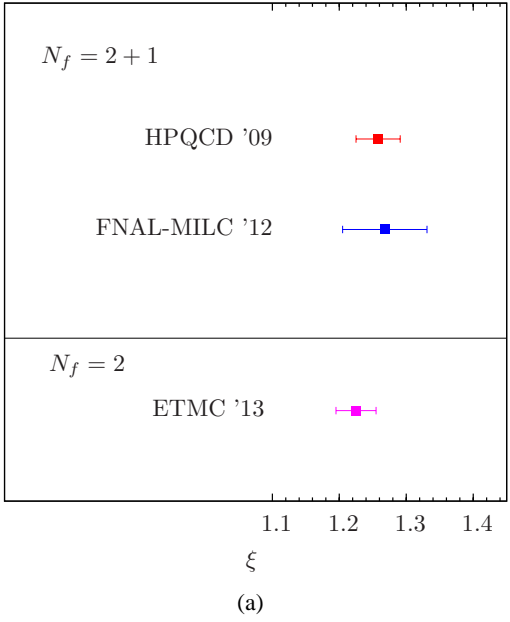}}
\subfigure[]{\includegraphics[scale=0.90,angle=-0]{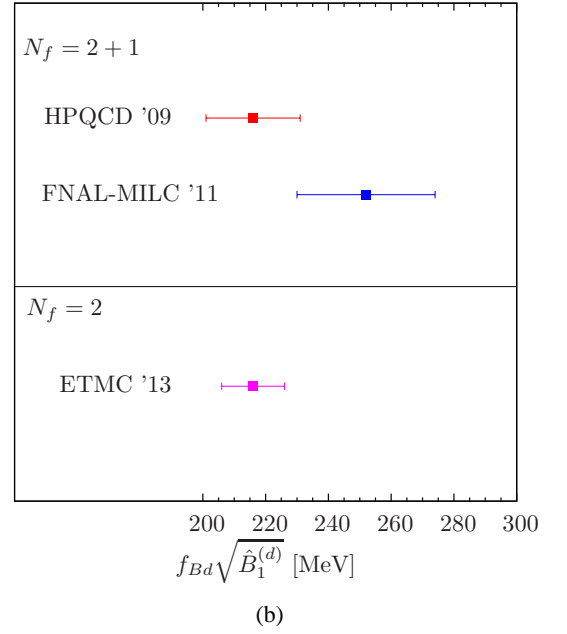}}
\subfigure[]{\includegraphics[scale=0.90,angle=-0]{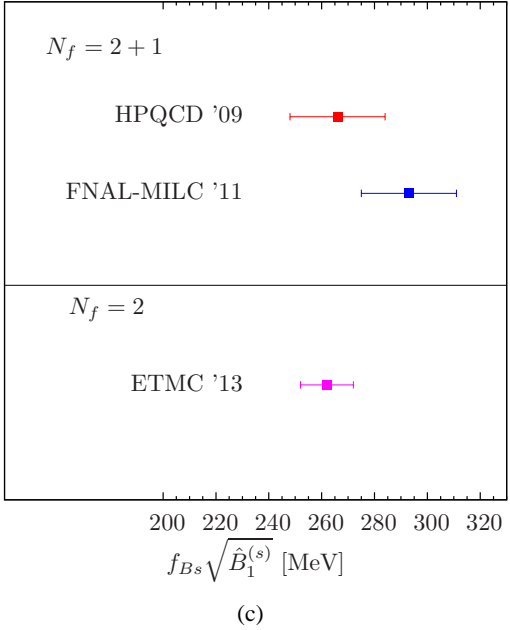}}
\begin{center}
\caption{\sl A comparison  of the available continuum extrapolated determinations of  $\xi$ panel (a), 
 $f_{Bd}\sqrt{\hat{B}_{1}^{(d)}}$ panel (b) and  $f_{Bs}\sqrt{\hat{B}_{1}^{(s)}}$ panel (c). 
Bag parameters are given in the RGI scheme.   
The results of the present work have been labeled as``ETMC '13". For the results of 
the other lattice groups we refer to (from top to bottom) :
(a) Refs.~\cite{Gamiz:2009ku, Bazavov:2012zs} 
(b) Refs.~\cite{Gamiz:2009ku, Bouchard:2011xj}; 
(c) Refs.~\cite{Gamiz:2009ku, Bouchard:2011xj}. 
  }
\label{fig:fB_res}
\end{center}
\end{figure}

In Fig.~\ref{fig:fB_res} we compare our results for $\xi$, $f_{Bd}\sqrt{\hat{B}_{1}^{(d)}}$ and $f_{Bs}\sqrt{\hat{B}_{1}^{(s)}}$, with bag-parameters expressed in the RGI scheme, with the ones obtained by other lattice collaborations.

In conclusion, in this work using smeared interpolating operators in computing the 2- and 3-point correlation functions: \\ 
(i) we  update our older continuum limit results~\cite{Dimopoulos:2011gx} for the b-quark mass and 
the pseudoscalar decay constants for the $B$ 
and $B_s$ pseudoscalar mesons for which  local interpolating
operators had been used in computing the 2-point correlation functions. 
Our final errors now are smaller though they are principally determined  by uncertainties
in the renormalisation constant for the quark mass and the lattice scale. 
Besides that, the fitting procedure of our ratios against 
the inverse heavy quark mass towards the exactly known static limit has gained more accuracy 
with respect to our previous work thanks to
the higher precision in calculating the ratios; \\
(ii) we have presented results, extrapolated in the 
continuum limit and to the physical light quark mass point, 
about the bag-parameters for the full
four-fermion operator basis that control the neutral $B$-meson 
oscillations, as well as results for the parameter $\xi$  and
the quantities $f_{Bd} \sqrt{B_i^{(d)}}$,
$f_{Bs} \sqrt{B_i^{(s)}}$ ($i=1, \ldots, 5$). 

Finally, preliminary results for the $b$-quark mass employing the ratio method on 
$N_f=2+1+1$ dynamical quark configurations generated by ETMC have already been presented~\cite{LAT2013-2+1+1}. ETMC will present  
in the near future a complete analysis of the physical observables studied in the present paper but employing $N_f=2+1+1$  gauge ensembles.

\newpage
\noindent{\bf Acknowledgements}\\
We thank Guido Martinelli for useful discussions and for comments on the manuscript.
We also thank Aida El-Khadra and Nazario Tantalo for very helpful discussions on
alternative ways to define ratios well suited for evaluating $B$-meson matrix
elements, as illustrated in this work for the case of $f_{B_{s}}$. We are grateful to Filippo Sala
for spotting a numerical error in a previous version of this manuscript.
A.S. acknowledges interesting discussions with P. Fritzsch on the role of smearing for heavy-light systems.
P.D. thanks Aida El-Khadra, M. Papinutto and A. Vladikas for useful discussions.
G. H. acknowledges the support by DFG (SFB 1044).
We acknowledge computer time made available to us on the Altix system at the HLRN supercomputing service in Berlin
under the project "B-physics from lattice QCD simulations". Part of this work has been completed thanks to allocation
of CPU time on BlueGene/Q -Fermi based on the agreement between INFN and CINECA and the specific initiative INFN-RM123.
We acknowledge partial support from ERC Ideas Starting Grant n.~279972 ``NPFlavour'' and ERC
Ideas Advanced Grant n.~267985 ``DaMeSyFla". \\

\clearpage
\begin{appendices}
\vspace{1.2cm}
\section{Optimised smeared interpolating operators}
\label{APP_optimised_smeared}

Optimised two-point correlators can be built using optimized
fields $\Phi^{\textrm{opt}}$ as suggested in Eq.~(\ref{optsource})
which we report here for the reader's convenience
\begin{equation} \label{eq:app_optsource}
\Phi^{\rm{opt}} \sim \, {\rm w}\,  \Phi^{\rm{S}} + (1-\rm{w})\, \Phi^{\rm{L}}. 
\end{equation} 
Without loss of generality we consider in practice correlators with 
an optimized field at the source and a local field at the sink, which read\footnote{We should note that it would also be possible to 
construct an optimised correlator formed by a local source $\Phi^{{\rm L}}$ and a sink given by the operator defined in 
Eq.~(\ref{optsource}).}
\begin{equation}\label{eq:C2opt}
C_{2}^{\textrm{opt-L}}(x_0)={\rm w}_{0}^{\textrm{}}\, C_{2}^{({\rm SL})}(x_0)+\tilde{\rm w}_0\, C_{2}^{({\rm LL})}(x_0)
\end{equation}
where we have defined $C_{2}^{({\rm SL})}(x_0) \equiv \langle \Phi^{\rm{S}\, \dagger} \Phi^{\rm{L}} \rangle(x_0)$ and 
$C_{2}^{({\rm LL})}(t) \equiv \langle \Phi^{\rm{L}\, \dagger} \Phi^{\rm{L}} \rangle(x_0)$. We have also set

\begin{equation}\label{eq:w0}
{\rm w}_{0}\equiv {\rm w}^{\textrm{}}/C_{2}^{({\rm SL})}(\overline x_{0}), \,\,\,\,\, \tilde{\rm w}_0 
\equiv (1-{\rm w})/C_{2}^{({\rm LL})}(\overline x_{0})
\end{equation}
The choice of the normalization time $\overline x_{0}$, where by construction
$C_{2}^{\textrm{opt-L}}(\overline x_0) = 1$, is in principle arbitrary and only 
affects the optimal value of w and its behaviour as a function of the mass 
parameters. The $\overline x_{0}$ values used at each $\beta$ in this study
are gathered in Table \ref{tab:time-interval}. They have been chosen so as to
yield a reasonably smooth behaviour of the optimal w--value as a function
of the heavy quark mass $\overline \mu_h$, which in turn eases its numerical
search. We recall that here $\Phi^{{\rm L}}$ and 
$\Phi^{{\rm S}}$ always carry the quantum numbers of a pseudoscalar density.

The optimal values of ${\rm w}$ are those for which earliest Euclidean 
time projection on the ground state can be achieved. We now describe the procedure we have followed. 
We consider the correlator given in  Eq.~(\ref{eq:C2opt})
being computed for a set of ${\rm w}$ values and for each one of them we get estimates of the pseudoscalar meson mass 
on several Euclidean time intervals~\footnote{For instance,
at $\beta=3.80$ there are five time intervals: $(\Delta x_0)^{(1)} =
[5:8]$, \dots \, , $(\Delta x_0)^{(5)} = [9:12]$.}, 
$(\Delta x_0)^{(j)}=\left[x_{0}^{\rm{min}},x_0^{\rm{max}}\right]^{(j)}$. In this way we 
obtain a set of estimators for the pseudoscalar meson mass, namely $M_{ps}^{(j)}$.
In Table \ref{tab:time-interval} we collect the Euclidean time 
intervals $(\Delta x_0)^{(j)}$ that we used at each $\beta$.
Given these time intervals, at each $\beta$ and each $\overline \mu_h$ 
we vary w with a sufficiently fine resolution (typically
around 0.05), then for all w-values we compute the maximal
spread among the pseudoscalar mass estimates, $M_{ps}^{(j)}$, corresponding
to the various intervals $(\Delta x_0)^{(j)}$: $(\Delta M_{ps}) =
\Delta M_{ps}({\rm w}; \overline \mu_h, \beta) \equiv
{\rm max}_{\rm pairs~j,j'} \left| M_{ps}^{(j)} - 
M_{ps}^{(j')} \right|_{\overline \mu_h, \beta}$.
The (almost) optimal value of ${\rm w}$ is of course depending on
$\beta$ and  $\overline \mu_h$ and is taken as the one for which
(i) $(\Delta M_{ps})$ attains a minimum and (ii) the values
of $M_{ps}^{(j)}$ display an oscillating (i.e.\ non-monotonic)
behaviour as a function of $j$ (i.e.\ $(\Delta x_0)^{(j)}$).
The latter condition strongly restricts the values of $x_0^{{\rm min}}$
to be considered in practice for the intervals $(\Delta x_0)^{(j)}$,
the upper end of which, $x_0^{{\rm max}}$, is taken at some larger
Euclidean time (but not too large so as to avoid introducing too
much statistical noise).

\begin{table}[!h]
\begin{center}
\begin{tabular}{|c|c|c|}
\hline
$\beta$ & $\left[x_0^{{\rm min}}/a,x_0^{{\rm max}}/a\right]^{(j)}$  & $\overline x_{0}$\tabularnewline
\hline
\hline 
3.80 & 5:8, 6:9, 7:10, 8:11, 9:12 & 2\tabularnewline
\hline 
3.90 & 8:11, 9:12, 10:13, 11:14, 12:15 & 4\tabularnewline
\hline 
4.05 & 9:14, 10:15, 11:16, 12:17, 13:18, 14:19 & 6\tabularnewline
\hline 
4.20 & 10:16, 11:17, 12:18, 13:19, 14:20 & 7\tabularnewline
\hline 
\end{tabular}
\caption{ $(\Delta x_0)^{(j)}=\left[x_{0}^{\rm{min}},x_0^{\rm{max}}\right]^{(j)}$ Euclidean time intervals 
used in order to get the optimal value of w at each $\beta$. $\overline x_0$ denotes the normalization
time, see Eq.~(\ref{eq:w0}).   
} 
\label{tab:time-interval}
\end{center}
\end{table}   

\begin{figure}[!h]
\vspace*{0.5cm}
\begin{center}
\includegraphics[width=0.55\textwidth]{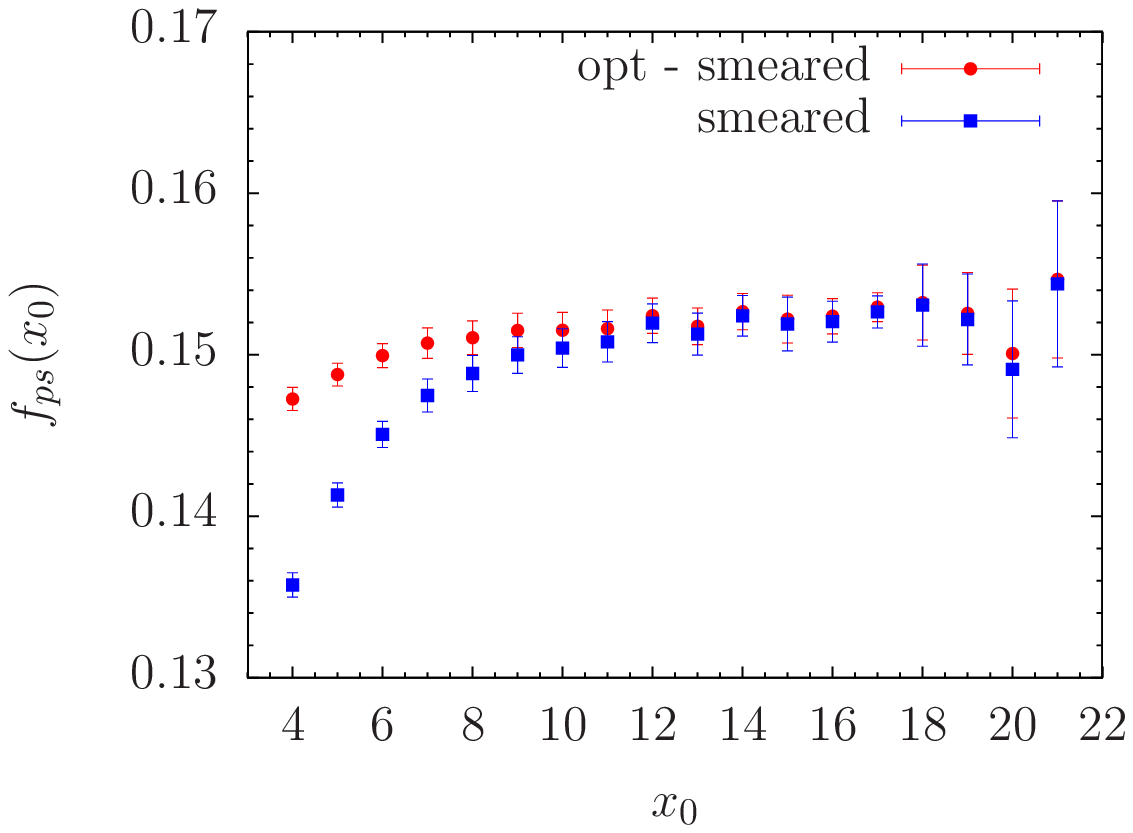} 
\caption{\sl Pseudoscalar decay constant computation at $\beta=3.80$ 
and $\left(a\mu_{\ell}; a\mu_{s}, a\mbox{\ensuremath{\mu}}_{h}\right)=\left(0.0080; 0.0194, 0.5246\right)$
employing smeared fields (blue squares) compared to the one where optimised fields (red circles) are used.  
\label{fig:F-wdep}}
\end{center}
\end{figure}

Using maximally twisted mass lattice fermions, the decay constant of a pseudoscalar
(non-singlet) meson made out of two valence quarks with masses $\mu_1$ and $\mu_2$,
regularized with Wilson parameters $r_1 = -r_2$, is evaluated via the (Ward-identity
based) formula~(\cite{Frezzotti:2000nk, FrezzoRoss1, Blossier:2009bx}:
\begin{equation}
f_{ps}=\dfrac{\left(\mu_{1}+\mu_{2}\right)}{M_{ps}\sinh M_{ps}}\langle 0|P|PS \rangle, 
\end{equation}
where we have set $a=1$ and $P=\bar q_1 \gamma_5 q_2$ is the pseudoscalar density.
Using optimised interpolating
fields, $f_{ps}$ is  extracted from the following formula (in lattice units)

\begin{equation}
f_{ps}(x_0)=\dfrac{\left(\mu_{1}+\mu_{2}\right)}{M_{ps} \sinh (M_{ps})}\dfrac{\sqrt{M_{ps} e^{M_{ps}T/2}}}
{\sqrt{\cosh\left[M_{ps}\left(\frac{T}{2}-x_0\right)\right]}}
\dfrac{C_{2}^{\textrm{opt-L}}(x_0)}
{\sqrt{C_{2}^{\textrm{opt-opt}}(x_0)}}, 
\label{eq:f-WL-WW}
\end{equation}
where the correlator $C_{2}^{{\rm opt-opt}}$ is given by

\begin{equation}
C_{2}^{{\rm opt-opt}}  =  \langle \Phi^{\rm{opt}~\dagger} \Phi^{\rm{opt}} \rangle (x_0) .
\end{equation}
In Fig.~\ref{fig:F-wdep} 
we illustrate an example on the pseudoscalar decay
constant computation using optimised and smeared interpolating operators.

We have investigated the possibility of using improved interpolating operators in the computation
of $B$-parameters. Optimised bare $B_{i}$ are evaluated from the ratio
of the double-optimised 3-point correlation function defined as 
\begin{equation}
C_{3;i}^{\textrm{opt-opt}}(x_0) = \langle \Phi^{{\rm opt}}\, O_i\, \Phi^{{\rm opt}} \rangle (x_0), \,\,\, i = 1, \ldots 5
\end{equation}
 and the two 2-point correlation functions, $C_{2}^{\textrm{opt-L}}(x_0)$ and $C_{2}^{'\textrm{opt-L}}(x_0)$, 
 ({\it cf.} Eq.~(\ref{BBratio}))
 
\begin{equation}\label{eq:EBi_opt} 
E[B_{i}^{\textrm{opt-opt}}](x_0) \,= \, \dfrac{C_{3;i}^{\textrm{opt-opt}}(x_0)}{C_{2}^{\textrm{opt-L}}(x_0) \,\, 
C_{2}^{' \textrm{opt-L}}(x_0)}, 
\quad  i=1, \ldots, 5 .
\end{equation} 

This computation implies the employment of optimised 
$\langle P^{\dagger} A_{0} \rangle$ and $\langle P^{\dagger} P \rangle$ 2-point correlation functions, for $i=1$ and $i=2,3,4,5$,
respectively.

\begin{figure}[!h]
\vspace*{0.3cm}
\begin{center}
\subfigure[]{\includegraphics[scale=0.65,angle=-0]{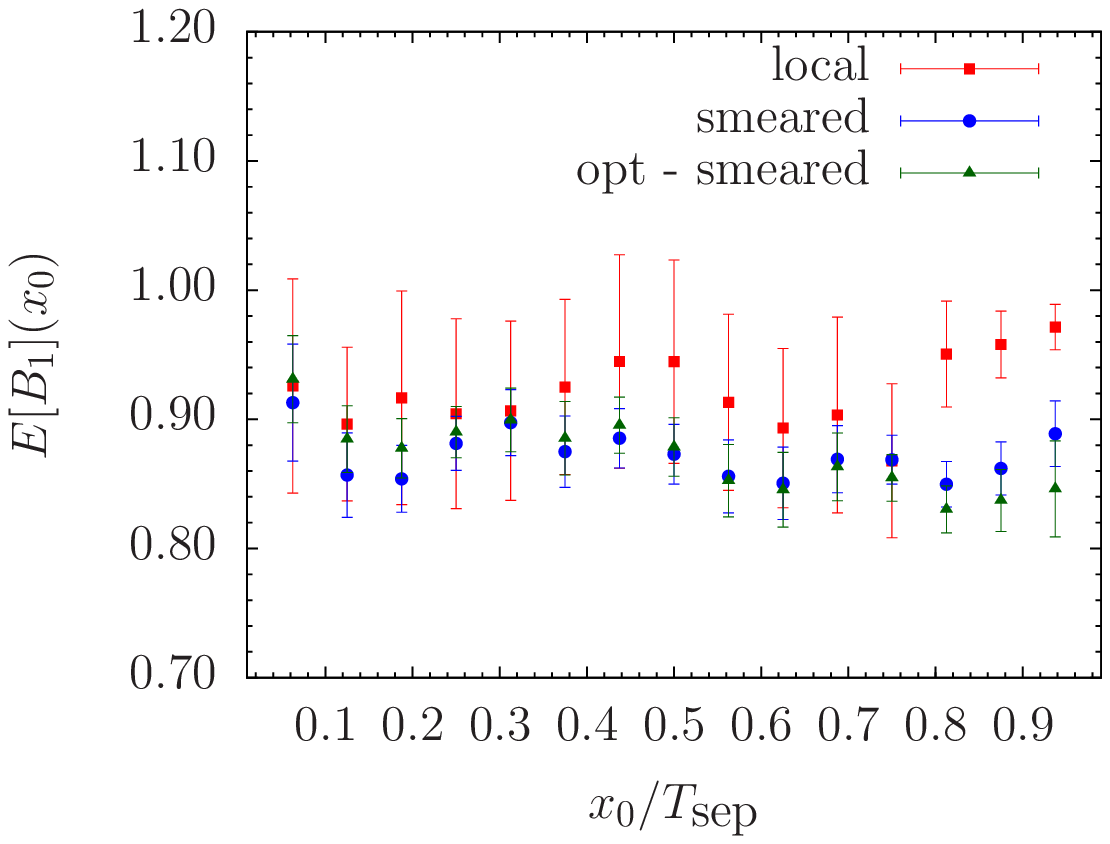}}
\hspace*{0.5cm}
\subfigure[]{\includegraphics[scale=0.65,angle=-0]{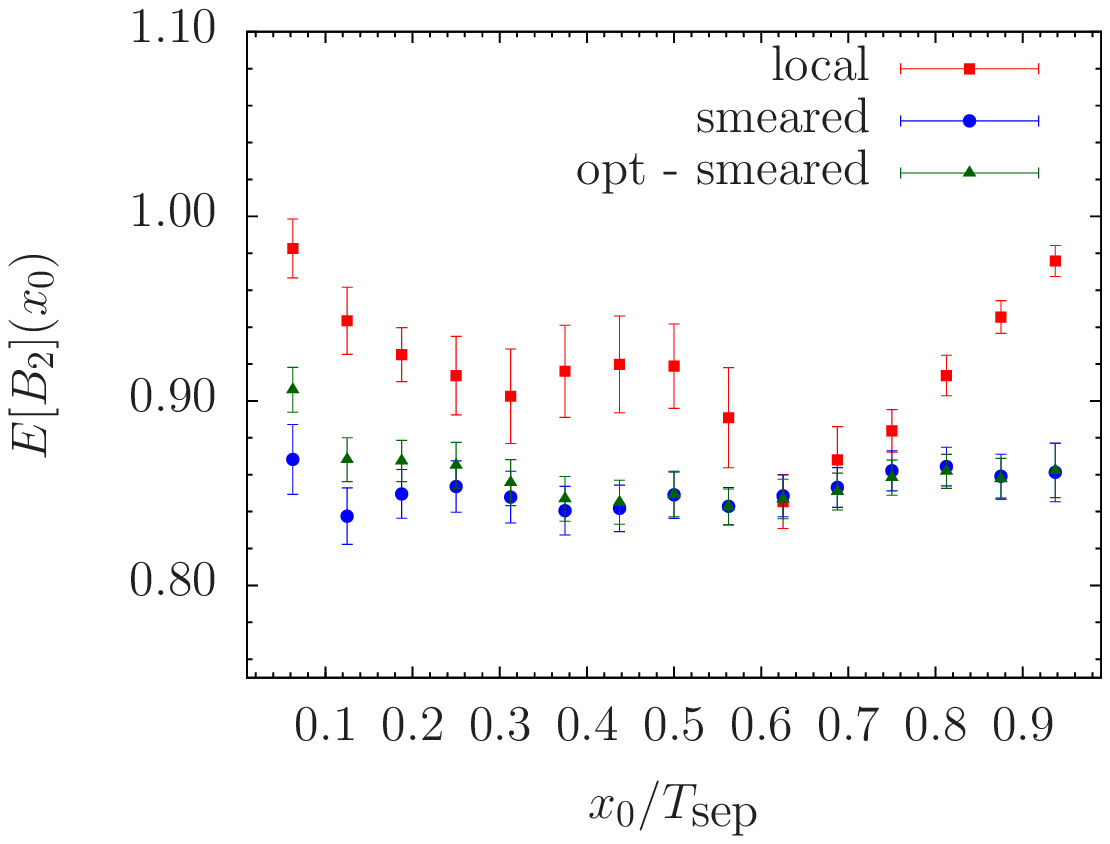}}
\caption{\sl Plateau quality for $E\left[B_{1}\right]$ (left)
and $E\left[B_{2}\right]$ (right) at $\beta=3.80$ and for 
$\left(a\mu_{\ell},a\mbox{\ensuremath{\mu}}_{h}\right)=\left(0.0080,0.5246\right)$
obtained from local (red squares), smeared (blue circles) and optimised
interpolating fields (green triangles).
  }
\label{fig:B-wdep}
\end{center}
\end{figure}

At $\beta=3.80$ we have compared results for the (bare) bag parameters $B_i$ ($i=1, \ldots, 5$) obtained through  
Eq.~(\ref{eq:EBi_opt}) with the ones coming through the use of smeared interpolating operators. We have found out that  
within our current statistical errors it is hardly noticed any difference on the plateau values of the bag parameters.  
In the two panels of Fig.~\ref{fig:B-wdep} we illustrate two examples supporting the above numerical observation for the cases of 
$B_i$ with $i=1, 2$, respectively. 
A similar behaviour has been observed also for the bag
parameters $B_3$, $B_4$ and $B_5$.
In the two figures we have also included data corresponding to the case where local sources have been employed. In this last case, 
as it might be expected, the plateau quality results problematic if heavy quark masses are to be employed
as we do in the present study. 
Based on the above observations we have thus opted for using smeared interpolating fields in the computation of the 
bag parameters throughout this work.

Finally, we note that statistical errors on pseudoscalar meson masses, pseudoscalar decay constants as well as 
on four-fermion operator matrix elements have been evaluated using the jackknife procedure. 
With 16 jackknife bins for each configuration ensemble we have verified that autocorrelations are well
under control. In order to  take into proper account the cross correlations we compute statistical errors on the fit results
using the bootstrap method and for that we employ 1000 bootstrap samples produced by independent gauge configuration ensembles.

\clearpage
\vspace{1.2cm}
\section{QCD-HQET matching of $B$-parameters}
\label{APP_QCD-HQET}

According to the HQET scaling laws, each $B$-parameter scales with the
inverse heavy quark mass $\overline\mu_{h}$ as a constant up to some perturbative logarithmic 
corrections of order $\sim1/\log\left(\overline\mu_{h}/\Lambda_{\textrm{QCD}}\right)$.
These corrections are expected to be tiny in the range of heavy quark masses
we are dealing with. This point is checked, and the corresponding systematic
uncertainty is quantified, by matching at different
(TL, LL, NLL) perturbative orders the QCD $B$-parameters to their counterparts
in HQET, thereby yielding $B$-parameters with a well-defined static limit and
correspondingly smaller ( O($ \alpha(\overline\mu_{h})$), 
O($ \alpha(\overline\mu_{h})^2$), O($ \alpha(\overline\mu_{h})^3$) )
logarithmic corrections to their leading $1/\overline \mu_h$--behaviour.
It turns out (see the error budget discussion in Section~\ref{sec:summary_final_results}) that the impact of
logarithmic corrections to the power-like $1/\overline \mu_h$--behaviour
on the chain equations~(\ref{eq:omega_d_chain}) and~(\ref{eq:omega_s_chain}), through which $B$-parameters are
determined, is small (O(1\%)) compared to other uncertainties in the calculation.
It is only to this well controlled extent that a perturbative QCD-to-HQET 
matching of the quantities of interest, carried out by using the formulae 
given in the present Appendix, enters in our ratio method computation.

In this Appendix, we consider the QCD four-fermion operators in the SUSY basis of Eq.~(\ref{def_Oi}). 
The corresponding HQET operators have exactly the same form but with
the relativistic heavy quark field $h$ 
replaced by the (infinitely) heavy quark field of HQET. 
In the HQET due to the heavy quark spin symmetries, the operator ${\cal O}_3$ is related to ${\cal O}_1$ 
and ${\cal O}_2$ by the relation ${\cal O}_3=-{\cal O}_2-1/2 \,{\cal O}_1$. Nevertheless, we find it 
convenient to work with the redundant basis of five operators, which includes ${\cal O}_3$, 
in order to deal with squared 5x5 evolution and matching matrices in HQET as in QCD.

The relation between QCD $B$-parameters evaluated at the heavy quark mass 
$\overline \mu_{h}$ and their counterparts in HQET
can be expressed as:
\be
\tilde{B}_{i}\left(\overline{\mu}_{h};\mu^{*}\right)=W_{ij}\left(\mu^{*},\overline{\mu}_{h},\mu\right)B_{j}
\left(\overline{\mu}_{h};\mu\right)
\ee
where $B_{i}\left(\overline{\mu}_{h};\mu\right)$ denotes the $B$-parameters in QCD renormalized at the scale 
$\mu$ and $\tilde{B}_{i}\left(\overline{\mu}_{h};\mu^{*}\right)$ are the HQET $B$-parameters renormalized at the scale 
$\mu^{*}$. The latter satisfy the heavy quark scaling laws with logarithmic corrections dictated by their 
renormalization group evolution, which are therefore easy to take into account when applying the ratio method. 

The matrix $W\left(\mu^{*},\overline{\mu}_{h},\mu\right)$ can be decomposed as follows:
\be
\mathbf{W}\left(\mu^{*},\overline{\mu}_{h},\mu\right) =
\tilde{\mathbf{U}}\left(\mu^{*},\overline{\mu}_{h}\right)
\mathbf{C}\left(\overline{\mu}_{h}\right)^{-1}
\mathbf{U}\left(\overline{\mu}_{h},\mu\right)
\ee
The matrix $\mathbf{U}\left(\overline{\mu}_{h},\mu\right)$ encodes the full QCD evolution from the scale $\mu$ to the scale $\overline{\mu}_{h}$ 
for the five $\Delta B=2$ $B$-parameters. $\tilde{\mathbf{U}}\left(\mu^{*},\overline{\mu}_{h}\right)$ is the 
corresponding evolution matrix in HQET, from $\overline{\mu}_{h}$ to $\mu^{*}$.
Finally, $\mathbf{C}\left(\overline{\mu}_{h}\right)$ provides the matching from HQET to QCD at the common scale $\overline{\mu}_{h}$.

At next-to-leading-log (NLL), the explicit expression of $\mathbf{W}$ reads
\be
\begin{array}{cl}
\mathbf{W}^{\mbox{NLL}}\left(\mu^{*},\overline{\mu}_{h},\mu\right)=\mathcal{C}^{-1} & 
\left\{ \left[1-\dfrac{\alpha\left(\mu^{*}\right)}{4\pi}
\mathbf{\tilde{J}}_{B}^{T}\right]\left[\dfrac{\alpha\left(\mu^{*}\right)}{\alpha\left(\overline{\mu}_{h}\right)}
\right]^{\frac{\tilde{\mathbf{\gamma}}_{B}^{(0)}}{2\beta_{0}}}
\left[1+\dfrac{\alpha\left(\overline{\mu}_{h}\right)}{4\pi}\mathbf{\tilde{J}}_{B}^{T}\right]\right.\\
\\
 & \times\left[1-\mathbf{c}_{B}^{(1)}\dfrac{\alpha\left(\overline{\mu}_{h}\right)}{4\pi}\right]\\
\\
 & \left.\times\left[1-\dfrac{\alpha\left(\overline{\mu}_{h}\right)}{4\pi}\mathbf{J}_{B}^{T}\right]
 \left[\dfrac{\alpha\left(\overline{\mu}_{h}\right)}{\alpha\left(\mu\right)}\right]^{\frac{\mathbf{\gamma}_{B}^{(0)}}{2\beta_{0}}}
 \left[1+\dfrac{\alpha\left(\mu\right)}{4\pi}\mathbf{J}_{B}^{T}\right]\right\} \mathcal{C}
 \label{eq:WNLL}
\end{array}
\ee
where the superscript $T$ stands for ``transposed". The corresponding leading-log (LL) expression of $\mathbf{W}$ 
is obtained by 
setting $\mathbf{J}_{B} = \mathbf{\tilde J}_{B} = \mathbf{c}_{B}^{(1)} =0$.
Moreover, in Eq.~(\ref{eq:WNLL}):
\begin{itemize}

\item $\beta_{0}$ is the leading order coefficient of the QCD beta function, $\beta_{0}=11-2N_{f}/3$

\item $\mathcal{C}=\mbox{diag}\left\{ 8/3,-5/3,1/3,2,2/3\right\} $ 

\item $\gamma_{B}^{(0)}$ is the scheme-independent one-loop anomalous dimension matrix (ADM) of the QCD $B$-parameters. 
It is obtained by combining the one-loop ADM of the corresponding four-fermion operator $\gamma^{(0)}$ with the one-loop 
ADM of the pseudoscalar density $\gamma_{P}^{(0)}$ as: 
\be
\left\{ \mathbf{\gamma}_{B}^{(0)}\right\} _{ij}=\gamma_{ij}^{(0)}-2\gamma_{P}^{(0)}{\displaystyle \sum_{k=2}^{k=5}\delta_{ik}\delta_{jk}}
\ee
where $\gamma_{P}^{(0)}=-8$~\cite{Chetyrkin:1999pq}   and the expression of $\gamma^{(0)}$ in the SUSY basis of Eq.~(\ref{def_Oi})
reads~\cite{Ciuchini:1997bw}
\be
\gamma^{(0)}=\left(\begin{array}{ccccc}
4 & 0 & 0 & 0 & 0\\
0 & -28/3 & 4/3 & 0 & 0\\
0 & 16/3 & 32/3 & 0 & 0\\
0 & 0 & 0 & -16 & 0\\
0 & 0 & 0 & -6 & 2
\end{array}\right)
\ee

\item $\tilde{\gamma}_{B}^{(0)}$ is the scheme-independent one-loop ADM of the HQET $B$-parameters. 
It can be obtained from the ADMs of the static-light four-fermion 
operator $\tilde{\gamma}^{(0)}$ and the static-light axial
current $\tilde{\gamma}_{A}^{(0)}$ as~\footnote{In HQET
the static-light axial current and pseudoscalar density operators
happen to coincide and carry a non-zero anomalous dimension.}
\be
\left\{ \tilde{\mathbf{\gamma}}_{B}^{(0)}\right\} _{ij}=\tilde{\gamma}_{ij}^{(0)}-2\tilde{\gamma}_{A}^{(0)}\delta_{ij}
\ee
where $\tilde{\gamma}_{A}^{(0)}=-4$~\cite{Gimenez:1991bf}  and $\tilde{\gamma}^{(0)}$ is given 
by~\cite{Bpar:SPQR1} 
\be
\tilde{\gamma}^{(0)}=\left(\begin{array}{ccccc}
-8 & 0 & 0 & 0 & 0\\
0 & -16/3 & -8/3 & 0 & 0\\
0 & -8/3 & -16/3 & 0 & 0\\
0 & 0 & 0 & -7 & -3\\
0 & 0 & 0 & -3 & -7
\end{array}\right)
\ee

\item $\mathbf{J}_{B}$ is the scheme-dependent two-loop ADM of the $B$-parameters in QCD. 
It is obtained from the ADMs of the four-fermion operator $\mathbf{J}$ and of the pseudoscalar density $J_{P}$ through 
\be
\left\{ J_{B}\right\} _{ij}=J_{ij}-2J_{P}{\displaystyle \sum_{k=2}^{k=5}\delta_{ik}\delta_{jk}}
\ee
where in the $\rm\overline{MS}$ scheme with $N_{f}=2$ active 
flavors one obtains $J_{P}=8134/2523$.
The expression of $\mathbf{J}$
in the SUSY basis and in the $\rm\overline{MS}$ scheme of Ref.~\cite{mu:4ferm-nlo}, with $N_{f}=2$, is
\be
J=\left(\begin{array}{ccccc}
\frac{9875}{5046} & 0 & 0 & 0 & 0\\
\\
0 & \frac{1318145}{310329} & \frac{1633930}{310329} & 0 & 0\\
\\
0 & \frac{33817}{310329} & -\frac{2024698}{310329} & 0 & 0\\
\\
0 & 0 & 0 & \frac{2882869}{565152} & \frac{3365}{188384}\\
\\
0 & 0 & 0 & \frac{627}{224} & -\frac{576173}{565152}
\end{array}\right)
\ee

\item $\mathbf{\tilde{J}}_{B}$ is the scheme-dependent two-loop ADM of the $B$-parameters in HQET. It is obtained from the ADMs of the 
four-fermion operator $\mathbf{\tilde{J}}$ and of the axial density $\tilde{J}_{A}$, both in HQET, as 
\be
\left\{ \tilde{J}_{B}\right\} _{ij}=\tilde{J}_{ij}-2\tilde{J}_{A}\delta_{ij}
\ee
where $\tilde{J}_{A}=1037/2523-28\pi^{2}/261$ in the $\rm\overline{MS}$ scheme with $N_{f}=2$.
The HQET two-loop ADM of the four-fermion operators has been computed in Ref.~\cite{Reyes:thesis}. In the SUSY basis, it reads
\be
\tilde{\mathbf{J}}=\left(\begin{array}{ccccc}
\frac{68\pi^{2}}{261}-\frac{943}{2523} & 0 & 0 & 0 & 0\\
\\
0 & \frac{74\pi^{2}}{261}-\frac{24509}{30276} & \frac{13193}{30276}-\frac{2\pi^{2}}{87} & 0 & 0\\
\\
0 & \frac{18239}{30276}-\frac{2\pi^{2}}{87} & \frac{74\pi^{2}}{261}-\frac{29555}{30276} & 0 & 0\\
\\
0 & 0 & 0 & \frac{74\pi^{2}}{261}-\frac{5819}{5046} & \frac{4375}{5046}-\frac{2\pi^{2}}{87}\\
\\
0 & 0 & 0 & \frac{4375}{5046}-\frac{2\pi^{2}}{87} & \frac{74\pi^{2}}{261}-\frac{5819}{5046}
\end{array}\right)
\ee

\item $\mathbf{c}_{B}^{(1)}$ is the LL order coefficient of the matching matrix between HQET and QCD. 
By combining the matching relations for the four-fermion operators, the axial and the pseudoscalar densities, one finds
\be
\left\{ c_{B}^{(1)}\right\} _{ij}=c_{ij}^{(1)}-2c_{A}^{(1)}
\delta_{i1}\delta_{j1}-2c_{P}^{(1)}{\displaystyle \sum_{k=2}^{k=5}\delta_{ik}\delta_{jk}}
\ee
where the axial and pseudoscalar coefficients take the values $c_{A}^{(1)}=-8/3$ and $c_{P}^{(1)}=8/3$~\cite{Gimenez:1991bf} 
and the matching matrix for the four-fermion operators is~\cite{Bpar:SPQR1}
\be
\mathbf{c}^{(1)}=\left(\begin{array}{ccccc}
-14 & -8 & 0 & 0 & 0\\
0 & 61/12 & -13/4 & 0 & 0\\
0 & -77/12 & -121/12 & 0 & 0\\
0 & 0 & 0 & 17/2 & -11/2\\
0 & 0 & 0 & 7/2 & -21/2
\end{array}\right)
\ee
\end{itemize}

\clearpage
\vspace{1.2cm}
\section{Ratio method and the static limit relation between the four-fermion operators $\mathbf{{\mathcal O}_1}$, 
$\mathbf{{\mathcal O}_2}$ \\ and $\mathbf{{\mathcal O}_3}$ }
\label{APP_O123}

In the static limit the equations of motion relate the four-fermion operators $\mathcal{O}_{1}^{(q)}$,  
$\mathcal{O}_{2}^{(q)}$ and $\mathcal{O}_{3}^{(q)}$ ($q = d, s$), at tree-level, via the relationship
\begin{equation}\label{eq:O123}
\mathcal{O}_{3}^{(q)}+\mathcal{O}_{2}^{(q)}+\dfrac{1}{2}\mathcal{O}_{1}^{(q)}=0, \,\,\,\, q = d, s
\end{equation}
In this appendix it will be shown that using the ratio method and data taken from relativistic quark simulations one can verify
the validity of Eq.~(\ref{eq:O123}) in the infinite heavy quark mass limit. 

Following the discussion in Ref.~\cite{Bertone:2012cu}, see Eq.~(4.16) and (4.19),  
we define the ratios of the three-point correlation 
functions
\begin{equation}\label{eq:E[Ri]}
E\left[R_{i}^{(q)}\right]\left(x_{0}\right)=\dfrac{C_{3;i}^{(q)}\left(x_{0}\right)}{C_{3;1}^{(q)}\left(x_{0}\right)},
\,\,\,\,\,\,\, i =2, 3\,\, {\rm and} \,\, q = d, s 
\end{equation}
in order to get lattice estimators for the ratios of the matrix elements, given by
\begin{equation}\label
{eq:Ri-def}
R_{i}^{(q)}=
\dfrac{\langle\overline{B}_{q}^{0}|O_{i}^{(q)}|B_{q}^{0}\rangle}{\langle\overline{B}_{q}^{0}|O_{1}^{(q)}|B_{q}^{0}\rangle},\,\,\,\, (i = 2,3 \,\, 
{\rm and} \,\, q = d, s). 
\end{equation}
We will show numerically, using just TL formulae for connecting matrix elements
in QCD to their HQET counterparts, that the following equation
\begin{equation}\label{eq:static-relation-R}
R_{3}^{(q)}+R_{2}^{(q)}+\dfrac{1}{2}=0, \,\,\,\,\,\,\, q = d, s
\end{equation}
is satisfied in the static limit with good precision.

Following the familiar procedure that we have used throughout 
this paper we define suitable ratios of $R_{i}^{(d)}$, or $R_{i}^{(s)}$
evaluated at nearby heavy quark mass values. With the aforementioned TL
approximation in the formulae for QCD-to-HQET connection they read

\begin{eqnarray}
r_{i}^{(d)}\left(\overline{\mu}_{h},\lambda; \overline{\mu}_{\ell},a\right) &=&
\dfrac{R_{i}^{(d)}\left(\overline{\mu}_{h},\overline{\mu}_{\ell},a\right)}{R_{i}^{(d)}\left(\overline{\mu}_
{h}/\lambda,\overline{\mu}_{\ell},a\right)}\,\,\,\,\,\,(i=2,3)\label{eq:ri-d} \\
r_{i}^{(s)}\left(\overline{\mu}_{h},\lambda; \overline{\mu}_{\ell}, \overline{\mu}_{s},a\right) &=& \dfrac{R_{i}^{(s)}\left(\overline{\mu}_{h},\overline{\mu}_{\ell},\overline{\mu}_{s}a\right)}{R
_{i}^{(s)}\left(\overline{\mu}_{h}/\lambda, \overline{\mu}_{\ell}, \overline{\mu}_{s}a\right)}\,\,\,\,(i=2,3) \label{eq:ri-s}
\end{eqnarray}
and they satisfy the asymptotic limit condition
\begin{equation}
\lim_{\overline{\mu}_{h}\rightarrow\infty}r_{i}^{(q)}\left(\overline{\mu}_{h}\right)=1 \,\,\,\,(i = 2,3 \,\, 
{\rm and} \,\, q = d, s).
\end{equation}

\begin{figure}[!h]
\vspace*{0.3cm}
\begin{center}
\subfigure[]{\includegraphics[scale=0.65,angle=-0]{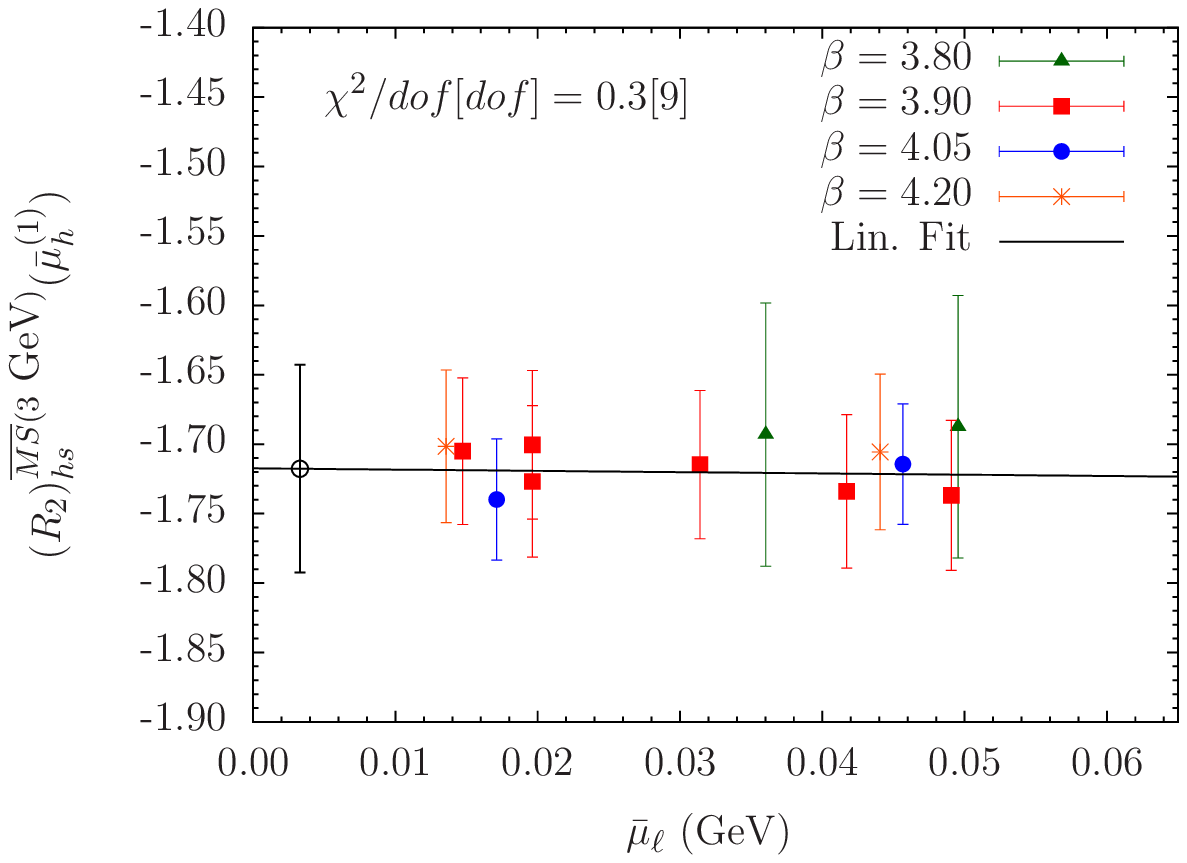}}
\hspace*{0.5cm}
\subfigure[]{\includegraphics[scale=0.65,angle=-0]{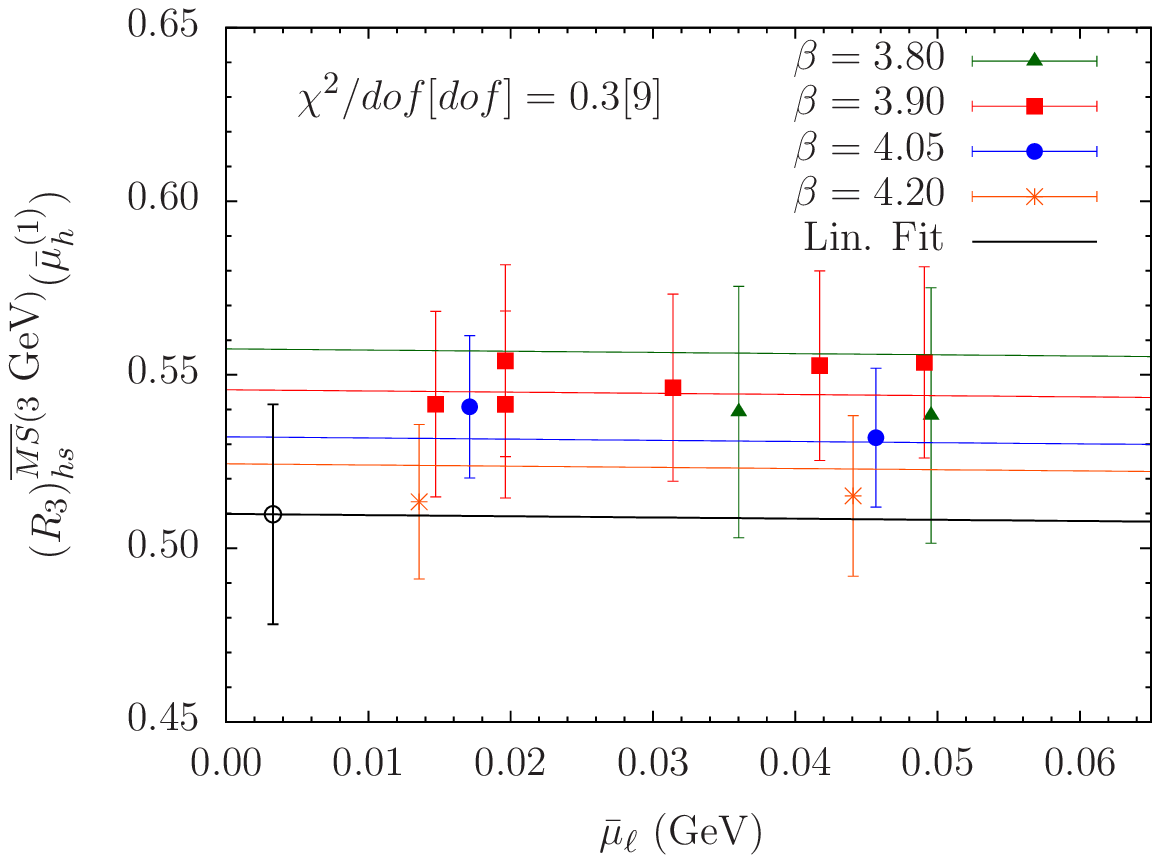}}
\caption{\sl Combined chiral and continuum limit
at the triggering point for $R_{2}^{(s)}$ (left) and $R_{3}^{(s)}$
(right), renormalised in the $\overline{\rm{MS}}$ scheme of Ref.~\cite{mu:4ferm-nlo} at 3 GeV. 
The fit ansatz is a linear fit in $\overline{\mu}_{\ell}$ and
in $a^{2}$.
  }
\label{fig:Ri_trig}
\end{center}
\end{figure}

In Figs~\ref{fig:Ri_trig}(a) and (b) we display the combined chiral-continuum fit against the light quark mass 
for $R_2$ and $R_3$, respectively, at the triggering point ($\overline \mu_{h}^{(1)}$). 
In both fits we have used a linear fit ansatz in $\overline \mu_{\ell}$ as well as  in $a^2$.

\begin{figure}[!h]
\vspace*{0.3cm}
\begin{center}
\subfigure[]{\includegraphics[scale=0.65,angle=-0]{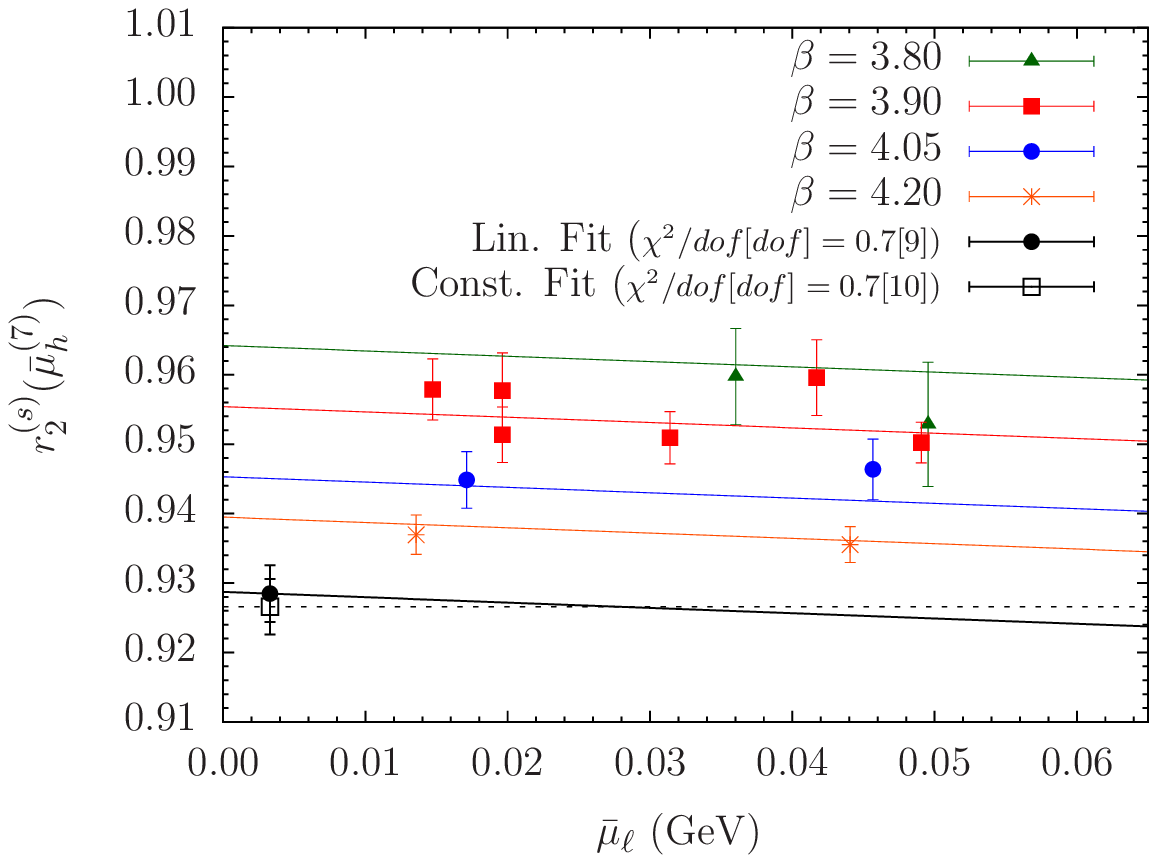}}
\hspace*{0.5cm}
\subfigure[]{\includegraphics[scale=0.65,angle=-0]{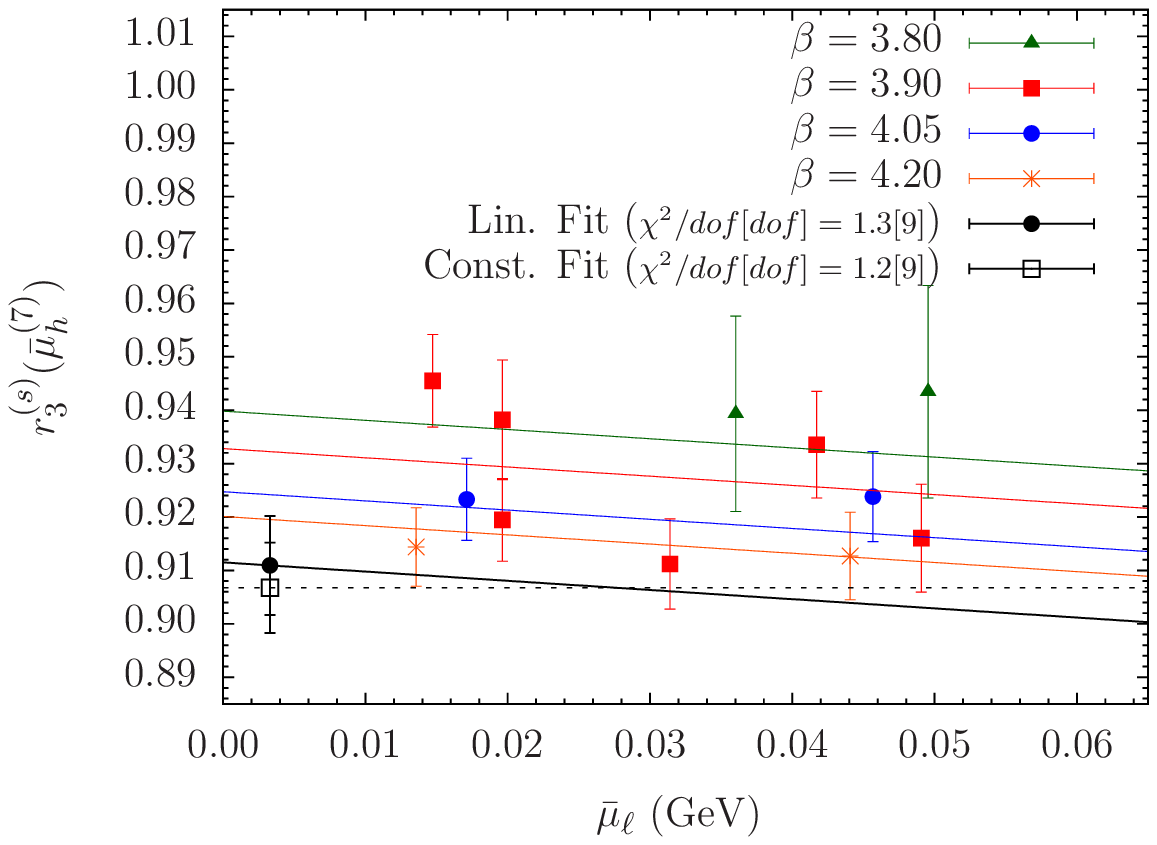}}
\caption{\sl Combined chiral and continuum fits
of the ratio $r_{2}^{(s)}\left(\overline{\mu}_{h}\right)$ (left panel)
and $r_{3}^{(s)}\left(\overline{\mu}_{h}\right)$ (right panel) against
$\overline{\mu}_{\text{\ensuremath{\ell}}}$ for the largest value of the
heavy quark mass considered in this work ($n=7$). The full black line corresponds to a
linear fit ansatz in $\overline{\mu}_{\ell}$ and $a^2$, while the dashed one corresponds
to the continuum limit curve in the case of a linear fit in $a^{2}$
without dependence on $\overline{\mu}_{\ell}$ (constant fit). In both panels colored lines correspond to 
the linear fit ansatz in $\overline{\mu}_{\ell}$ and $a^2$. 
  }
\label{fig:chiral-continuum-ri}
\end{center}
\end{figure}

For each pair of nearby heavy quark masses we get the continuum limit result at the physical point of the ratios 
defined in Eqs~(\ref{eq:ri-d}) and (\ref{eq:ri-s}) through 
a combined chiral-continuum extrapolation.      
In Figs~\ref{fig:chiral-continuum-ri}(a) and (b) we illustrate this extrapolation
for $r_{2}^{(s)}\left(\overline{\mu}_{h}^{(n)}\right)$ and $r_{3}^{(s)}\left(\overline{\mu}_{h}^{(n)}\right)$, respectively  at the largest
value of heavy quark mass used in this work ($n=7$).

\begin{figure}[!h]
\vspace*{0.3cm}
\begin{center}
\subfigure[]{\includegraphics[scale=0.65,angle=-0]{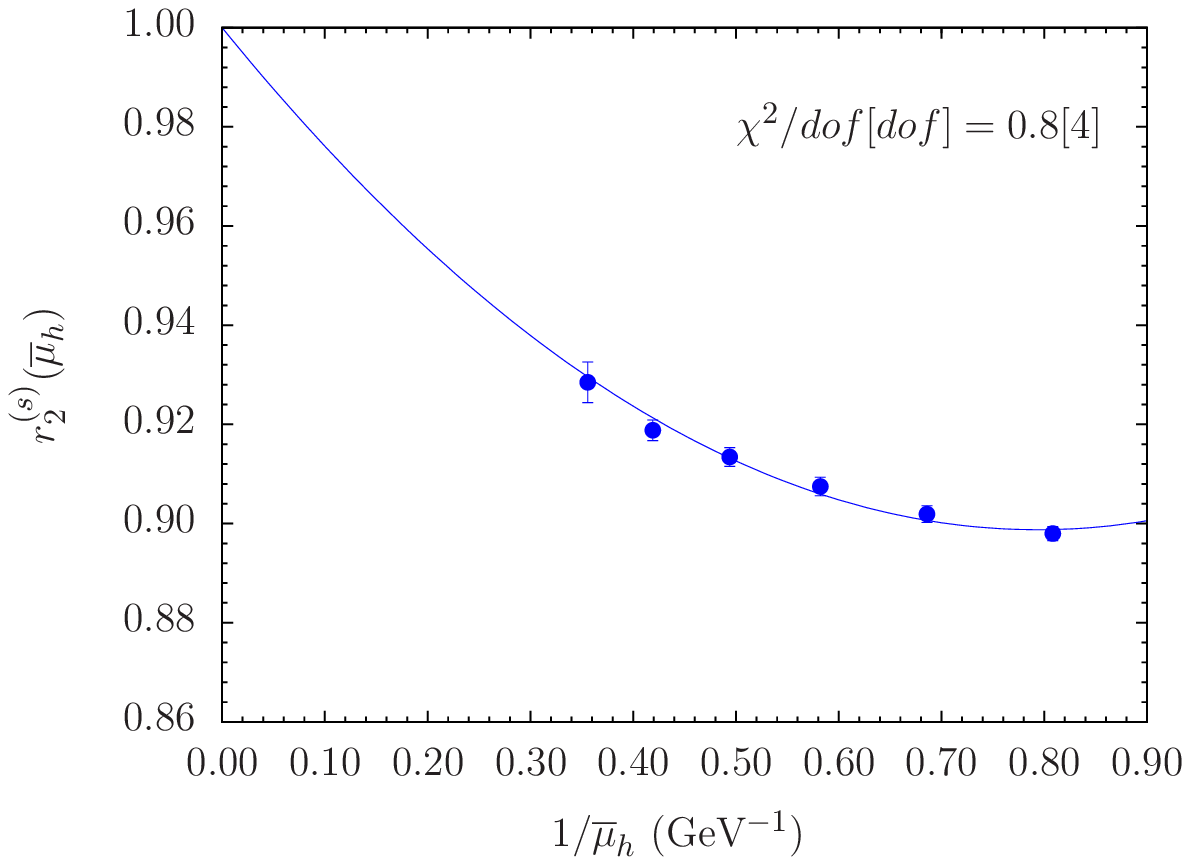}}
\hspace*{0.5cm}
\subfigure[]{\includegraphics[scale=0.65,angle=-0]{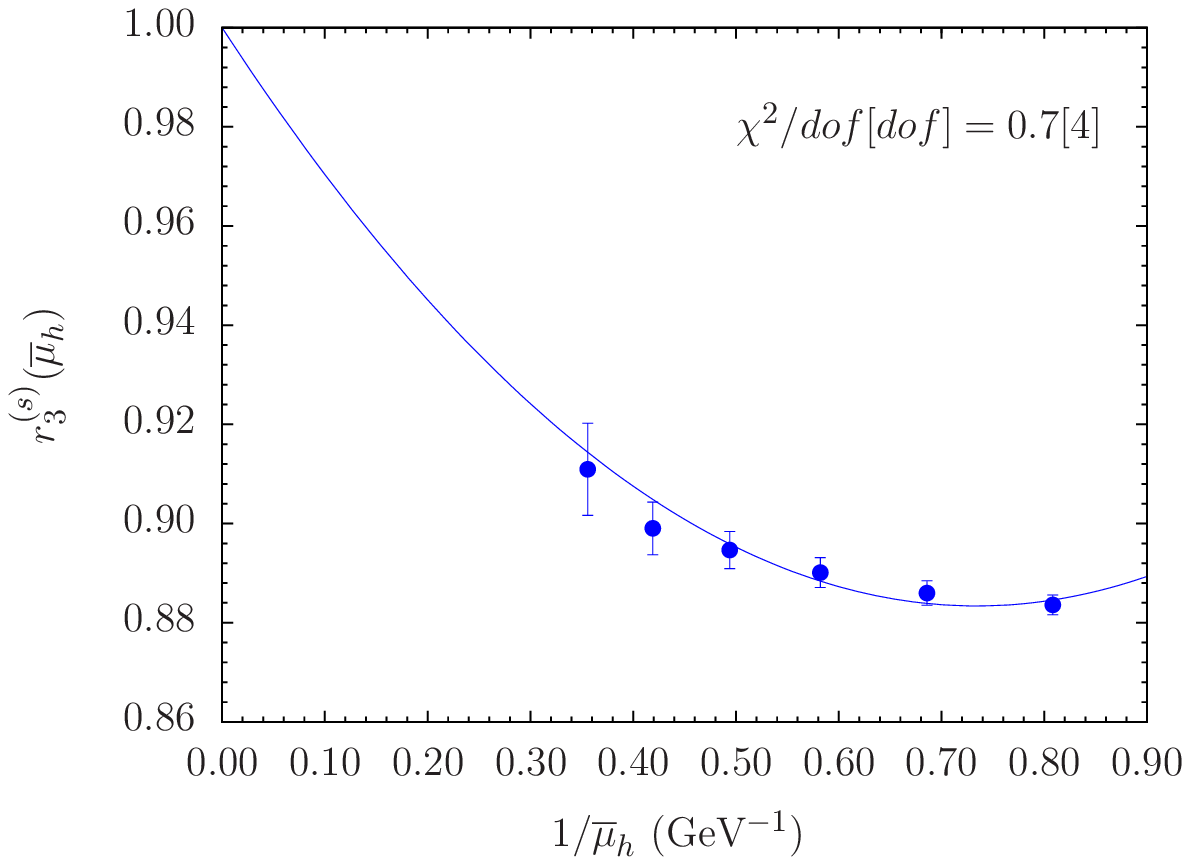}}
\caption{\sl $r_{2}^{(s)}(\overline \mu_h)$ and $r_{3}^{(s)}(\overline \mu_h)$ versus $1/\overline{\mu}_{h}$.
For both cases  a second order polynomial fit ansatz in $1/\overline{\mu}_{h}$ with the 
static condition at unity has been used.  
  }
\label{fig:fit-muh}
\end{center}
\end{figure}

In Fig.~\ref{fig:fit-muh} we show the dependence of $r_{2}^{(s)}(\overline \mu_h)$
and $r_{3}^{(s)}(\overline \mu_h)$ on the inverse heavy quark mass $1/\overline{\mu}_{h}$.
We fit data employing a second order polynomial fit function in $1/\overline{\mu}_{h}$
while the static condition at unity is explicitly imposed.

\begin{figure}[!h]
\vspace*{0.3cm}
\begin{center}
\subfigure[]{\includegraphics[scale=0.65,angle=-0]{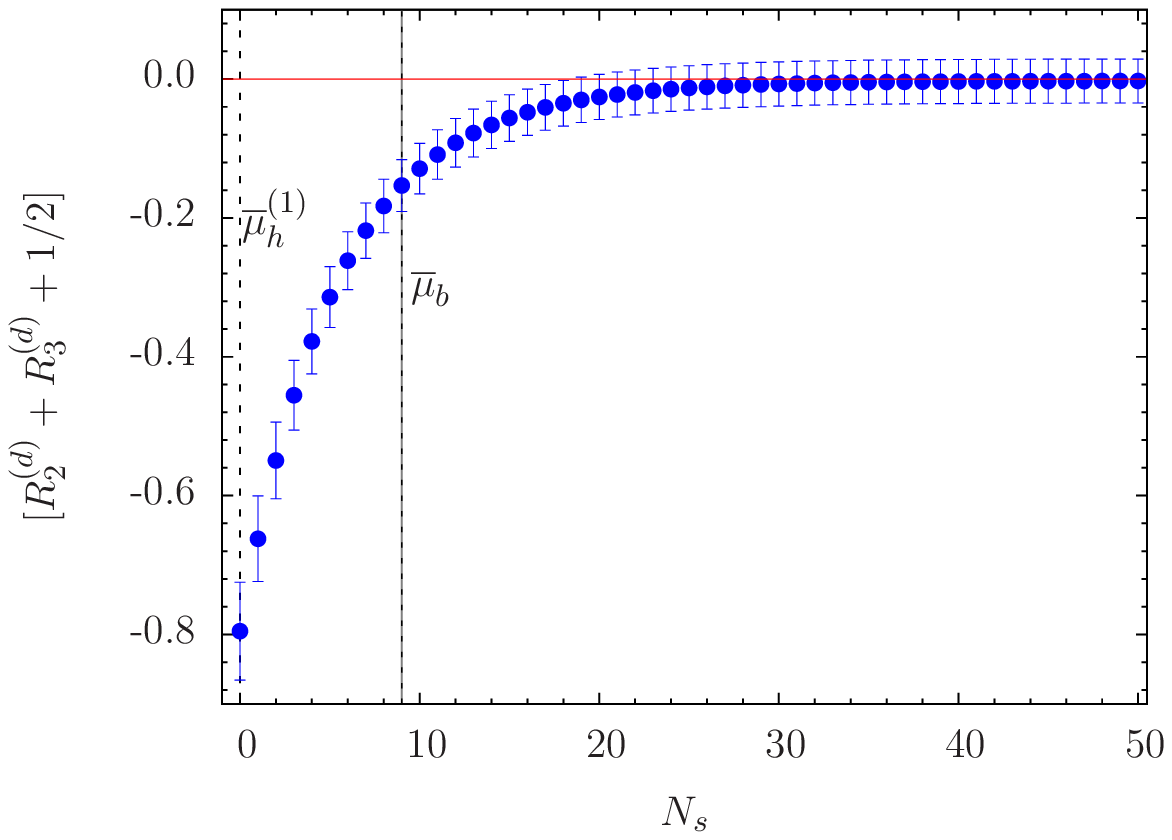}}
\hspace*{0.5cm}
\subfigure[]{\includegraphics[scale=0.65,angle=-0]{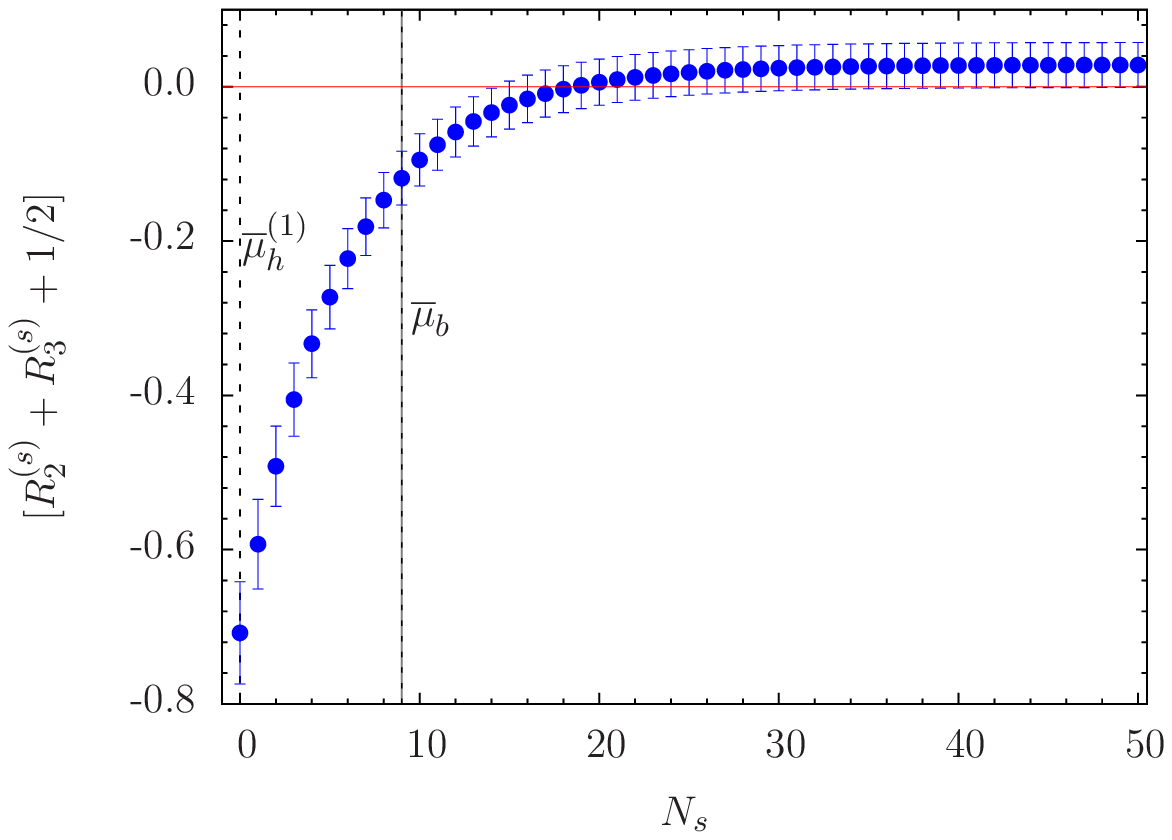}}
\caption{\sl The behaviour of $\left[ R_{3}^{(q)}+R_{2}^{(q)}+\dfrac{1}{2} \right]$ with the increasing number of $N_s$  for 
$q = d$ (left panel) and  $q = s$ (right panel). In each plot the vertical dashed line  indicates 
the triggering point $(\overline \mu_h^{(1)})$, while the dotted vertical one indicates the position of $\overline \mu_b$. 
  }
\label{fig:R23_vs_Ns}
\end{center}
\end{figure}

The determinations of $R_{i}^{(q)}$ in the infinite heavy quark mass limit are obtained
through the use of a chain equation assuming a high number of steps $N_s$ in terms of which we write  
$\overline{\mu}_{h}^{(N_{s})}=\lambda^{N_{s}}\overline{\mu}_{h}^{(1)}$. The chain equation reads 

\begin{equation}\label{eq:chain-eq}
R_{i}^{(q)}\left(\overline{\mu}_{h}^{(N_{s})}\right)=
\left[\prod_{k=2}^{k=N_{s}}r_{i}^{(q)}\left(\overline{\mu}_{h}^{(k)}\right)\right]R_{i}^{(q)}\left(\overline{\mu}_{h}^{
(1)}\right), \,\,\,\, q = d, s
\end{equation}
where  with $R_{i}^{(q)}\left(\overline{\mu}_{h}^{(1)}\right)$ we denote the triggering point estimate. 
We get the following results:

\begin{equation}
\lim_{N_{s}\rightarrow\infty}\left[ R_{3}^{(d)}+R_{2}^{(d)}+\dfrac{1}{2} \right] \left(\overline{\mu}_{h}^{(N_{s})}\right)=-0.003(32)
\end{equation}

\begin{equation}
\lim_{N_{s}\rightarrow\infty}\left[ R_{3}^{(s)}+R_{2}^{(s)}+\dfrac{1}{2} \right] \left(\overline{\mu}_{h}^{(N_{s})}\right)=0.028(29)
\end{equation}
In Fig.~\ref{fig:R23_vs_Ns}(a) and (b) we show the behaviour of the quantites $\left[ R_{3}^{(d)}+R_{2}^{(d)}+\dfrac{1}{2} \right]$ and 
$\left[ R_{3}^{(s)}+R_{2}^{(s)}+\dfrac{1}{2} \right]$, respectively, as the number of $N_s$ increases. We observe that for $N_{s} > 25$ the
asymptotic value compatible with zero within one standard deviation is reached.

\end{appendices}
\clearpage
\bibliography{lattice}

\end{document}